\documentclass[twoside,gimmsy2e]{article} 
\usepackage{amsmath}    
\usepackage{amsthm}     
\usepackage{amscd}      
\usepackage{amssymb}    
\usepackage{eucal}      
\usepackage{latexsym}   
\usepackage{graphicx}   
\usepackage{verbatim}   
\usepackage{gimmsy2e}   

\usepackage{epstopdf}
\makeatletter
\@addtoreset{equation}{subsection}
\def\thesubsection{\thesection\Alph{subsection}}
\def\thesection{\arabic{section}}

\def\theequation{\thesubsection.\arabic{equation}}

\makeatother

\def\intprod{\mathbin{\hbox to 6pt{%
                 \vrule height0.4pt width5pt depth0pt
                 \kern-.4pt
                 \vrule height6pt width0.4pt depth0pt\hss}}}

\let\hook\intprod

\newcommand{\ns}{\mspace{-1.5mu}}             
\newcommand{\ps}{\mspace{1.5mu}}              

\newcommand{\cc}{\mathcal C}

\newcommand{\ce}{\mathcal E}
\newcommand{\cf}{\mathcal F}
\newcommand{\cg}{\mathcal G}

\newcommand{\ci}{\mathcal I}
\newcommand{\cj}{\mathcal J}
\newcommand{\ck}{\mathcal K}
\newcommand{\cl}{\mathcal L}

\newcommand{\cn}{\mathcal N}

\newcommand{\cp}{\mathcal P}
\newcommand{\cq}{\mathcal Q}

\newcommand{\cs}{\mathcal S}

\newcommand{\cv}{\mathcal V}
\newcommand{\cw}{\mathcal W}

\newcommand{\cy}{\mathcal Y}
\newcommand{\cz}{\mathcal Z}

\newcommand{\eps}{\epsilon}
\newcommand{\varep}{\varepsilon}
\newcommand{\sig}{\sigma} 	
\newcommand{\Sig}{\Sigma}

	\newcommand{\Fe}{\mathfrak{E}}
	\newcommand{\Ff}{\mathfrak{F}}
\newcommand{\fg}{\mathfrak{g}}	
	\newcommand{\Fh}{\mathfrak{H}}
	
	\newcommand{\Fj}{\mathfrak{J}}

\newcommand{\fs}{\mathfrak{s}}

	\newcommand{\Fx}{\mathfrak{X}}

\begin{document}

\pagenumbering{roman}

\thispagestyle{empty}


\title
{\huge \bf Momentum Maps \\[.5ex] and \\[.5ex] Classical Fields \\[3pt]
{\Large\it Part II: Canonical Analysis of Field Theories}\\[1.5ex]}
\author{
{\bf Mark~J.~Gotay} 
\thanks{Research partially supported  
by NSF grants DMS 96-23083 and 00-72434.} 
\\[-2pt]
Department of Mathematics\\[-2pt]
University of Hawai`i\\[-2pt]
Honolulu, Hawai`i 96822, USA\\[-2pt]
gotay@math.hawaii.edu\\
\and 
{\bf James~Isenberg} 
\thanks{Partially supported by NSF grant PHY 00-99373.}
\\[-2pt]
Department of Mathematics\\[-2pt]
University of Oregon\\[-2pt]
Eugene, Oregon 97403, USA \\[-2pt]
jim@newton.uoregon.edu\\
\and 
{\bf Jerrold~E.~Marsden} 
\thanks{Partially supported by NSF grant DMS 02-04474.}
\\[-2pt]
Control and Dynamical Systems 107-81\\[-2pt] 
California Institute of Technology\\[-2pt]
Pasadena, California 91125, USA\\[-2pt]
marsden@cds.caltech.edu\\[12pt]
\and
\centerline{\rm With the collaboration of}
\and
{\bf Richard~Montgomery}
\and 
{\bf J\c edrzej  \'Sniatycki}
\and
{\bf Philip B.Yasskin}
}
\date{November 23, 2003; minor revisions, August 2004}

\thispagestyle{empty}

\maketitle

\newpage
\thispagestyle{empty}
\tableofcontents

\addtocontents{toc}{\protect\vspace{5ex}}

\newpage
\setcounter{page}{0}
\pagenumbering{arabic}
\pagestyle{myheadings}

 
\setcounter{section}{4}
\newpage

\newpage

\part*{\large\bfseries{\LARGE II}---{\LARGE C}ANONICAL 
{\LARGE A}NALYSIS OF {\LARGE F}IELD {\LARGE T}HEORIES}
\addcontentsline{toc}{part}{II---Canonical Analysis of Field
Theories}{\null}
\bigskip\bigskip

With the covariant formulation in hand from the first part of this
book, we begin in this second part to study the canonical (or
``instantaneous'') formulation of classical field
theories. The canonical formluation works with fields defined as
time-evolving cross sections of bundles over a Cauchy surface,
rather than as sections of bundles over spacetime as in the covariant
formulation. More precisely, for a given classical field theory, the
(infinite-dimensional) instantaneous configuration space consists of
the set $\mathcal{Y}_\Sigma$ of all smooth sections of a specified
bundle $Y_\Sigma$ over a Cauchy surface $\Sigma$, and a solution to
the field equations is represented by a trajectory in
$\mathcal{Y}_\Sigma$. As in classical mechanics, the Lagrangian
formulation of the field equations of a classical field theory is
defined on the tangent bundle $T\mathcal{Y}_\Sigma$, and the
Hamiltonian formulation is defined on the cotangent bundle
$T^{*}\mathcal{Y}_\Sigma$, which has a canonically defined symplectic
structure $\omega_\Sig$. 

To relate the canonical and the covariant approaches to
classical field theory, we start in Chapter 5 by discussing
embeddings
$\Sigma \rightarrow X$ of Cauchy surfaces in spacetime, and
considering the corresponding pull-back bundles $Y_\Sig \rightarrow
\Sigma$ of the covariant configuration bundle $Y \rightarrow X$. We go
on in the same chapter to relate the covariant multisymplectic
geometry of $(Z,\Omega)$ to the instantaneous symplectic geometry of
$(T^*\mathcal{Y}_\Sig, \omega_\Sig)$ by showing that the
multisymplectic form $\Omega$ on $Z$ naturally induces the symplectic
form $\omega_\Sig$ on $T^*\mathcal{Y}_\Sig$.

The discussion in Chapter 5 concerns primarily kinematical structures,
such as spaces of fields and their geometries, but does
not involve the action principle or the field equations for a given
classical field theory. In Chapter 6, we proceed to consider field
dynamics. A crucial feature of our discussion here is the degeneracy of
the Lagrangian functionals for the field theories of interest.  As a
consequence of this degeneracy, we have constraints on the choice of
initial data, and gauge freedom in the evolution of the fields.
Chapter 6 considers the role of initial value constraints and gauge
transformations in field dynamics. The
discussion is framed primarily in the Hamiltonian formulation of the
dynamics. 

One of the primary goals of this work is to show how momentum maps are
used in classical field theories which have both initial
value constraints and gauge freedom. In Chapter 7, we begin to do this
by describing how the covariant momentum maps defined on the multiphase
space $Z$ in Part I induce a generalization of momentum
maps---``energy-momentum maps''---on the instantaneous phase spaces
$T^*\mathcal{Y}_\Sig$. We show that for a group action which leaves the
Cauchy surface invariant, this energy-momentum map coincides with
the usual notion of a momentum map. We also show, when the gauge group
``includes'' the spacetime diffeomorphism group, that one of
the components of the energy-momentum map corresponding to spacetime
diffeomorphisms  can be identified (up to sign) with the Hamiltonian
for the theory.

\section[Symplectic Structures Associated with Cauchy 
Surfaces]{
Symplectic Structures Associated with\\ Cauchy 
Surfaces}

The transition from the covariant to the instantaneous formalism once a
Cauchy surface (or a  foliation by Cauchy surfaces) has been chosen is
a central ingredient of this work. It will eventually be used to  cast
the field dynamics into adjoint form and to determine when the first
class constraint set (in the  sense of Dirac) is the zero set of an
appropriate energy-momentum map.

\subsection{
Cauchy Surfaces and Spaces of Fields}

In any particular field theory, we assume there is singled out a class
of hypersurfaces  which we call {\bfi Cauchy surfaces\/}. We will not
give a precise definition here, but our usage of the  term is intended
to correspond to its meaning in general relativity (see, for instance,
Hawking  and Ellis [1973]).

Let $\Sig$ be a compact (oriented, connected) boundaryless
$n$-manifold. We denote by $\operatorname{Emb}(\Sig, X)$ the space of
all smooth embeddings of $\Sig$ into $X$. (If the ($n+1$)-dimensional
``spacetime'' $X$ carries a nonvariational  Lorentz metric,
we then understand $\operatorname{Emb}(\Sig, X)$ to be the
space of smooth {\it spacelike\/} embeddings of $\Sig$ into
$X$.)
As usual, many of the formal aspects of the constructions
also work in the noncompact context with asymptotic conditions
appropriate to the allowance of the necessary integrations by parts.
However, the analysis necessary to cover the noncompact case need not
be trivial; these considerations are important when dealing with
isolated systems or asymptotically flat spacetimes.  See Regge and
Teitelboim [1974], Choquet--Bruhat,  Fischer and Marsden [1979a],
\'Sniatycki [1988], and Ashtekar, Bombelli, and Reula [1991]. 

For $\tau \in \operatorname{Emb}(\Sig, X)$, let $\Sigma_\tau =
\tau(\Sig)$. The hypersurface $\Sigma_\tau$ will eventually be a
Cauchy surface for the dynamics; we view $\Sig$ as a  {\bfi
reference\/} or {\bfi model Cauchy surface\/}.  We will not need to
topologize $\operatorname{Emb}(\Sig, X)$ in this  paper; however, we
note that when completed in appropriate $C^k$ or Sobolev topologies,
$\operatorname{Emb}(\Sig,  X)$ and other manifolds of maps introduced
below are known to be smooth manifolds (see, for  example, Palais
[1968] and Ebin and Marsden [1970]).

\medskip If $\pi_{X\ns K} : K \to X$ is a fiber bundle over $X$,  then
the space of smooth sections  of the  bundle will be denoted by the
corresponding script letter, in this case $\ck$.  Occasionally, when
this notation might be confusing,  we will resort to the notation
$\Gamma(K)$ or $\Gamma(X, K)$. We let $K_\tau$ denote the restriction
of the bundle $K$ to $\Sigma_\tau \subset X$ and let the corresponding
script letter denote the space of its smooth sections, in this case
$\ck_\tau$. The collection of all $\ck_\tau$ as $\tau$ ranges over
$\operatorname{Emb}(\Sig, X)$ forms a bundle over
$\operatorname{Emb}(\Sig, X)$ which we will denote $\ck^\Sig$.

\begin{figure}[ht] \label{Gimmsy5-1}
\begin{center}
\includegraphics[scale=.95,angle=0]{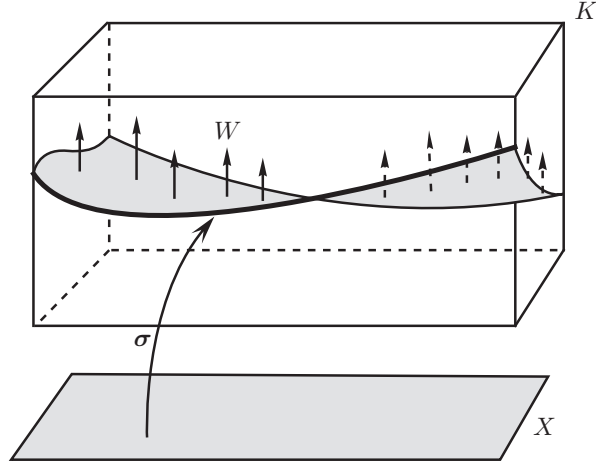}
\caption{A tangent vector $ W \in T _\sigma \mathcal{K} $}
\end{center}
\end{figure}

The {\bfi tangent space\/} to $\ck$ at a point $\sig$ is given by 
\begin{equation}\label{eqn5A:1}
T_\sigma\ck  =  \left\{W : X \to V\ns K  \bigm|  W  \text{ covers }
\sigma\right\}, 
\end{equation}
where $V\ns K$ denotes the vertical tangent bundle of $K$.  See Figure
5-1.
	       
Similarly, the {\bfi smooth cotangent space\/} to $\ck$ at $\sig$ is
\begin{equation}\label{eqn5A:2} 
T^*_\sigma \ck  =  \left\{\pi : X \to L(V\ns K, \Lambda^{n+1}X) 
\bigm| \pi \text{ covers } \sigma\right\},
\end{equation} 
where $L(V\ns K, \Lambda^{n+1}X)$ is the vector bundle over $K$ whose
fiber at $k \in K_x$ is the set of linear maps from $V_kK$ to
$\Lambda^{n+1}_xX$. The natural pairing  of $T^*_\sigma \ck$  with
$T_\sigma \ck$ is given by integration:
\begin{equation}\label{eqn5A:3}
\langle \pi, V \rangle  =  \int_X  \pi(V).  
\end{equation}  
One obtains similar formulas for $\ck_\tau$ from the above by
replacing $X$ with $\Sigma_\tau$ and $K$ with $K_\tau$ throughout
(and replacing $n + 1$ by $n$ in (\ref{eqn5A:2})).  See Figure 5-2.

\begin{figure}[ht]\label{Gimmsy5-2}
\begin{center}
\includegraphics[scale=1.05,angle=0]{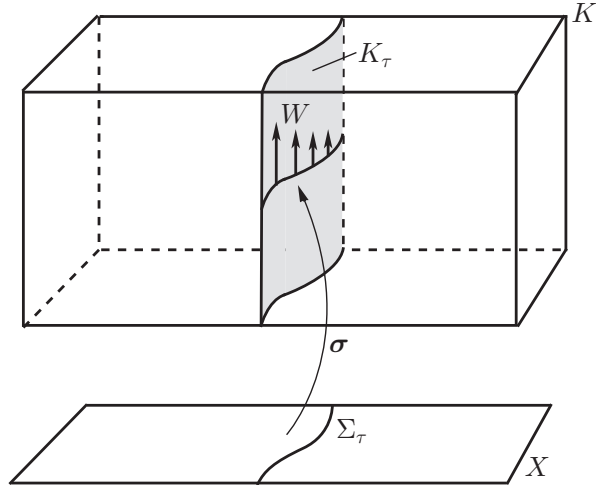}
\caption{A tangent vector $W \in T_\sigma \ck_\tau$}
\end{center}
\end{figure}	              

If $\xi_K$ is any $\pi_{X\ns K}$-projectable vector field on $K$, we
define the {\bfi Lie derivative\/} of $\sig \in \ck$ along $\xi_K$ to
be the element of $T_\sig\ck$ given by
\begin{equation}\label{eqn5A:5}
\pounds_{\xi_K}\sig = T\sig \circ \xi_X - \xi_K \circ \sig.
\end{equation} 
Note that $-\pounds_{\xi_K}\sig$ is exactly the vertical component of
$\xi_K \circ \sig$. In coordinates $(x^\mu,k^A)$ on $K$ we have
\begin{equation}\label{eqn5A:6} 
(\pounds_{\xi_K}\sig)^A = \sig^A{}_{,\mu} \xi^\mu - \xi^A \circ \sig,
\end{equation} 
where $\xi_K = (\xi^\mu,\xi^A).$

Finally, if $f$ is a map $\ck \rightarrow \cf(X)$ we define the
``formal'' partial derivatives $D_\mu f : \ck \rightarrow \cf(X)$ via
\begin{equation}\label{eqn5A:4}
D_\mu f (\sig) = f(\sig)_{,\mu}.
\end{equation}
Intrinsically, this is the coordinate representation of the 
differential of the real valued function $f (\sigma)$.

\subsection{
Canonical Forms on $T^*\cy_\tau$}

In the instantaneous formalism the configuration space at ``time" $\tau
\in \operatorname{Emb}(\Sig, X)$ will be  denoted $\cy_\tau$, hereafter
called the $\tau$-{\bfi configuration space\/}.  Likewise, the
$\tau$-{\bfi phase space\/} is$T^*\cy_\tau$, the smooth cotangent
bundle of $\cy_\tau$ with its canonical one-form $\theta_\tau$ and
canonical two-form $\omega_\tau$.  These forms are defined using the
same construction as for ordinary cotangent bundles (see Abraham and
Marsden [1978] or Chernoff and Marsden [1974]).  Specifically, we
define $\theta_\tau$ by
\begin{equation}
\theta_\tau(\varphi, \pi)(V)  =  \int_{\Sigma_\tau}
\pi(T\pi_{\cy_\tau,T^*\cy_\tau}\cdot V)  
\label{eqn5B:1}
\end{equation}
where $(\varphi, \pi)$ denotes a point in $T^*\cy_\tau$, $V \in
T_{(\varphi,\pi)} T^*\cy_\tau$ and
$\pi_{\cy_\tau,T^*\cy_\tau} : T^*\cy_\tau \to \cy_\tau$ is the
cotangent bundle projection.  We define 
\begin{equation}
\omega_\tau =  -\mathbf{d}\theta_\tau.  
\label{eqn5B:2}
\end{equation}

We now develop coordinate expressions for these forms. To this end
choose a chart$ \left(x^0, x^1, \dots,  x^n \right) $ on $X$  which
is {\bfi adapted\/} to $\tau$ in the sense that $\Sigma_\tau$ is
locally a level set of $x^0$.  Then an element $\pi \in T^*_\varphi
\cy_\tau$,  regarded as a map  $\pi~:~\Sigma_\tau~\to~L( V Y_\tau,
\Lambda^n\Sigma_\tau )$, is expressible as 
\begin{equation}
\pi  =  \pi_A\,dy^A \otimes d^{\ps n}\ns x_0,  
\label{eqn5B:3}
\end{equation}
so for the canonical one- and two-forms on $T^*\cy_\tau$ we get
\begin{equation}
\theta_\tau(\varphi, \pi)  =  \int_{\Sigma_\tau} \pi_A\,d\varphi^A 
\otimes d^{\ps n}\ns x_0
 \label{eqn5B:4}
\end{equation}
and
\begin{equation}
\omega_\tau(\varphi, \pi)  =  \int_{\Sigma_\tau} (d\varphi^A 
\wedge d\pi_A) \otimes  d^{\ps n}\ns x_0.  
\label{eqn5B:5}
\end{equation}
For example, if $V \in T_{(\varphi,\pi)}(T^*\cy_\tau) $ is given in
adapted coordinates by $V = (V^A, W_A)$, then we have
\begin{equation*}
\theta_\tau(\varphi, \pi)(V)  =  \int_{\Sigma_\tau} \pi_A V^A d^{\ps
n}\ns x_0.  
\label{eqn5B:6}
\end{equation*}

\subsection{
Presymplectic Structure on  $\cz_\tau$  }

To relate the symplectic manifold $T^*\cy_\tau$ to the multisymplectic
manifold $Z$, we first use the  multisymplectic structure on $Z$ to
induce a presymplectic structure on $\cz_\tau$ and then identify 
$T^*\cy_\tau$ with the quotient of $\cz_\tau$ by the kernel of this
presymplectic form.  Specifically, define the {\bfi canonical
one-form\/} $\Theta_\tau$ on $\cz_\tau$ by
\begin{equation}
\Theta_\tau(\sigma)(V)  = 
\int_{\Sigma_\tau}\sigma^*(\mathbf{i}_V\Theta),  
\label{eqn5C:1}
\end{equation}
where $\sigma \in \cz_\tau$, $V \in T_\sigma \cz_\tau$, and $\Theta$ is
the canonical $(n + 1)$-form on $Z$ given by (2B.9). The {\bfi
canonical two-form\/} $\Omega_\tau$ on $\cz_\tau$ is 
\begin{equation}
\Omega_\tau =  - \mathbf{d}\Theta_\tau.  
\label{eqn5C:2}
\end{equation}

\begin{lem}\label{lem5.1} 
At $\sigma \in \cz_\tau$ and with $\Omega$ given by 
{\rm  (2B.10)}, we have
\begin{equation} 
\Omega_\tau(\sigma)(V, W)  =  \int_{\Sigma_\tau}
\sigma^*(\mathbf{i}_W\mathbf{i}_V\Omega).  
\label{eqn5C:3}
\end{equation}
\end{lem}

\begin{proof} 
Extend $V, W$ to vector fields $\cv, \cw$ on $\cz_\tau$ by fixing
$\pi_{X\ns Z}$-vertical vector fields $v, w$ on $Z_\tau$ such that $V
= v \circ \sig$ and $W = w \circ \sig$ and letting $\cv(\rho) = v
\circ \rho$ and $\cw(\rho) = w \circ \rho$ for $\rho \in \cz_\tau$. 
Note that if $f_\lambda$ is the flow of $w$, $\cf_\lambda(\rho) =
f_\lambda \circ \rho$ is the flow of $\cw$. Then, from the definition 
of the bracket in terms of flows, one finds that
$$ 
[\cv, \cw](\rho)  = [v, w] \circ \rho.
$$

The derivative of $\Theta_\tau(\cv)$ along $\cw$ at $\sig$ is
$$
\begin{aligned}
\cw \left[ \Theta_\tau(\cv) \right](\sigma) 
& = \left. 
\frac{d}{d\lambda}
\left[\Theta_\tau (\cv) \circ 
\cf_\lambda(\sigma)\right]\right|_{\lambda=0} 
  = \left.\frac{d}{d\lambda} 
\left[ \int_{\Sigma_\tau}
\cf_\lambda(\sigma)^*(\mathbf{i}_v\Theta)\right]\right|_{\lambda=0} 
\\[2ex]
& = \left.\frac{d}{d\lambda} \left[
\int_{\Sigma_\tau} 
\sigma^*f^*_\lambda(\mathbf{i}_v\Theta)\right]\right|_{\lambda=0} 
=
\int_{\Sigma_\tau} \sigma^*[\pounds_w\mathbf{i}_v\Theta].
\end{aligned}
$$ 
Thus, at $\sigma \in \cz_\tau$,
$$
\begin{aligned}
\mathbf{d}\Theta_\tau(\cv,\cw) 
& = \cv\left[\Theta_\tau(\cw)] - \cw[\Theta_\tau(\cv)] -
    \Theta_\tau([\cv, \cw\right]) \\[2ex] 
& = \int_{\Sigma_\tau} \sigma^*\left[\pounds_v\mathbf{i}_w\Theta - 
        \pounds_w\mathbf{i}_v\Theta - \mathbf{i}_{[v, w]}
        \Theta\right]  \\[2ex]  
& =  \int_{\Sigma_\tau} \sigma^*(- \mathbf{d}\mathbf{i}_w\mathbf{i}_v
      \Theta + \mathbf{i}_w\mathbf{i}_v \mathbf{d}\Theta),
\end{aligned}
$$ 
and the first term vanishes by the definitions of $Z$ and $\Theta$, as
both $v,w$ are $\pi_{X\ns Z}$-vertical.\footnote{\ This term also
vanishes by Stokes' theorem, but in fact (\ref{eqn5C:3}) holds
regardless of whether $\Sig_\tau$ is compact and boundaryless.}
\end{proof}

\medskip 
The two-form $\Omega_\tau$ on $\cz_\tau$ is closed, but it has a
nontrivial kernel, as the following  development will show.
	
\subsection{
Reduction of $\cz_\tau$ to $T^*\cy_\tau$}

Our next goal is to prove that $\cz_\tau /\ker \Omega_\tau $ is
canonically isomorphic to $T^*\cy_\tau$ and that  the inherited
symplectic form on the former is isomorphic to the canonical one on the
latter.  To do  this, define a vector bundle map $R_\tau : \cz_\tau \to
T^*\cy_\tau$ over $\cy_\tau$ by
\begin{equation} 
\left\langle R_\tau(\sigma), V \right\rangle =  \int_{\Sigma_\tau}
\varphi^*(\mathbf{i}_V
\sigma),  
\label{eqn5D:1}
\end{equation}
where $\varphi  =  \pi_{Y\! Z} \circ \sig$ and $V \in
T_\varphi\cy_\tau$;  the integrand in (\ref{eqn5D:1}) at a point
$x \in \Sigma_\tau$ is the interior product of $V(x)$ with $\sig(x)$,
resulting in an $n$-form on $Y$,  which is then pulled back along
$\varphi$ to an $n$-form on $\Sigma_\tau$ at $x$. Interpreted as a map
of $\Sigma_\tau$ to $L(VY_\tau, \Lambda^n\Sigma_\tau )$ which covers
$\varphi$, $R_\tau(\sigma)$ is given by
\begin{equation} 
\left\langle R_\tau(\sigma)(x), v \right\rangle  = 
\varphi^*\mathbf{i}_v\sig(x),  
\label{eqn5D:2}
\end{equation}
where $v \in V_{\varphi(x)}Y_\tau$.  In adapted coordinates, 
$\sigma \in \cz_\tau$ takes the form
\begin{equation}
(p_A{}^\mu \circ \sig)\,dy^A \wedge d^{\ps n}\ns x_\mu + (p \circ
\sig)\,d^{\ps n+1}\ns x,  
\label{eqn5D:3}
\end{equation}
and so we may write
\begin{equation} 
R_\tau(\sigma)  =  (p_A{}^0 \circ \sig)\,dy^A \otimes d^{\ps n}\ns x_0.  
\label{eqn5D:4}
\end{equation}
Comparing (\ref{eqn5D:4}) with (\ref{eqn5B:3}), we see that the {\bfi
instantaneous momenta} $\pi_A$ correspond to the temporal  components
of the multimomenta $p_A{}^\mu$.  Moreover, $R_\tau$ is obviously a
surjective submersion with
\pagebreak
\begin{equation*}
\ker R_\tau = \left\{ \sig \in \cz_\tau \mid p_A{}^0 \circ \sig = 0
\right\}.  
\label{eqn5D:5}
\end{equation*}
\begin{remark}[Remark]
Although we have defined $R_\tau$ as a map on
sections from $\cz_\tau$ to $T^*\cy_\tau$, in actuality $R_\tau$ is a
pointwise operation. We may in fact write (\ref{eqn5D:2}) as
$R_\tau(\sigma)  =  r_\tau \circ \sigma  $, where
\begin{equation*} 
r_\tau : Z_\tau \to V^*Y_\tau \otimes \Lambda^n\Sigma_\tau 
\label{eqn5D:7}
\end{equation*}
is a bundle map over $Y_\tau$.  From (\ref{eqn5D:3}) and
(\ref{eqn5D:4}), we see that in coordinate form $r_\tau(p,p_A{}^\mu) 
=  p_A{}^0$ with
\begin{equation*}
\ker r_\tau = \left\{p_A{}^i  dy^A \otimes d^{\ps n}\ns x_i +
p \,d^{\ps n+1}\ns x \in Z_\tau\right\}.  
\label{eqn5D:5}\tag*{$\blacklozenge$}
\end{equation*}
\renewcommand{\lozsymbol}{}
\end{remark}
\begin{prop}\label{prop5.2}
We have
\begin{equation}
R^*_\tau\theta_\tau = \Theta_\tau. 
\label{eqn5D:9}
\end{equation}
\end{prop}

\begin{proof} 
Let $V \in T_\sigma\cz_\tau$.  By the definitions of pull-back
and the canonical one-form,
$$ 
\left\langle (R^*_\tau\theta_\tau) (\sigma), V \right\rangle =
\left\langle\theta_\tau(R_\tau(\sigma)), TR_\tau\cdot V\right\rangle =
\langle R_\tau(\sigma), T\pi_{\cy_\tau,T^*\cy_\tau}
\cdot TR_\tau\cdot V\rangle.
$$ 
However, since $R_\tau$ covers the identity,
$$
\pi_{\cy_\tau,T^*\cy_\tau} \circ R_\tau =  \pi_{\cy_\tau,\cz_\tau} 
$$ 
and so
\begin{equation*}
T\pi_{\cy_\tau,T^*\cy_\tau} \cdot TR_\tau\cdot V  = 
T\pi_{\cy_\tau,\cz_\tau}\cdot V  =  T\pi_{Y\! Z} \circ V.
\end{equation*}
Thus by (\ref{eqn5D:1}), with $\varphi = \pi_{Y\! Z} \circ \sig$,
\begin{align*}
\left\langle R^*_\tau\theta_\tau (\sigma), V \right\rangle
& = \langle R_\tau(\sigma),
T\pi_{Y\! Z}
\circ V\rangle = \int_{\Sigma_\tau}
\varphi^*((T\pi_{Y\! Z} \circ V)  \hook \sigma) \\[2ex] 
& =  \int_{\Sigma_\tau}
\sigma^*\pi^*_{Y\! Z}((T\pi_{Y\! Z} \circ V) \hook \sigma) 
\\[2ex]
&=  \int_{\Sigma_\tau}
\sigma^*(V \hook\,\pi^*_{Y\! Z}\sigma). \\[-1.5ex]
\end{align*}
However, by (2B.7) and (2B.9), $\pi^*_{Y\! Z}\sigma = \Theta \circ
\sig$.  Thus by (\ref{eqn5C:1}),
\begin{equation}
\left\langle R^*_\tau\theta_\tau (\sigma), V \right\rangle
=  \left\langle\Theta_\tau(\sigma), V\right\rangle.
\tag*{\qed}
\end{equation}
\renewcommand{\qedsymbol}{}
\end{proof}
\pagebreak
\begin{cor}\label{cor5.3}\mbox{}
\begin{enumerate}
\renewcommand{\labelenumi}{\mbox{\rm(\roman{enumi})}}
\item 
 $R^*_\tau \omega_\tau = \Omega_\tau$.
\item 
 $\ker T_\sigma R_\tau = \ker \Omega_\tau(\sigma)$.
\item 
 The induced quotient map $\cz_\tau / \ker R_\tau = \cz_\tau / 
\ker \Omega_\tau \to T^*\cy_\tau$ is a symplectic diffeomorphism.
\end{enumerate}
\end{cor}

\begin{proof} 
(i) follows by taking the exterior derivative of (\ref{eqn5D:9}).  
(ii) follows from (i),  the (weak) nondegeneracy of
$\omega_\tau$, the definition of pull-back and the fact that $R_\tau$
is a submersion. Finally, (iii)   follows from  (i), (ii), and the fact
that $R_\tau$ is a surjective vector bundle map between vector bundles
over $\cy_\tau$. 
\end{proof}

Thus, for each Cauchy surface $\Sigma_\tau$, the multisymplectic
structure $\Omega$ on $Z$ induces a  presymplectic structure
$\Omega_\tau$ on $\cz_\tau$, and this in turn induces the canonical
symplectic structure $\omega_\tau$ on the instantaneous phase space
$T^*\cy_\tau$. Alternative constructions of $\Theta  _\tau $ and $
\omega _\tau $ are given in  Zuckerman [1986], Crnkovi\'{c} and Witten
[1987],  and Ashtekar, Bombelli, and Reula [1991].  
	
\vskip 6pt
\startrule
\vskip-12pt
            \addcontentsline{toc}{subsection}{Examples}
\begin{examples}
\mbox{}

\paragraph{\bf a\ \; Particle Mechanics.}\enspace 
For particle mechanics $\Sig$ is a point, and $\tau$ maps $\Sig$ to
some $t \in \mathbb R$.  We identify $\cy_\tau$ with $Q$ and
$\cz_\tau$ with $\mathbb R \times T^*Q$, with coordinates $(q^A, p,
p_A)$.  The one-form $\theta_\tau$ is $\theta_\tau = p_Adq^A$ and
$R_\tau$ is given by $(q^A, p, p_A) \mapsto (q^A, p_A)$.  Thus  the
$\tau$-phase space is just $T^*Q$, and the process of reducing the
multisymplectic formalism to the  instantaneous formalism in particle
mechanics is simply reduction to the autonomous case.

\paragraph{\bf b\ \; Electromagnetism.}\enspace 
In the case of electromagnetism, $\Sig$ is a 3-manifold and $\tau \in 
\operatorname{Emb}(\Sig, X)$ is a 
parametrized spacelike hypersurface.  The space $\cy_\tau$ consists of 
fields $A_\nu$ over $\Sigma_\tau$, $T^*\cy_\tau$ consists of fields
and their conjugate momenta $(A_\nu, \Fe^\nu)$ on $\Sigma_\tau$, 
while the space $\cz_\tau$ consists of fields and multimomenta fields
$(A_\nu, p, \Ff^{\nu\mu})$ on $\Sigma_\tau$.  In adapted coordinates
the map $R_\tau$ is given by
\begin{equation} 
(A_\nu, p, \Ff^{\nu\mu}) \mapsto  (A_\nu, \Fe^\nu),   
\label{ex5D:b1} 
\end{equation}
where $\Fe^\nu = \Ff^{\nu 0}$. The canonical momentum $\Fe^\nu$ can
thus be identified with the negative of the electric field density.
The symplectic structure on $T^*\cy_\tau$ takes the form
\begin{equation}
\omega_\tau(A, \Fe )  =  \int_{\Sigma_\tau} (dA_\nu \wedge d\Fe^\nu) 
\otimes d^{\ps 3}\ns x_0.   
\label{ex5D:b2} 
\end{equation}
\medskip

When electromagnetism is parametrized, we simply append the metric $g$
to to the other field variables as a parameter.  Let $ S _2 ^{\ps 3,
1}(X,\Sig_\tau)$ denote the subbundle of $S _2 ^{\ps 3, 1}(X)$
consisting of Lorentz metrics relative to which $\Sig_\tau$ is
spacelike. Thus we replace 
$ \cy _\tau$ by 
$$ 
\tilde{\cy} _\tau = \cy _\tau \times \cs _2 ^{3, 1}(X,\Sig_\tau)
_\tau ,
$$ 
which consists of sections $ (A; g) $ of $ \tilde{Y} = Y
\times S _2 ^{\ps 3, 1}(X,\Sig_\tau) $ over $ \Sigma _\tau $.
Similarly, we replace $\cz _\tau$ by 
$$ 
\cz_\tau \times \cs _2 ^{3, 1}(X,\Sig_\tau) _\tau, \
$$ 
etc. The metric just gets carried along by $R _\tau $ in
(\ref{ex5D:b1}), and the expression (\ref{ex5D:b2}) for $ \omega _\tau
$ remains unaltered.

\paragraph{\bf c\ \; A Topological Field Theory.}\enspace 
Since in a topological field theory there is no metric on
$X$, it does not make sense to speak of ``spacelike hypersurfaces''
(although we shall continue to informally refer to
$\Sigma_{\tau}$ as a ``Cauchy surface''). Thus we may take $\tau$ to
be {\em any\/} embedding of $\Sigma$ into $X$.

Other than this, along with the fact that $\Sigma$ is 2-dimensional,
Chern--Simons theory is much the same as electromagnetism.
Specifically, ${\mathcal Y}_{\tau}$ consists of fields $A_{\nu}$ over
$\Sigma_{\tau}$, $T^*{\mathcal Y}_{\tau}$ consists of fields and their
conjugate momenta $(A_{\nu},\pi^{\nu})$ over $\Sigma_{\tau}$, and
${\mathcal Z}_{\tau}$ consists of fields and their multimomenta
$(A_{\nu},p,p^{\nu\mu})$ over $\Sigma_{\tau}$. Then
$R_{\tau}$ and $\omega_{\tau}$ are given by
\begin{equation}
(A_{\nu},p,p^{\nu\mu}) \mapsto (A_{\nu},\pi^{\nu})
\label{eq:ex5D:d1}
\end{equation}
and
\begin{equation}
\omega_{\tau}(A_{\nu},\pi^{\nu}) = \int_{\Sigma_{\tau}}(dA_{\nu}
\wedge d\pi^{\nu}) \otimes d^{\ps 2}\ns x_0
\label{eq:ex5D:d2}
\end{equation}
respectively, where $\pi^{\nu} = p^{\nu 0}$.

\paragraph{\bf d\ \; Bosonic Strings.}\enspace 
  Here $\Sig$ is a 1-manifold and $\tau \in
\operatorname{Emb}(\Sig, X)$ is  a param\-etr\-ized curve in
$X$.  Since $Y = (X \times M) \times_X S^{1,1}_2(X)$,
$\cy_\tau$ consists of fields $(\varphi^A, h_{\sigma\ns\rho})$ over
$\Sigma_\tau$, $T^*\cy_\tau$ consists of fields and their conjugate
momenta
$(\varphi^A, h_{\sigma\ns\rho}, \pi_A, \rho^{\sigma\ns\rho})$, and
$\cz_\tau$ consists of  fields and their multimomenta 
$(\varphi^A,
h_{\sigma\ns\rho}, p, p_A{}^\mu, q^{\sigma\ns\rho \mu})$. In adapted
coordinates, the map
$R_\tau$ is
\begin{equation}\label{ex5D:e1} 
(\varphi^A, h_{\sigma\ns\rho}, p, p_A{}^\mu, q^{\sigma\ns\rho \mu}) 
\mapsto  (\varphi^A, h_{\sigma\ns\rho}, \pi_A,
\rho^{\sigma\ns\rho})   
\end{equation}
where $\pi_A = p_A{}^0$ and $\rho^{\sigma\ns\rho} = q^{\sigma\ns\rho \ps
0}$.  The symplectic form on $T^*\cy_\tau$ is then
\begin{equation}\label{ex5D:e2} 
\quad\qquad \omega_\tau(\varphi, h, \pi,\rho)  =  \int_{\Sigma_\tau}
(d\varphi^A \wedge d\pi_A + dh_{\sigma\ns\rho} \wedge
d\rho^{\sigma\ns\rho}) \otimes d^{\ps 1}\ns x_0. \qquad \blacklozenge
\end{equation}
\renewcommand{\lozsymbol}{}
\end{examples}

\section{
Initial Value Analysis of Field Theories}

In the previous chapter we showed how to space + time decompose
multisymplectic  structures.  Here we perform a similar decomposition
of dynamics using the notion of slicings.   This material puts the
standard initial value analysis into our context, with a few
clarifications  concerning how to intrinsically split off the time
derivatives of fields in the passage from the  covariant to the
instantaneous pictures.  A main result of this chapter is that the
dynamics is  compatible with the space + time decomposition in the
sense that Hamiltonian dynamics in the  instantaneous formalism
corresponds directly to the covariant Lagrangian dynamics of Chapter 3;
see \S 6D.  We also discuss a symplectic version of the Dirac--Bergmann
treatment of degenerate  Hamiltonian systems, initial value
constraints, and gauge transformations in \S 6E.

\subsection{
Slicings}

To discuss dynamics, that is, how fields evolve in time, we define a
global notion of ``time."   This is accomplished by introducing
``slicings" of spacetime and the relevant bundles over it.

A {\bfi slicing} of an $(n + 1)$-dimensional spacetime $X$ consists of
an $n$-dimensional manifold $\Sig$ (sometimes known as a {\bfi
reference Cauchy surface\/}) and a diffeomorphism
$$
\fs_X : \Sig \times \mathbb R  \to  X. 
$$ 
For $\lambda \in \mathbb R$, we write $\Sigma_\lambda = \fs_X(\Sig
\times \{\lambda\})$ and $\tau_\lambda : \Sig \to \Sigma_\lambda
\subset X$ for the embedding defined by $\tau_\lambda(x) =
\fs_X(x,\lambda) $.  See Figure 6-1. The slicing parameter
$\lambda$ gives rise to a global notion of ``time" on $X$ which need
not coincide with locally defined coordinate time, nor with proper time
along the curves $\lambda \mapsto \fs_X(x,\lambda) $.  The {\bfi
generator\/} of $\fs_X$ is the vector field $\zeta_X$ on $X$ defined by
$$
\frac{\partial}{\partial\lambda} \fs_X(x,\lambda)  
= \zeta_X(\fs_X(x,\lambda) ).
$$ 
Alternatively, $\zeta_X$ is the push-forward by $\fs_X$ of the
standard vector field ${\partial}/{\partial\lambda}$ on $\Sig \times
\mathbb R$;  that is,
\begin{equation} \label{eqn6A:1} 
\zeta_X = T\fs_X\cdot \frac{\partial}{\partial\lambda}. 
\end{equation}

\begin{figure}[ht]\label{Gimmsy6-1}
\begin{center}
\includegraphics[scale=1.0,angle=0]{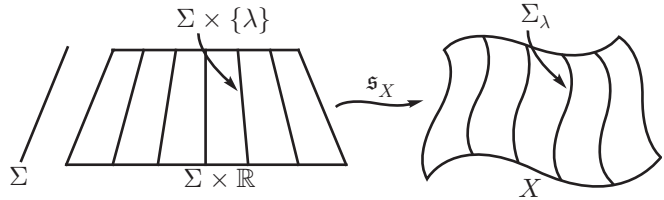}
\caption {A slicing of spacetime}
\end{center}
\end{figure}	 

Given a bundle $K \to  X$ and a slicing $\fs_X$ of $X$,   a {\bfi
compatible slicing\/} of $K$ is a  bundle $K_\Sig  \to  \Sig$  and a
bundle diffeomorphism $\fs_K : K_\Sig \times \mathbb R \to K$ such
that the diagram
\begin{equation} 
\label{eqn6A:2} 
\begin{CD} 
K_\Sig \times \mathbb R @>{\fs_K}>> K \\ @VVV @VVV \\
\Sig \times \mathbb R @>{\fs_X}>> X
\end{CD}
\end{equation}
commutes, where the vertical arrows are bundle projections.  We write
$K_\lambda = \fs_K(K_\Sig \times \{\lambda\})$ and $\fs_\lambda :
K_\Sig \to K_\lambda \subset K$ for the embedding defined by
$\fs_\lambda(k) = \fs_K(k,\lambda) $, as in Figure 6-2.  The
generating vector field $\zeta_K$ of $\fs_K$ is defined by a formula
analogous to (6A.1).  Note that $\zeta_K$ and $\zeta_X$ are complete
and everywhere transverse to the slices $K_\lambda$ and
$\Sigma_\lambda$, respectively.

\begin{figure}[ht] \label{Gimmsy6-2}
\begin{center}
\includegraphics[scale=1,angle=0]{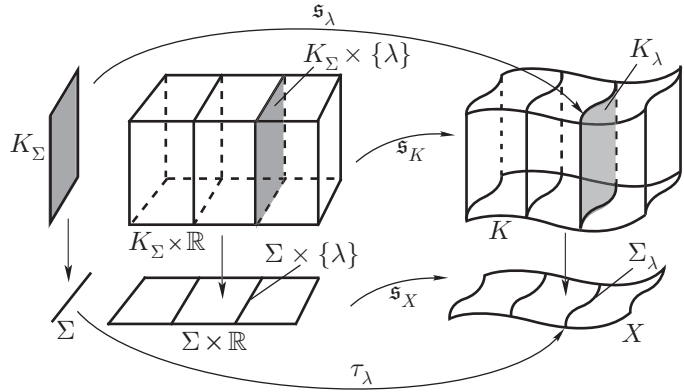}
\caption{A slicing of the bundle $K$}
\end{center}
\end{figure}	 

Every compatible slicing $(\fs_K, \fs_X)$ of $K \to X$ defines a
one-parameter group of bundle  automorphisms:  the flow $f_\lambda$ of
the generating vector field $\zeta_K$, which is given by
\begin{equation*} \label{eqn6A:3}
f_\lambda(k) = \fs_K(\fs_K^{-1}(k) + \lambda) , 
\end{equation*}
where ``$+ \;\lambda$" means addition of $\lambda$ to the second
factor of $K_\Sig \times \mathbb R$.  This flow is fiber-preserving
since $\zeta_K$ projects to $\zeta_X$.  Conversely, let $f_\lambda$
be a fiber-preserving flow on $K$ with generating vector field
$\zeta_K$.  Then $\zeta_K$ along with a choice of Cauchy surface
$\Sigma_\tau$ such that $\zeta_X \pitchfork \Sigma_\tau$ determines
(at least in a neighborhood of $K_\tau$ in $K$) a slicing $\fs_K :
K_\tau \times \mathbb R \to K$ according to $\fs_K(k,\lambda)  =
f_\lambda(k)$.  Any other slicing corresponding to the above data
differs from this $\fs_K$ by a diffeomorphism.

\medskip Slicings of bundles give rise to trivializations of associated
spaces of sections. Given $K \to  X$,  recall from  \S 5A that we have
the bundle
$$
\ck^\Sig = \bigcup_{\tau \in \operatorname{Emb}(\Sig,X)}  \ck_\tau 
$$ 
over $\operatorname{Emb}(\Sig, X)$, where $\ck_\tau$ is the space of
sections of $K_\tau = K\ns\! \bigm| \!\Sigma_\tau$.  Let
$\ck^\tau$  denote the portion of $\ck^\Sig$ that lies over the curve
of embeddings $\lambda \mapsto \tau_\lambda$, where  $\lambda \in
\mathbb R$.   In other words, 
\begin{equation*} \label{eqn6A:4}
\ck^\tau =  \bigcup_{\lambda \in \mathbb R} \ck_\lambda. 
\end{equation*}
The slicing $\fs_K : K_\Sig \times \mathbb R \to K$ induces a
trivialization $\fs_\ck : \ck_\Sig \times \mathbb R \to \ck^\tau$
defined by
\begin{equation}\label{eqn6A:5}  
\fs_\ck (\sig_\Sig,\lambda)   =  \fs_\lambda \circ \sig_\Sig \circ 
\tau_\lambda^{-1}.
\end{equation}
Let $\zeta_\ck$ be the pushforward of
${\partial}/{\partial\lambda}$ by means of this trivialization;
then from (\ref{eqn6A:5}),
\begin{equation} \label{eqn6A:6} 
\zeta_\ck(\sigma)  =  \zeta_K \circ \sig.  
\end{equation}
See Figure 6-3.
\begin{figure}[ht] \label{Gimmsy6-3}
\begin{center}
\includegraphics[scale=1.1,angle=0]{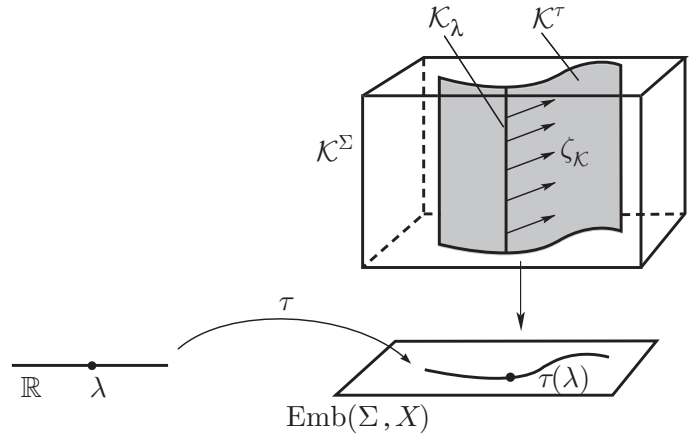}
\caption {Bundles of spaces of sections}
\end{center}
\end{figure}	 

\begin{remarks}[Remarks \ 1.]
A slicing $\fs_X$ of $X$ gives rise to at least one compatible
slicing $\fs_K$ of any bundle $K \to X$, since $X \approx \Sig \times
\mathbb R$ is then homotopic to $\Sig$.

\paragraph{\bf 2.}
 In many examples, $Y$ is a tensor bundle over $X$, so $\fs_Y$
can naturally be induced by a  slicing $\fs_X$ of $X$.  Similarly, in
Yang--Mills theory, slicings of the connection bundle are  naturally
induced by slicings of the theory's principal bundle.

\paragraph{\bf 3.}
 Slicings of the configuration bundle $Y \to X$ naturally induce
slicings of certain bundles  over it.  For example, a slicing $\fs_Y$
of $Y$ induces a slicing $\fs_Z$ of $Z$ by push-forward; if  
$\zeta_Y$ generates $\fs_Y$,  then $\fs_Z$ is generated by the
canonical lift $\zeta_Z$ of $\zeta_Y$  to $Z$.  (As a consequence,
$\pounds_{\zeta_Z}\Theta = 0$.)  Likewise, a slicing of $J^{1}Y$  is
generated by the jet prolongation $\zeta_{J^1 Y }= j^1\zeta_Y$ of
$\zeta_Y$  to $J^{1}Y$.

\paragraph{\bf 4.}
 When considering certain field theories, one may wish to modify these
constructions  slightly. In gravity, for example, one considers only
those pairs of metrics and slicings for  which each $\Sigma_\lambda$ is
spacelike. 
This is an open and invariant condition  and so the
nature of the construction is not materially changed.  

\paragraph{\bf 5.}
 It may happen that $X$ is sufficiently complicated topologically that
it cannot be globally  split as $\Sig \times \mathbb R$ for any
$\Sig$.  In such cases one can only slice portions of spacetime and
our constructions must be understood in a restricted sense.  However,
for globally hyperbolic  spacetimes, a well-known result of Geroch
(see Hawking and Ellis [1973]) states that $X$  is indeed
diffeomorphic to $\Sig \times \mathbb R$.   

\paragraph{\bf 6.}
 Sometimes one wishes to allow curves of embeddings that are not
slicings. (For instance, one could allow two embedded hypersurfaces to
intersect.) It is known by  direct calculation that the  adjoint
formalism (see Chapter 13) is valid even for curves of  embeddings that
are associated with maps $\fs$ that need not be diffeomorphisms. See,
for  example, Fischer and Marsden [1979a].

\paragraph{\bf 7.}
 In the instantaneous formalism, dynamics is usually studied relative
to a fixed slicing of  spacetime and the bundles over it.  It is
important to know to what extent the dynamics is  the ``same" for all
possible slicings.  To this end we introduce in Part IV  fiducial 
models of  all relevant objects which are {\it universal\/} for all
slicings in the sense that one can work  abstractly on the fixed model
objects and then transfer the results to the spacetime context  by
means of a slicing.  This provides a natural mechanism for comparing
the results  obtained by using different slicings.

\paragraph{\bf 8.}
 In practice, the one-parameter group of automorphisms of the
configuration bundle $Y$  associated to a slicing is often induced by a
one-parameter subgroup of the gauge group $\cg$   of the theory;  let
us call such slicings $\cg$-{\bfi slicings\/}. In fact, later we will
focus on slicings which arise in this way via the gauge group action. 
For $\cg$-slicings we have $\zeta_Y = \xi_Y$ for some $\xi \in \fg$. 
This provides a crucial link between dynamics and the gauge group, and
will ultimately enable us in \S 7F to correlate the Hamiltonian with
the energy-momentum  map for the gauge group action.  For classical
fields propagating on a fixed background  spacetime, it is necessary
to treat the background metric parametrically---so that $\mathcal{G}$
projects  {\it onto\/} $\operatorname{Diff}(X)$---to obtain such
slicings. (See Remark 1 in \S 8A.)

\paragraph{\bf 9.}
 For some topological field theories, there is a subtle interplay
between the existence of a slicing of spacetime and that of a
symplectic structure on the space of solutions of the field
equations.  See Horowitz [1989] for a discussion.


\paragraph{\bf 10.}
 Often slicings  of $X$ are arranged to implement certain ``gauge
conditions" on the fields.  For example, in  Maxwell's theory one may
choose a slicing relative to which the Coulomb gauge
condition $\nabla {\bf \cdot} {\bf A} = 0$ holds.  In general
relativity, one often chooses a slicing of a given spacetime so that
each  hypersurface
$\Sigma_\lambda$ has constant mean curvature.  
This can be accomplished by solving the adjoint
equations (1.3) together with the gauge conditions, which  will
simultaneously generate a slicing of spacetime and a
solution of the field equations, with the solution 
``hooked'' to the slicing via the gauge condition. Note that in this
case the slicing is not predetermined (by specifying the atlas fields
$\alpha_i(\lambda)$ in advance), but rather is determined implicitly
(by fixing the $\alpha_i(\lambda)$ by means of the adjoint equations
together with the gauge conditions.)

\paragraph{\bf 11.} In principle slicings can be choosen
arbitrarily, not necessarily according to a given \emph{a priori} rule.
For example, in numerical relativity, to achieve certain accuracy
goals, one may wish to choose slicings that focus on those regions in
which the fields that have been computed up to that point have large
gradients, thereby effectively using the slicing to produce an
adaptive numerical method. In this case, the slicing is determined
``on the fly'' as opposed to being fixed \emph{ab initio}. Of
course, after a piece of spacetime is constructed, the slicing produced
is consistent with our definitions. 
\end{remarks}

For a given field theory, we say that a slicing $\fs_Y$ of the
configuration bundle $Y$ is  {\bfi Lagrangian\/}  if the Lagrangian
density  $\cl$ is equivariant with respect to the one-parameter 
groups of automorphisms associated to the induced slicings of $J^{1}Y$ 
and $\Lambda^{n+1}X$.  Let $f_\lambda$ be the flow of $\zeta_Y$ so that
$j^1\! f_\lambda $ is the flow of $\zeta_{J^1Y}$; then equivariance
means
\begin{equation}\label{eqn6A:7} 
\cl(j^1\! f_\lambda (\gamma)) = (h^{-1}_\lambda) ^*\cl(\gamma) 
\end{equation}
for each $\lambda \in \mathbb R$ and $\gamma \in J^1Y$,  where
$h_\lambda$ is the flow of
$\zeta_X$. Throughout the rest of this paper we will assume:

\paragraph{\bf A2 \ Lagrangian Slicings}\enspace
{\it For a given configuration bundle $Y$ and a given
Lagrangian density $\cl$ on $Y$, there exists
a Lagrangian slicing of $Y$.}\\

From now on ``slicing'' will mean ``Lagrangian slicing''.  In practice
there are usually many  such slicings.  For example, in tensor
theories, slicings of $X$ induce slicings of $Y$ by pull-back;  these
are automatically Lagrangian as long as a  metric $g$ on spacetime is
included as a field variable (either variationally or parametrically).
For theories on a fixed spacetime background, on the other hand, a
slicing of $Y$ typically will be Lagrangian only if the flow generated
by $\zeta_X$ consists of isometries of $(X, g)$.  Since $(X, g)$ need
not have any continuous isometries, it may be necessary to treat $g$
parametrically to satisfy {\bf A2}.  Note that by virtue of the
covariance assumption {\bf A1}, $\mathcal{G}$-slicings are
automatically Lagrangian. (See, however, Example \textbf{c} 
following.) This requirement will play a key role in establishing the
correspondence between dynamics in the covariant and
$(n + 1)$-formalisms.
\medskip

For certain constructions we require only the notion of an {\bfi
infinitesimal slicing\/} of a spacetime $X$. This consists of a Cauchy
surface $\Sigma_\tau$ along with a  spacetime vector field $\zeta_X$
defined over $\Sigma_\tau$ which is everywhere transverse to
$\Sigma_\tau$.  We think of  $\zeta_X$ as defining a ``time direction"
along $\Sigma_\tau$.  In the same vein, an  {\bfi infinitesimal
slicing\/} of a bundle $K \to X$ consists of $K_\tau$ along with a
vector field $\zeta_K$ on $K$ defined over $K_\tau$ which is
everywhere transverse to $K_\tau$.  The infinitesimal slicings
$(\Sigma_\tau, \zeta_X)$ and $(K_\tau,\zeta_K)$ are called {\bfi
compatible\/} if $\zeta_K$ projects to $\zeta_X$; we shall always
assume this is the case. See Figure 6-4.

\begin{figure}[ht] \label{Gimmsy6-4}
\begin{center}
\includegraphics[scale=.8,angle=0]{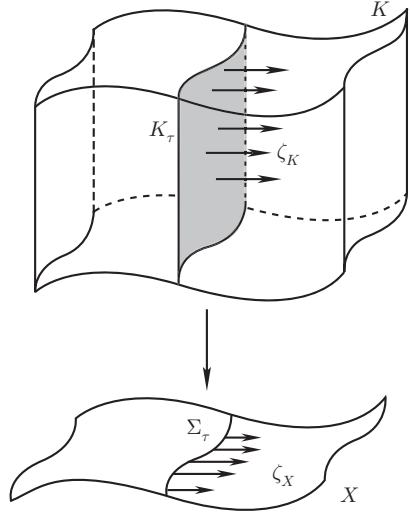}
\caption{Infinitesimal slicings}
\end{center}
\end{figure}	 

An important special case arises when the spacetime $X$ is endowed
with a Lorentzian metric $g$. Fix a spacelike hypersurface
$\Sigma_\tau \subset X$  and let $e_{\perp}$ denote the
future-pointing timelike unit normal vector field on $\Sigma_\tau$;
then $(\Sigma_\tau,e_\perp)$ is an infinitesimal slicing of $X$. In
coordinates adapted to $\Sigma_\tau$ we expand
\begin{equation}\label{ex6A: l&s}
\frac{\partial}{\partial x^0} = N e_\perp + 
M^i \frac{\partial}{\partial x^i},
\end{equation}
where $N$ is a function on $\Sigma_\tau$ (the {\bfi lapse\/}) and
${\bf M} = M^i{\partial}/{\partial x^i}$ is a vector field tangent to
$\Sigma_\tau$ (the {\bfi shift\/}). It is often useful to refer an 
arbitrary infinitesimal slicing $\zeta_X = \zeta^\mu\partial/\partial
x^\mu$ to the frame $\{e_\perp,\partial_i\}$, relative to which we
have
\begin{equation}\label{ex6A:zeta}
\zeta_X = \zeta^0N e_\perp + (\zeta^0M^i +
\zeta^i)\frac{\partial}{\partial x^i}.
\end{equation}
We remark that, in general, neither $\partial/\partial x^0$
nor $\zeta_X$ need be timelike.

In both our and ADM's (Arnowitt, Deser, and Misner [1962]) formalisms,
these lapse and shift functions play a key role. For instance, in the
construction of spacetimes from initial data (say, using a computer),
they are used to control the choice of slicing. This can be seen most
clearly by imposing the ADM coordinate condition that $\partial /
\partial x^0$ coincide with $\zeta_X$, in which case
(\ref{ex6A:zeta}) reduces simply to
\begin{equation}  \label{ex6A:c4} 
\zeta_X = Ne_{\perp} + {\mathbf M}.  
\end{equation}

\startrule
\vskip-12pt
            \addcontentsline{toc}{subsection}{Examples}
\begin{examples}
\mbox{}

\paragraph{\bf a\ \; Particle Mechanics.}\enspace   
Both $X = \mathbb R$ and $Y = \mathbb R \times Q$ for particle
mechanics are  ``already sliced" with $\zeta_X  = {d}/{dt}$ and
$\zeta_Y = {\partial}/{\partial t}$ respectively.  From the
infinitesimal equivariance equation (4D.2), it follows that this
slicing is  Lagrangian relative to $\cl = L(t, q^A, v^A) dt$  iff
${\partial L}/{\partial t} = 0$,  that is, $L$ is time-independent. 

One can consider more general slicings of $X$, interpreted as
diffeomorphisms $\fs_X : \mathbb R \to \mathbb R$. The induced slicing
$\fs_Y : Q \times \mathbb R  \to Y$ given by
$\fs_Y(q^1, \dots, q^N ,t) = (q^1, \dots, q^N,\fs_X(t))$ will be
Lagrangian if $\cl$ is time reparametrization-invariant.
\medskip

We can be substantially more explicit for the relativistic free
particle.  Consider an arbitrary  slicing $Q \times \mathbb R \to Y$
with generating vector field
\begin{equation} \label{ex6A:a1} 
\zeta_Y  =  \chi \frac{\partial}{\partial t} + \zeta^A
\frac{\partial}{\partial q^A}.   
\end{equation}
From (4D.2) we see that the slicing is 
Lagrangian relative to  (3C.8) iff
\begin{equation} \label{ex6A:a2} 
g_{BC,A} v^Bv^C\zeta^A + g_{AC}v^C \left( \frac{\partial
\zeta^A}{\partial t} + v^B
\frac{\partial \zeta^A}{\partial q^B} \right) =  0.   
\end{equation}
(The terms involving $\chi$  drop out as $\cl$ is time
reparametrization-invariant.)  But (\ref{ex6A:a2}) holds for all
$\mathbf v$ iff $\partial \zeta^A/\partial t = 0$ and
$$   
0 = g_{BC,A}v^Bv^C\zeta^A + g_{AC}v^C v^B \frac{\partial
\zeta^A}{\partial q^B} = v^Av^B\zeta_{(A;B)}.
$$
Thus  $\zeta^A \partial/\partial q^A$ must be a Killing vector
field.  It follows that the most general Lagrangian slicing consists of
time reparametrizations horizontally and isometries vertically.

\paragraph{\bf b\ \; Electromagnetism.}\enspace
Any slicing of the spacetime $X$ naturally induces a slicing of the
bundle ${\tilde Y} = \Lambda^1X \times S^{\ps 3,1}_2(X)$ by
push-forward. If $\zeta_X = \zeta^\mu \partial/\partial x^\mu$, the
generating vector field of this induced slicing is
$$
\zeta_{\tilde {Y}} = \zeta^\mu
\frac{\partial}{\partial x^\mu}  - A_\nu
\zeta^{\nu}_{\:\:\: ,\alpha}\frac{\partial}{\partial A_\alpha} -
(g_{\sigma \mu}\zeta^{\mu}_{\:\:\: ,\rho} + g_{\rho
\mu}\zeta^{\mu}_{\:\:\: ,\sigma})\frac{\partial}{\partial g_{\sigma
\rho}}.
$$
The most general slicing of $\tilde Y$ replaces the coefficients of the
second and third terms by $\chi_\alpha$ and $\chi_{\sigma \rho}$,
respectively, where the $\chi$s are any functions on $\tilde Y$.

The restriction to $\mathcal G$-slicings, with $\mathcal G =
\mbox{Diff}(X) \; \circledS \; \mathcal F(X)$ as in Example {\bf b}
of \S 4C, is not very severe for the parametrized version of Maxwell's
theory. Any complete vector field $\zeta_X = \zeta^\mu
\partial/\partial x^\mu$ may be used as the generator of the spacetime
slicing; then for the slicing of $\tilde Y$ we have the generator
\begin{equation}
\zeta_{\tilde Y} = \zeta^\mu \frac{\partial}{\partial x^\mu} +
(\chi_{,\alpha} - A_\nu \zeta^{\nu}_{\:\:\:
,\alpha})\frac{\partial}{\partial A_\alpha} - (g_{\sigma
\mu}\zeta^{\mu}_{\:\:\: ,\rho} + g_{\rho\mu} \zeta^{\mu}_{\:\:\:
,\sigma})\frac{\partial}{\partial g_{\sigma\ns\rho}},
\label{eq:ex6A:b2}
\end{equation}
where $\chi$ is an arbitrary function on $X$ (generating a Maxwell
gauge transformation).  A more general Lagrangian slicing (which,
however, is not a $\mathcal G$-slicing) is obtained from this upon
replacing $\chi_{,\alpha}$ by the components of a closed 1-form on
$X$.
\medskip

On the other hand, if we work with electromagnetism on a fixed
spacetime background, the $\zeta_X$ must be a Killing
vector field of the background metric $g$, and $\zeta_Y$ is of the
form (\ref{eq:ex6A:b2}) with this restriction on $\zeta^\mu$ (and
without the term in the direction $\partial/\partial g_{\sigma
\rho}$.) If the background spacetime is Minkowskian, then
$\zeta_X$ must be a generator of the Poincar\'e group. For a
generic background spacetime, there are no Killing vectors, and
hence no Lagrangian slicings. (This leads one to favor the
parametrized theory.)

\paragraph{\bf c\ \; A Topological Field Theory.}\enspace
With reference to Example {\bf b} above, we see that with $\mathcal G
= \mbox{Diff}(X) \; \circledS \; \mathcal F(X)$, a $\mathcal
G$-slicing of $Y = \Lambda^1X$ is generated by
\begin{equation}
\zeta_Y = \zeta^\mu
\frac{\partial}{\partial x^\mu}  + (\chi_{,\alpha} - A_\nu
\zeta^{\nu}_{\:\:\: ,\alpha})\frac{\partial}{\partial A_\alpha}.
\label{eq:ex6A:d1}
\end{equation}
Note that (\ref{eq:ex6A:d1}) does not generate a
Lagrangian slicing unless $\chi = 0$, since the replacement $A \mapsto
A + \mbox{\bf d}\chi$ does not  leave the Chern--Simons Lagrangian
density invariant (cf. \S4D).

\paragraph{\bf d\ \; Bosonic Strings.}\enspace 
In this case the configuration bundle  
$$ 
Y = (X \times M) \times_X S^{1,1}_2(X)
$$ 
is already sliced with $\zeta_X = {\partial}/{\partial x^0}$ and
$\zeta_Y = {\partial}/{\partial x^0}$.  More generally, one can
consider slicings with  generators of the form
\begin{equation}  \label{ex6A:e1}
\zeta^\mu \frac{\partial}{\partial x^\mu}
+ \zeta^A
\frac{\partial}{\partial \phi^A} + \zeta_{\sigma p}
\frac{\partial}{\partial h_{\sigma\ns\rho}}.
\end{equation}
Such a slicing will be Lagrangian relative to the Lagrangian
density (3C.23) iff $\zeta^A{\partial}/{\partial \phi^A}$ is a
Killing vector field of $(M, g)$ (this works much the same way as
Example \textbf{a}) and
\begin{equation}  \label{ex6A:e2}
\zeta_{\sigma\ns\rho}= -(h_{\sigma\alpha}\zeta^\alpha{}{}_{,\rho} +
h_{\rho \alpha}\zeta^\alpha{}_{,\sigma}) + 2\lambda h_{\sigma\ns\rho}  
\end{equation}
for some function $\lambda$ on $X$.  The first two terms in this
expression represent that ``part" of the  slicing which is induced by
the slicing $\zeta_X$ of $X$ by push-forward, and the last term
reflects the freedom to conformally rescale $h$ while leaving the
harmonic map Lagrangian invariant.  The  slicing represented by
(\ref{ex6A:e1}) will be a $\cg$-slicing,  with $\cg =
\operatorname{Diff}(X) \;\circledS\;\mbox{Con}^{1,1}_2(X)$, iff
$\zeta^A = 0$. \hfill $\blacklozenge$
\end{examples}

\subsection{
Space + Time Decomposition of the Jet Bundle}

In Chapter 5 we have space + time decomposed the multisymplectic
formalism relative to a fixed  Cauchy surface $\Sigma_\tau \in X$ to
obtain the associated $\tau$-phase space $T^*\cy_\tau$ with its
symplectic structure $\omega_\tau = -\mathbf{d}\theta_\tau$. Now we
show how to perform a similar decomposition of the jet bundle
$J^{1}Y$  using the notion of an infinitesimal slicing.  Effectively,
this enables us to invariantly separate the temporal from the spatial
derivatives of the fields.

Fix an infinitesimal slicing $(Y_\tau, \zeta := \zeta_Y)$ of $Y$ and
set
$$
\varphi  :=  \phi \! \bigm| \! \Sigma_\tau \quad \text{and}\quad
\dot\varphi :=  \pounds_\zeta\phi\! \bigm| \! \Sigma_\tau,
$$ 
so that in coordinates
\begin{equation} \label{eqn6B:2}
\varphi^A = \phi^A \! \bigm| \! \Sigma_\tau \quad \text{and}
\quad\dot\varphi^A  =  (\zeta^\mu
\phi^A{}_{,\mu} - \zeta^A
\circ
\phi)\!\bigm|\!\Sig_\tau.
\end{equation}
Define an affine bundle map $\beta_\zeta : (J^1Y)_\tau \to J^1(Y_\tau) 
\times VY_\tau$ over $Y_\tau$ by
\begin{equation}  \label{eqn6B:4} 
\beta_\zeta(j^1\ns \phi(x))  =  (j^1\ns\varphi(x), \dot\varphi(x))  
\end{equation}
for $x \in \Sig_\tau$. In coordinates adapted to $\Sigma_\tau$, 
(\ref{eqn6B:4}) reads
\begin{equation}  \label{eqn6B:5} 
\beta_\zeta(x^i, y^A, v^A{}_\mu)  =  (x^i, y^A, v^A{}_j, \dot y^A). 
\end{equation}
Furthermore, if the coordinates on $Y$ are arranged so that 
$$
\frac{\partial}{\partial x^0}\! \bigm|\!
Y_\tau = \zeta,\quad\text{ then }\quad \dot y^A = v^A{}_0.
$$ 

This last observation establishes:

\begin{prop}\label{prop6.1}
If $\zeta_X$ is transverse to $\Sigma_\tau$,
then $\beta_\zeta$ is an isomorphism.
\end{prop}

The bundle isomorphism $\beta_\zeta$ is the {\bfi jet decomposition
map\/} and its inverse the {\bfi jet  reconstruction map\/}.  Clearly,
both can be extended to maps on sections; from (\ref{eqn6B:4}) we have
\begin{equation}  \label{eqn6B:6} 
\beta_\zeta(j^1\ns\phi \circ i_\tau)   =  (j^1\ns \varphi, \dot\varphi)  
\end{equation}
where $i_\tau : \Sigma_\tau \to X$ is the inclusion.  In fact:

\begin{cor}\label{cor6.2}
$\beta_\zeta$ induces an isomorphism of $(j^1\cy
)_\tau$ with $T\cy_\tau$, where
$(j^1\cy )_\tau$ is the collection of restrictions of holonomic sections
of $J^{1}Y \to X$  to $\Sigma_\tau$.\footnote {\ 
$(j^1\cy )_\tau$ should not be confused with the collection of holonomic
sections of $J^1(Y_\tau) \to \Sig_\tau$,
since the former contains information about temporal derivatives that is
not included in the latter.}
\end{cor}

\begin{proof} 
Since $\dot\varphi$ is a section of $VY_\tau$
covering
$\varphi$,  by (\ref{eqn5A:1}) it defines an element of
$T_\varphi\cy_\tau$.  The result now follows from the previous 
Proposition and the comment afterwards. 
\end{proof}

\medskip
One may wish to decompose $Y$, as well as
$J^{1}Y$, relative to a slicing.  This is done so that one works with
fields that are spatially covariant rather than spacetime covariant. 
For  example, in electromagnetism, sections of $Y = \Lambda^1X$ are
one-forms $A = A_\mu dx^\mu$ over spacetime and sections of $Y_\tau =
\Lambda^1X \! \bigm| \!\Sigma_\tau$ are spacetime one-forms restricted to
$\Sigma_\tau$. One may split
\begin{equation}  \label{eqn6B:7} 
Y_\tau =  \Lambda^1\Sigma_\tau\times_{\Sigma_\tau} \Lambda^0\Sigma_\tau,  
\end{equation}
so that the instantaneous configuration space consists of spatial
one-forms $\mathbf{A} = A_mdx^m$ together with spatial scalars $a$. 
The map
$\Lambda^1X \! \bigm| \!\Sigma_\tau\to \Lambda^1\Sigma_\tau
\times_{\Sigma_\tau}
\Lambda^0\Sigma_\tau  $ which effects this split takes the form
\begin{equation}  \label{eqn6B:8}
 A \mapsto (\mathbf{A}, a)  
\end{equation}
where $a = i_\tau^*(\zeta_X \hook A)$ and $\mathbf{A} = i_\tau^*A$.

One particular case of interest is that of a metric tensor $g$ on
$X$. Let $S^{\ps n,1}_{2}(X,\Sig_\tau)$ denote the subbundle of $S^{\ps
n,1}_{2}(X)$ consisting of those Lorentz metrics on $X$ with respect
to which $\Sig_\tau$ is spacelike. We may space $+$ time split
\begin{equation}
S^{\ps n,1}_{2}(X,\Sig_\tau) \! \bigm| \!\Sigma_\tau = S^{\ps
n}_2(\Sigma_\tau) \times_{\Sigma_\tau} T\Sigma_\tau
\times_{\Sigma_\tau} \Lambda^0\Sigma_\tau
\end{equation}

\noindent as follows (cf. \S 21.4 of Misner, Thorne,
and Wheeler [1973]). Let $e_\perp$ the the
forward-pointing unit timelike normal to $\Sigma_\tau$, and let $N, \ps {\bf
M}$ be the lapse and shift functions defined via (\ref{ex6A: l&s}).
Set $\gamma =
i_\tau^*g$, so that $\gamma$ is a Riemannian metric on $\Sigma_\tau.$
Then the decomposition $g \mapsto (\gamma,{\bf M},N)$ with respect to the
infinitesimal slicing
$(\Sigma_\tau,e_\perp)$ is given by
$$g = \gamma_{jk} (dx ^j + M ^j dt)(dx ^k + M ^k dt) - N ^2 dt ^2$$
or, in terms of matrices,
\begin{equation}\label{eqn6B9}
\left(
\begin{array}{cc}
g_{00} & g_{0i} \\
 & \\
g_{i0} & g_{jk}
\end{array} \right) = \left(
\begin{array}{cc}
M_kM^k - N^2 & M_i \\
 & \\
M_i & \gamma_{jk}
\end{array} \right).
\end{equation}
This decomposition has the corresponding contravariant form
$$g ^{-1} = \gamma^{jk} \partial_j \partial_k -
         \frac{1 }{N ^2} ( \partial _t - M ^j \partial _j) 
 ( \partial _t - M ^k \partial _k)$$
or, in terms of matrices,

\begin{equation}\label{eqn6B10}
\left(
\begin{array}{cc}
g^{00} & g^{0i} \\
 & \\
g^{i0} & g^{jk}
\end{array}
\right) = \left(
\begin{array}{cc}
-1/N^2 & M^i/N^2 \\
 & \\
M^i/N^2 & \gamma^{jk} - M^jM^k/N^2
\end{array} \right)
\end{equation}
\vskip 6pt
\noindent where $M_i = \gamma_{ij}M^j.$ Furthermore, the metric volume
$\sqrt{-g}$ decomposes as
\begin{equation}\label{eqn6B11}
\sqrt{-g} = N\sqrt{\gamma}.
\end{equation}


The dynamical analysis can by carried out whether or not these splits of
the configuration  space are done; it is largely a matter of taste. 
Later, in Chapters 12 and 13 when we discuss dynamic fields and atlas
fields, these types of splits will play a key role.

\subsection{
The Instantaneous Legendre Transform}

Using the jet reconstruction map we may space + time split the
Lagrangian as follows.   Define
$$
\cl_{\tau,\zeta}: J^1(Y_\tau)  \times VY_\tau \to \Lambda^n\Sigma_\tau 
$$ 
by
\begin{equation}  \label{eqn6C:1} 
\cl_{\tau,\zeta}(j^1\ns\varphi(x), \dot\varphi(x))  
=  i^*_\tau \mathbf{i}_{\zeta_X}\cl(j^1\ns\phi(x)),  
\end{equation}
where $j^1\ns\phi \circ i_\tau$ is the reconstruction of
$(j^1\ns\varphi, \dot\varphi)$. The {\bfi instantaneous Lagrangian\/}
$L_{\tau,\zeta}: T\cy_\tau \to \mathbb R$ is defined by
\begin{equation}
L_{\tau,\zeta}(\varphi, \dot\varphi)  =  \int_{\Sigma_\tau}
\cl_{\tau,\zeta}(j^1\ns\varphi, \dot\varphi)  
  \label{eqn6C:2} 
\end{equation}
for $(\varphi, \dot\varphi) \in T\cy_\tau$ 
(cf. Corollary~\ref{cor6.2}).  
In coordinates adapted to $\Sigma_\tau$ this becomes, with the aid of
(\ref{eqn6C:1}) and (3A.1),
\begin{equation} \label{eqn6C:2b} 
L_{\tau,\zeta}(\varphi,\dot\varphi) = \int_{\Sigma_\tau} L(j^1\ns\varphi,
\dot\varphi)\zeta^0 d ^{\ps n}\ns x_0.
\end{equation}

The instantaneous Lagrangian $L_{\tau,\zeta}$ defines an {\bfi instantaneous
Legendre transform\/}
\begin{equation} \label{eqn6C:3} 
\mathbb FL_{\tau,\zeta} : T\cy_\tau \to T^*\cy_\tau; \quad (\varphi,
\dot\varphi) \mapsto (\varphi, \pi)  
\end{equation}
in the usual way (cf. Abraham and Marsden [1978]).  In
adapted coordinates
$$
\pi  =  \pi_A\,dy^A \otimes d^{\ps n}\ns x_0 
$$ 
and (\ref{eqn6C:3}) reads
\begin{equation}
\pi_A  =  \frac{\partial \cl_{\tau,\zeta}}{\partial \dot y^A}.  
  \label{eqn6C:4} 
\end{equation}
We call
\begin{equation*}
\cp_{\tau,\zeta} = \operatorname{ im} 
\mathbb FL_{\tau,\zeta}  \subset  T^*\cy_\tau   
  \label{eqn6C:5} 
\end{equation*}
the {\bfi instantaneous\/} or $\tau$-{\bfi primary constraint set\/}.  

\paragraph{\bf A3 \ Almost Regularity}\enspace
{\it Assume that
$\cp_{\tau,\zeta}$ is a smooth, closed, submanifold of
$T^*\cy_\tau$ and that $\mathbb FL_{\tau,\zeta}$ is a submersion with
connected fibers. 
}

\begin{remarks}[Remarks\ 1.]
Assumption {\bf A3} is 
satisfied in cases of interest.

\paragraph{\bf 2.}
We shall see momentarily that $\cp_{\tau,\zeta}$ is independent
of $\zeta$.

\paragraph{\bf 3.}
In obtaining (\ref{eqn6C:4}) we use the fact that $\cl$ is first order.
See Gotay [1991] for a treatment of the higher order case. 
\end{remarks}

We now investigate the relation between the covariant and instantaneous
Legendre  transformations.  Recall that over $\cy_\tau$ we have the
symplectic bundle map $R_\tau : (\cz_\tau, \Omega_\tau)  \to (T^*\cy_\tau,
\omega_\tau) $ given by
$$  
\langle R_\tau(\sigma), V\rangle  =  \int_{\Sigma_\tau}
\varphi^*(\mathbf{i}_V\sigma)
$$  
where $\varphi = \pi_{Y\! Z} \circ \sig$ and $V \in T_\varphi\cy_\tau$. 

\begin{prop}\label{prop6.3}
Assume $\zeta_X$ is transverse to
$\Sigma_\tau$.  Then the following diagram commutes:
\begin{equation}
\begin{CD} 
(j^1\cy)_\tau  @>{\mathbb F \cl}>> \cz_\tau \\ 
@V{\beta_\zeta}VV @VV{R_\tau}V \\ 
T\cy_\tau @>>{\mathbb FL_{\tau,\zeta}}> T^* \cy_\tau
\end{CD}   
  \label{eqn6C:6} 
\end{equation}
\end{prop}

\begin{proof} 
 Choose adapted coordinates in which $\partial_0 \! \bigm| \! Y_\tau =
\zeta$.  Since $R_\tau$ is given by $\pi_A = p_A{}^0\circ \sig$, going
clockwise around the diagram we obtain
$$  
R_\tau\!\left(\mathbb F\cl(j^1\ns\phi \circ i_\tau) \right) = \frac{\partial
L}{\partial v^A{}_0}(\phi^B, \phi^B{}_{,\mu}) \ps dy^A \otimes d^{\ps n}\ns
x_0.
$$  
This is the same as one gets going counterclockwise, taking into
account (\ref{eqn6B:5}), (\ref{eqn6C:4}) and the  fact that $\mathbb
FL_{\tau,\zeta}$ is evaluated at $\dot\varphi^A  = \phi^A{}_{,0}$. 
\end{proof}

We define the {\bfi covariant primary constraint set\/} to be
$$ 
N  =  \mathbb F\cl(J^1Y) \subset Z 
$$ 
and with a slight abuse of notation, set
\begin{equation*}
\cn_\tau =  \mathbb F\cl\left((j^1\cy )_\tau\right)  
\subset \cz_\tau.   
  \label{eqn6C:7} 
\end{equation*}

\begin{cor}\label{cor6.4}
If $\zeta_X$ is transverse to
$\Sigma_\tau$, then
\begin{equation}
R_\tau(\cn_\tau)  = \cp_{\tau,\zeta}.   
  \label{eqn6C:8} 
\end{equation}
In particular, $\cp_{\tau,\zeta}$ is independent of $\zeta$, and so
can be denoted simply $\cp_\tau$.
\end{cor}

\begin{proof} 
 By Corollary~\ref{cor6.2}, $\beta_\zeta$ is onto $T\cy_\tau$.  The
result now follows from the commutative diagram (\ref{eqn6C:6}). 
\end{proof}

Denote by the same symbol $\omega_\tau$ the pullback of the symplectic
form on $T^*\cy_\tau$ to the  submanifold $\cp_\tau$. When there is any
danger of confusion we will write $\omega_{T^*\cy_\tau}$ and
$\omega_{\cp_\tau}$. In general $(\cp_\tau, \omega_\tau) $ will be merely
presymplectic. However, the fact that 
$\mathbb FL_{\tau,\zeta}$ is fiber-preserving together with the almost
regularity assumption {\bf A3} imply that $\ker \omega_\tau$ {\it is a
regular distribution\/} on $\cp_\tau$ (in the sense that it defines a
subbundle of $T\cp_\tau$).
\medskip

As always, the {\bfi instantaneous Hamiltonian} is given by
\begin{equation}
H_{\tau,\zeta}(\varphi, \pi)  =  \langle \pi, \dot\varphi\rangle -
L_{\tau,\zeta}(\varphi, \dot\varphi)  
  \label{eqn6C:9} 
\end{equation}
and is defined only on $\cp_\tau$.  The density for $H_{\tau,\zeta}$ is
denoted by $\Fh_{\tau,\zeta}$.  We remark that to determine a
Hamiltonian, it is essential to specify a time direction $\zeta$ on
$Y$.  This is sensible, since the system cannot evolve without knowing
what ``time" is.  For $\zeta_Y =
\xi_Y$, where $\xi \in \fg$, the Hamiltonian will turn out to be the
negative of the energy-momentum map induced on $\cp_\tau$ (cf. \S 7F).  A
crucial step in establishing this relationship is the following result:

\begin{prop}\label{prop6.5}
Let $(\varphi, \pi) \in \cp_\tau$.  Then for
any holonomic lift $\sig$ of 
$(\varphi, \pi)$, 
\begin{equation}
H_{\tau,\zeta}(\varphi, \pi)  
=  - \int_{\Sigma_\tau} \sigma^*(\mathbf{i}_{\zeta_Z}\Theta).   
  \label{eqn6C:10} 
\end{equation}
\end{prop}

 Here $\zeta_Z$ is the canonical lift of $\zeta$ to $Z$
(cf. \S 4B).  By a {\bfi holonomic lift\/} of
$(\varphi, \pi)$ we mean any element $\sigma \in R^{-1}_\tau\{(\varphi,
\pi)\}
\cap \cn_\tau$.  Holonomic lifts of elements of $\cp_\tau$ always exist by
virtue of Proposition~\ref{prop6.3}.

\begin{proof} 
  We will show that (\ref{eqn6C:10}) holds on the level of
densities; that is,
\begin{equation}
\Fh_{\tau,\zeta}(\varphi, \pi)  =  - \sigma^*(\mathbf{i}_{\zeta_Z}\Theta).   
  \label{eqn6C:11} 
\end{equation}
Using adapted coordinates, (2B.11) yields
\begin{multline*}
\sigma^*(\mathbf{i}_{\zeta_Z}\Theta)  = \\[1.5ex]
\left\{(p_A{}^0 \circ \sig) \left(\zeta^A \circ \sig - \zeta^\mu
\sigma^A{}_{,\mu}\right) +
\left(p\circ \sig + (p_A{}^\mu\circ \sig)
\sigma^A{}_{,\mu}\right)\zeta^0\right\}d^{\ps n}\ns x_0 
\end{multline*} 
for any $\sigma \in \cz_\tau$.  Now suppose that $(\varphi, \pi) \in
\cp_\tau$, and let $\sig$ be any lift of $(\varphi, \pi)$ to $\cn_\tau$.
Thus, there is a $\phi \in \mathcal{Y}$ with $\mathbb F\cl\circ j^1\ns
\phi\circ i_\tau = \sig$.  Then, using  (3A.2), (\ref{eqn6B:2}),
(\ref{eqn5D:4}) and (\ref{eqn6C:1}), the above becomes
\begin{equation}
\sigma^*(\mathbf{i}_{\zeta_Z}\Theta) = - \pi(\dot\varphi) +
L(j^1\ns\phi)
\zeta^0\, d^{\ps n}\ns x_0= - \pi(\dot\varphi) + \cl_{\tau,\zeta}(\varphi,
\dot\varphi). 
\tag*{\qed}
\end{equation}
\renewcommand{\qed}{}
\end{proof}

%
Notice that (\ref{eqn6C:10}) and (\ref{eqn6C:11}) are manifestly linear in
$\zeta_Z$.  This linearity foreshadows the  linearity of the Hamiltonian (1.2) in the
``atlas fields"  to which we alluded in the introduction.
\pagebreak
\startrule
\vskip-12pt
            \addcontentsline{toc}{subsection}{Examples}
\begin{examples}
\mbox{}

\paragraph{\bf a\ \; Particle Mechanics.}\enspace 
 First consider a nonrelativistic particle
Lagrangian of the  form
$$ 
L(q, v) = \frac12 g_{AB}(q)v^Av^B + V(q).
$$ 
Taking $\zeta = {\partial}/{\partial t}$, the Legendre
transformation gives $\pi_A = g_{AB}(q)v^B$. If $g_{AB}(q)$ is
invertible for all $q$, then $\mathbb FL_t$ is onto for each $t$ and there
are no primary constraints.
\medskip

For the relativistic free particle, the covariant primary constraint
set $N \subset Z$ is determined  by the constraints
\begin{equation}
g^{AB} p_A p_B  =  - m^2 \qquad \text{and} \qquad  p = 0,   
  \label{6Cex:a1} 
\end{equation}
which follow from (3C.10).

Now fix any infinitesimal slicing 
$$
\left(Y_t, \zeta = \chi \frac{\partial}{\partial t} + \zeta^A
\frac{\partial}{\partial q^A}\right)
$$ 
of $Y$.  Then we may identify $(J^{1}Y)_t$ with $TQ$ according to 
(6B.2); that is,
$$ 
(q^A, v^A) \mapsto (q^A, \dot q^A)
$$ 
where $\dot q^A = \chi v^A - \zeta^A$.  The instantaneous Lagrangian
(6C.2) is then
\begin{equation}
L_{t,\zeta}(q, \dot q )  = - m  \| \dot{\bold q}+ {\boldsymbol\zeta}
\|   
  \label{6Cex:a2} 
\end{equation}
(provided we take $\chi > 0$).  The instantaneous Legendre transform
(6C.4) gives
\begin{equation}
\pi_A  = \frac{mg_{AB} (\dot q^B + \zeta^B)}{\| \dot{\bold q} +
{\boldsymbol\zeta} \|}.   
  \label{6Cex:a3} 
\end{equation}
The $t$-primary constraint set is then  defined by the ``mass
constraint"
\begin{equation}
g^{AB} \pi_A \pi_B  =  -m^2.   
  \label{6Cex:a4} 
\end{equation}
Comparing (\ref{6Cex:a4}) with (\ref{6Cex:a1}) we verify that $\cp_t =
R_t(\cn_t)$ as predicted by (\ref{eqn6C:8}).  Using (\ref{6Cex:a4}) and
(\ref{eqn6C:9})  we compute
\begin{equation}
H_{t,\zeta}(q, \pi) = -\zeta^A\pi_A.   
  \label{6Cex:a5} 
\end{equation}
Looking ahead to Part III (cf. also the Introduction and
Remark 9 of \S6E), it may seem curious that
$H_{t,\zeta}
$ does not vanish identically, since after all the relativistic free
particle is a param\-etr\-ized system.  This is because the slicing
generated by (\ref{ex6A:a1}) is {\it not\/} a $\cg$-slicing unless
$\zeta^A=0$, in which case the Hamiltonian {\it does\/} vanish.

\paragraph{\bf b\ \; Electromagnetism.}\enspace
 First we consider the parametrized case.  Let $\Sigma_\tau$  be
a spacelike hypersurface locally given by $x^0=$ constant, and
consider the infinitesimal $\mathcal G$-slicing $({\tilde
Y}_\tau,\zeta)$ with $\zeta$ given by (\ref{eq:ex6A:b2}):
\begin{equation*}
\hspace{28pt}
\zeta_{\tilde Y} = \zeta^\mu
\frac{\partial}{\partial x^\mu} + (\chi_{,\alpha} - A_\nu
\zeta^{\nu}_{\:\:\: ,\alpha})\frac{\partial}{\partial A_\alpha} -
(g_{\sigma \mu}\zeta^{\mu}_{\:\:\: ,\rho} + g_{\rho
\mu}\zeta^{\mu}_{\:\:\: ,\sigma})\frac{\partial}{\partial g_{\sigma
\rho}}.
\end{equation*}

We construct the instantaneous Lagrangian $L_{\tau,\zeta}$.
From (\ref{eqn6B:2}) we have
\begin{equation}
\dot A_{\mu}= \zeta^0 A_{\mu,0} + \zeta^i A_{\mu ,i}
-(\chi_{,\mu} - A_\nu \zeta^{\nu}_{\:\:\: ,\mu}),
  \label{6Cex:b2} 
\end{equation}
and so (3C.13) gives in particular
\begin{equation}\label{6Cex:b2b}
F_{0i} = \frac{1}{\zeta^0}\left(\dot A_i
- \zeta^k A_{i,k} +\chi_{,i}-A_{\nu}\zeta^{\nu}_{\:\:\:,i} - \zeta^0
A_{0,i}\right).
\end{equation}

\noindent Substituting this into (3C.12), (\ref{eqn6C:2b}) yields
\begin{eqnarray}  \label{6Cex:b3}
L_{\tau,\zeta} (A, \dot A;g) & = & \int_{\Sigma_\tau}
 \bigg [ \frac{1}{2\zeta^0}(g^{i0}g^{j0}-g^{ij}g^{00})
\nonumber
\\
  & & \hspace{4ex}\mbox{} \times  (\dot A_i
- \zeta^k A_{i,k} +\chi_{,i}-A_{\nu}\zeta^{\nu}_{\:\:\:,i} - \zeta^0
A_{0,i})
\nonumber \\
 & & \hspace{4ex} \mbox{}  \times  (\dot A_j  - \zeta^m A_{j,m}
+\chi_{,j}-A_{\rho}\zeta^{\rho}_{\:\:\:,j} - \zeta^0
A_{0,j})
\nonumber
\\
 & & \hspace{4ex} \mbox{} +  g^{ik}g^{0m}(\dot A_i  - \zeta^k A_{i,k}
+\chi_{,i}-A_{\nu}\zeta^{\nu}_{\:\:\:,i} - \zeta^0
A_{0,i})F_{km} \nonumber
\\  & & \hspace{4ex} \mbox{} -  \frac14
g^{ik}g^{jm}F_{ij}F_{km}\zeta^0
\bigg ]
\sqrt{-g}\,d^{\ps 3}\ns x_0.
\end{eqnarray}

The corresponding instantaneous Legendre transformation $\mathbb F
L_{\tau,\zeta}$ is defined by
\begin{eqnarray}
{\mathfrak E}^i & = &
\bigg (\frac{1}{\zeta^0}(g^{i0}g^{j0}-g^{ij}g^{00})   (\dot A_j  - \zeta^m
A_{j,m}
+\chi_{,j}-A_{\rho}\zeta^{\rho}_{\:\:\:,j} -
\zeta^0 A_{0,j}) \nonumber \\[2ex] 
& & \hspace{2ex}\mbox{} +
g^{ik}g^{0m}F_{km}
\bigg )\sqrt{-g}
  \label{6Cex:b4} 
\end{eqnarray}
and
\begin{equation}
{\mathfrak E}^0 =0.
  \label{6Cex:b5} 
\end{equation}
This last relation is the sole primary constraint in the Maxwell
theory.  Thus the $\tau$-primary constraint set is
\begin{equation}
\tilde{\mathcal P}_\tau = \big\{(A,\mathfrak E;g)\in T^*{{\mathcal
Y}}_\tau
\times \big({\mathcal S}^{3,1}_2\big)_\tau \, \big| \,
\mathfrak E^0=0 \big\}.
  \label{6Cex:b6} 
\end{equation}
It is clear that the almost regularity assumption {\bf A3} is satisfied
in this case, and that $\tilde{\mathcal P}_\tau$ is indeed independent
of the choice of
$\zeta$ as required by Corollary~\ref{cor6.4}.
Using (3C.14) and
(\ref{ex5D:b1}), one can also verify that (\ref{6Cex:b4}) and
(\ref{6Cex:b5}) are consistent with the covariant Legendre transformation.
In particular, the primary constraint $\mathfrak E^0=0$ is a consequence of
the relation
$\mathfrak E^{\nu} =
\mathfrak F^{\nu 0}$ together with the fact that $\mathfrak F^{\nu\mu}$
is antisymmetric on $N$.

Taking (\ref{6Cex:b5}) into account, (\ref{ex5D:b2}) yields the presymplectic
form
\begin{equation}
\omega_\tau(A,\mathfrak E;g)=\int_{\Sigma_\tau}(dA_i\wedge
d\mathfrak E^i)\otimes d^{\ps 3}\ns x_0
  \label{6Cex:b7} 
\end{equation}
on $\tilde{\mathcal P}_\tau$.  The Hamiltonian on $\tilde{\mathcal
P}_\tau$ is obtained by solving (\ref{6Cex:b4}) for $\dot A_i$ and substituting into
(\ref{eqn6C:9}). After some effort, we obtain
\begin{align}  \label{6Cex:b8}
H_{\tau,\zeta}(A,\mathfrak E;g)  &=  \int_{\Sigma_\tau}
 \bigg[\zeta^0N\gamma^{-1/2}\Big(\frac{1}{2} \gamma_{ij}\mathfrak E^i
\mathfrak E^j + \frac{1}{4N^2} \gamma^{ik} \gamma^{jm}\mathfrak F_{ij}
\mathfrak F_{km}\Big) \nonumber \\
&\qquad 
\mbox{} + \frac{1}{N\sqrt{\gamma}}(\zeta^0 M^i +
\zeta^i)\mathfrak E^j\mathfrak F_{ij} +
(\zeta^\mu A_\mu - \chi)_{,i} \mathfrak E^i
\bigg] d^{\ps 3}\ns x_0
\end{align}
where we have made use of the splitting (\ref{eqn6B9})--(\ref{eqn6B11}) 
of the metric
$g$.
%
Note the appearance of
the combination $\zeta^\mu A_\mu -\chi$ in (\ref{6Cex:b8}).
Later we will recognize this as the
``atlas field'' for the parametrized version of Maxwell's theory. Note
also the presence of the characteristic combinations $\zeta^0N$ and
$(\zeta^0 M^i + \zeta^i)$ originating from (\ref{ex6A:zeta}).
\medskip

For electromagnetism on a fixed spacetime background, the preceding
computations must be modified slightly. For definiteness, we assume
that $(X,g)$ is Minkowski spacetime $(\mathbb R^4,\eta)$, and that
$\Sigma_\tau$ is a spacelike hyperplane $x^0 =$ constant. The main
difference is that we must now require $\zeta_X$ to be a
Poincar\'e generator. Again for definiteness, we suppose that
$\zeta_X = \partial /\partial x^0$. Thus the slicing generator
$\zeta_{\tilde Y}$ is replaced by
\begin{equation}\label{6Cex:b26}
\zeta_{Y} =
\frac{\partial}{\partial x^0} +
\chi_{,\alpha}\frac{\partial}{\partial A_\alpha}.
\end{equation}

\noindent The computations above remain valid upon replacing
$(\zeta^0,\mbox{\boldmath{$\zeta$}})$ by $(1,{\bf 0})$. The Hamiltonian
in this case reduces to
%
\begin{equation}
H_{\tau,(1,{\bf 0})}(A,\mathfrak E)  =  \int_{\Sigma_\tau}
 \bigg[\frac12 \mathfrak E_i \mathfrak
E^i + \frac14 \mathfrak F_{ij}
\mathfrak F^{ij}  + ( A_0 - \chi)_{,i} \mathfrak
E^i
\bigg] d^{\ps 3 }\ns x_0.
\end{equation}

\paragraph{\bf c\ \;%
        A Topological Field Theory.}\enspace
Let $\Sigma_\tau$ be any compact surface in $X$, and fix the Lagrangian
slicing
\begin{equation}
\zeta = \zeta^\mu
\frac{\partial}{\partial x^\mu} - A_\nu
\zeta^{\nu}_{\:\:\: ,\alpha}\frac{\partial}{\partial A_\alpha}
\label{eq:ex6A:d2}
\end{equation}
as in Example {\bf c} of \S 6A. The computations are similar those in
Example {\bf b} above. In particular, (\ref{6Cex:b2}) and (\ref{6Cex:b2b}) remain
valid (with $\chi = 0$). Together with (3C.18), these yield
\begin{multline}\label{eqn6C:c20}
L_{\tau,\zeta}(A,\dot A)  = \\[2ex] \int_{\Sigma_\tau}
\epsilon^{0ij}\bigg((\dot A_i  - \zeta^k A_{i,k}
-A_{\nu}\zeta^{\nu}_{\:\:\:,i} -
\zeta^0 A_{0,i})A_j 
 +  \frac12  F_{ij}A_0 \zeta^0\bigg)d^{\ps 2}\ns x_0.
\end{multline}

The instantaneous Legendre transformation is
\begin{equation}\label{6C38}
\pi^i  =  \epsilon^{0ij}A_j \qquad \mbox{and} \qquad
\pi^0  =  0;
\end{equation}

\noindent compare (3C.19). In contrast to electromagnetism, all of these
relations are primary constraints. Thus the instantaneous primary
constraint set is
\begin{equation}
\mathcal P_\tau =\left\{(A,\pi)\in T^*{{\mathcal Y}}_\tau \mid
\pi^0=0 \:\mbox{ and }\: \pi^i  =  \epsilon^{0ij}A_j\right\}.
  \label{6C39} 
\end{equation}
Again we see that the regularity assumption {\bf A3} is satisfied. From
(\ref{6C38}) and (\ref{eq:ex5D:d2}) we obtain the presymplectic form on $\mathcal P_\tau$,
\begin{equation}\label{6C40}
\omega_\tau(A,\pi) = \int_{\Sigma_\tau} \big(\epsilon^{0ij} dA_i
\wedge dA_j\big) \otimes d^{\ps 2}\ns x_0.
\end{equation}

\noindent The Chern-Simons Hamiltonian is
\begin{equation}\label{6C41}
H_{\tau,\zeta}(A,\pi) = \int_{\Sigma_\tau}
\epsilon^{0ij}\bigg(\zeta^k F_{ki}A_j - \frac12 \zeta^0 F_{ij}A_0 +
(\zeta^\mu A_\mu)_{,i}A_j\bigg) d^{\ps 2}\ns x_0,
\end{equation}
\noindent which is consistent with (\ref{eqn6C:10}).

\paragraph{\bf d\ \; Bosonic Strings.}\enspace 
 Consider an infinitesimal slicing $(\Sigma_\tau,
\zeta)$ as in (\ref{ex6A:e1}), with $\zeta^A=0$.  (Here we must also
suppose that the pull-back of $h$ to $\Sigma_\tau$ is
positive-definite.)   Using (\ref{eqn6B:2}) and
(3C.23) the instantaneous Lagrangian turns out to be
\begin{align}
 L_{\tau,\zeta} (\varphi,h,\dot\varphi , \dot h) =  -\frac12
\int_{\Sigma_\tau} \sqrt{|h|}\, g_{AB}\! &\left(
\frac{1}{\zeta^0}h^{00}(\dot\varphi^A\right. -\zeta^1\partial\varphi^A)
(\dot\varphi^B-\zeta^1\partial\varphi^B) 
\nonumber \\[1.5ex]
   &\quad + 2 h^{01}
(\dot\varphi^A-\zeta^1\partial\varphi^A)\partial\varphi^B 
\nonumber \\[1.5ex]
  &\quad +
\zeta^0 h^{11}\biggl.\partial\varphi^A \partial\varphi^B \biggl)
d^{\ps 1}\ns x_0,  
  \label{6Cex:e1} 
\end{align}
 where we have set $\partial\varphi^A : =
\varphi^A{}_{,1}$. From this it follows that the instantaneous momenta
are
\begin{gather}
\pi_A  = {-}  \sqrt{|h|} \, g_{AB} \bigg( \frac{1}{\zeta^0}
h^{00}(\dot\varphi^B-\zeta^1\partial\varphi^B)+ h^{01}\partial\varphi^B
\bigg)
\label{6Cex:e2}
\\[1.5ex]
\rho^{\sigma\ns\rho}=0.
\label{6Cex:e3}
\end{gather}
Thus
\begin{equation}  \label{6Cex:e4} 
\cp_\tau =\left\{(\varphi,h,\pi,\rho)\in
T^*\cy_\tau \bigm|\rho^{\sigma\ns\rho}=0\right\}.  
\end{equation}
 This is consistent with (3C.24) and (3C.25) 
 via (\ref{ex5D:e1}). A
short computation then gives
\pagebreak
\begin{multline*}  
 H_{\tau,\zeta}(\varphi,h,\pi ,\rho)  =  \\[1.5ex]
-\int_{\Sigma_\tau}\left(\frac12|h|^{-1/2}\frac1{h^{00}}\zeta^0(\pi^2
+\partial\varphi^2)+ \left(\frac{h^{01}}{h^{00}}
\zeta^0-\zeta^1\right) (\pi\cdot\partial\varphi)\right)  d^{\ps 1}\ns x_0
\end{multline*}
 for the instantaneous Hamiltonian on $\cp_\tau$, where we have used
the abbreviations $\pi^2 :=g^{AB}\pi_A\pi_B$ and $\pi\cdot
\partial\varphi: =\pi_A\partial\varphi^A$, etc.  
If we space $+$ time split the metric $h$ as in
(\ref{eqn6B9})--(\ref{eqn6B11}), then the Hamiltonian becomes simply
\begin{multline}  \label{6Cex:e7}
 H_{\tau,\zeta} (\varphi , h,\pi , \rho) = \\[1.5ex]
 \int_{\Sigma_\tau}\left(\frac{1}{2\sqrt{\gamma}} \zeta^0 N(\pi^2
+\partial\varphi^2) + (\zeta^0M +
\zeta^1)(\pi\cdot\partial\varphi)\right) d^{\ps 1}\ns  x_0.
\end{multline}
 This expression should be compared with its counterpart in ADM
gravity, cf. Interlude III and Arnowitt, Deser, and Misner [1962].  In
\S 12C we will identify $\zeta^0N$ and $\zeta^0M +
\zeta^1$ as the ``atlas
fields'' for the bosonic string.

Finally, using (\ref{6Cex:e2}) and (\ref{6Cex:e3}) 
in (\ref{ex5D:e2}), 
the
presymplectic structure on $\cp_\tau$ is 
\begin{equation}  \label{6Cex:e8}
\quad\qquad\qquad\qquad \omega_\tau(\varphi, h, \pi ,
\rho)=\int_{\Sigma_\tau}(d\varphi^A\wedge d\pi_A)\otimes d^{\ps 2}\ns x_0. 
\qquad \qquad \blacklozenge 
\end{equation}
\renewcommand{\lozsymbol}{}
\end{examples}

\subsection{
Hamiltonian Dynamics}

We have now gathered together the basic ingredients of Hamiltonian
dynamics:  for each  Cauchy surface $\Sigma_\tau$, we have the
$\tau$-primary constraint set $\cp_\tau$,  a presymplectic structure
$\omega_\tau$ on
$\cp_\tau$, and a Hamiltonian $H_{\tau,\zeta}$ on $\cp_\tau$ relative to a
choice of evolution direction $\zeta$.  If we think of some fixed
$\Sigma_\tau$ as the ``initial time," then fields $(\varphi, \pi) \in
\cp_\tau$ are candidate initial data for the $(n + 1)$-decomposed field
equations; that is, Hamilton's equations.  To evolve this initial data,
we slice spacetime and the bundles over it into global moments of time
$\lambda$.

To this end, we regard $\operatorname{Emb}(\Sig, X)$ as the space of
all (param\-etr\-ized) Cauchy surfaces in  the $(n + 1)$-dimensional
``spacetime" $X$.  The arena for Hamiltonian dynamics in the
instantaneous  or $(n + 1)$-formalism is the ``instantaneous primary
constraint bundle" $\cp^\Sig$ over
$\operatorname{Emb}(\Sig, X)$ whose fiber above $\tau \in
\operatorname{Emb}(\Sig, X)$ is $\cp_\tau$.

Fix compatible slicings $\fs_Y$ and $\fs_X$ of $Y$ and $X$ with
generating vector fields $\zeta$ and 
$\zeta_X$, respectively.  As in \S 6A, let $\tau : \mathbb R \to
\operatorname{Emb}(\Sig, X)$ be the curve of embeddings defined by
$\tau(\lambda) (x) = \fs_X(x,\lambda) $.

Let $\cp^\tau$ denote the portion of $\cp^\Sig$ lying over the image of
$\tau$ in $\operatorname{Emb}(\Sig, X)$.  Dynamics relative to the chosen
slicing takes place in $\cp^\tau$; we view the $(n + 1)$-evolution of
the fields as being given by a curve
$$
 c(\lambda)   =  (\varphi(\lambda) , \pi(\lambda) )
$$
 in $\cp^\tau$  covering $\tau(\lambda) $.  All this is illustrated in
Figure 6-5.

\begin{figure}[ht] \label{Gimmsy6-5}
\begin{center}
\includegraphics[scale=1,angle=0]{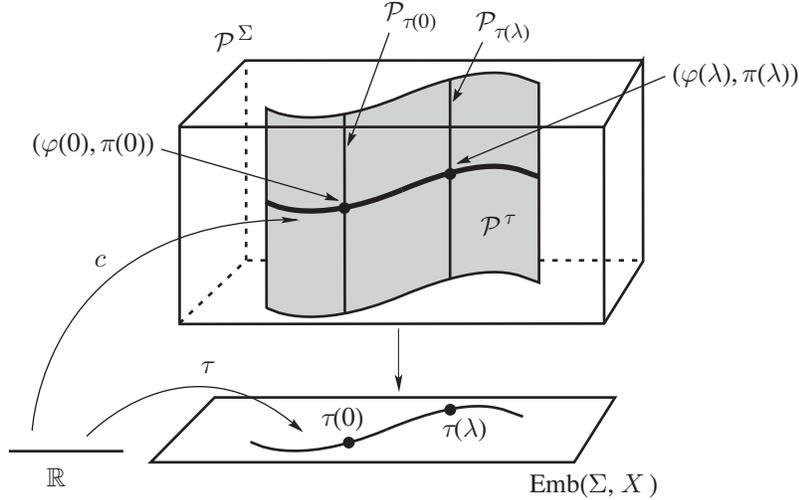}
\caption{Instantaneous Dynamics}
\end{center}
\end{figure}

Our immediate task is to obtain the $(n + 1)$-decomposed field
equations on  $\cp^\tau$,  which  determine the curve $c(\lambda) $.  This
requires setting up a certain amount of notation.

Recall from \S 6A that the slicing $\fs_Y$ of $Y$ gives rise to a
trivialization $\fs_{\cy}$ of
$\cy^\tau$, and hence induces trivializations $\fs_{j^1\ns\cy}$ of
$(j^1\cy )^\tau$ by jet prolongation and $\fs_{\cz}$ of $\cz^\tau$ and
$\fs_{T^*\ns \cy}$ of $T^*\cy^\tau$ by pull-back. These latter
trivializations are therefore \textit{presymplectic} and
\textit{symplectic};  that is, the associated flows restrict to
presymplectic and symplectic isomorphisms on fibers respectively.
Furthermore, the reduction maps $R_{\tau(\lambda) }: \cz_{\tau(\lambda) } \to
T^*\cy_{\tau(\lambda) }$ intertwine the trivializations $\fs_{\cz}$ and
$\fs_{T^*\cy}$ in the obvious sense.

Assume \textbf{A2}, viz.,  the slicing $\fs_Y$ of $Y$ is Lagrangian.  
From Proposition 4.6(i) 
$\mathbb F\cl : (j^1\cy )^\tau \to \cz^\tau$, regarded as a map on
sections, is equivariant with respect to the (flows corresponding to the)
induced trivializations of these spaces.  (Infinitesimally, this is
equivalent to the statement
$T\mathbb F\cl \cdot \zeta_{j^1\ns\cy} = \zeta_\cz$ where
$\zeta_{j^1\ns\cy}$ and $\zeta_\cz$ are the generating vector fields of
the trivializations.)  This observation, combined with the  above
remarks on reduction,  Proposition~\ref{prop6.3}, and  assumption {\bf
A3}, show that $\cp^\tau$ really  is a subbundle of $T^*\cy^\tau$, and
that the symplectic trivialization $\fs_{T^*\ns\cy}$ on $T^*\cy^\tau$
restricts to a presymplectic trivialization $\fs_\cp$ of $\cp^\tau$. 
We use this trivialization to coordinatize
$\cp^\tau$ by 
$(\varphi, \pi,\lambda) $.  The vector field $\zeta_\cp$ which generates this
trivialization is transverse to the fibers of $\cp^\tau$ and satisfies
${\zeta_\cp} \hook\,d\lambda = 1$.  To avoid a plethora of indices (and in
keeping with the notation of \S 6A), we will henceforth denote the
fiber $\cp_{\tau(\lambda) }$ of
$\cp^\tau$ over $\tau(\lambda)  \in \operatorname{Emb}(\Sig, X)$ simply by
$\cp_\lambda$, the presymplectic form $\omega_{\tau(\lambda) }$ by  
$\omega_\lambda$, etc.

Using $\zeta_\cp$, we may extend the forms $\omega_\lambda$ along the fibers
$\cp_\lambda$ to a (degenerate)  2-form $\omega$ on $\cp^\tau$ as
follows.  At any point $(\varphi, \pi) \in \cp_\lambda$, set
\begin{align}
          &\omega(\cv, \cw)  = \omega_\lambda(\cv, \cw),  
  \label{eqn6D:1} 
  \\[1.5ex]
&\omega(\zeta_\cp , \cdot )	 =  0,            
  \label{eqn6D:2} 
\end{align}
where $\cv, \cw$ are vertical vectors on $\cp^\tau$ (i.e., tangent to
$\cp_\lambda$) at $(\varphi, \pi)$.  Since $\cp_\lambda$ has codimension one
in
$\cp^\tau$, (\ref{eqn6D:1}) and (\ref{eqn6D:2}) uniquely determine
$\omega$.  It is closed since
$\omega_\lambda$ is and since the trivialization generated by $\zeta_\cp$ is
presymplectic (in other words, $\pounds_{\zeta_\cp}\omega = 0$;  cf.
Gotay, Lashof, \'Sniatycki, and Weinstein [1983]).    

Similarly, we define the function $H_\zeta$ on $\cp^\tau$ by
\begin{equation}  \label{eqn6D:3} 
 H_\zeta(\varphi, \pi,\lambda)  
 =  H_{\lambda, \zeta}(\varphi, \pi).     
\end{equation}
 Tracing back through the definitions (\ref{eqn6C:1}) and (\ref{eqn6C:2})
of the instantaneous Lagrangian $L_{\lambda, \zeta}$, we find 
the condition that the slicing be Lagrangian  guarantees 
that the function $L_\zeta : T\cy^\tau \to \mathbb R$ defined by
$$
 L_\zeta(\varphi, \dot\varphi,\lambda)  = L_{\lambda, \zeta}(\varphi, \dot\varphi)
$$
 is independent of $\lambda$. Therefore, if  {\bf A2} holds, 
(\ref{eqn6C:9}) implies that $\zeta_\cp [H_\zeta] = 0$. 

Consider the 2-form $\omega + \mathbf{d}H_\zeta \wedge
\mathbf{d}\lambda$ on $\cp^\tau$. By construction,
\begin{equation}  \label{eqn6D:4} 
\pounds_{\zeta_\cp}(\omega 
+ \mathbf{d}H_\zeta \wedge \mathbf{d}\lambda)  = 0.
\end{equation}

We say that a curve $c : \mathbb R \to \cp^\tau$ is a {\bfi dynamical
trajectory\/} provided $c(\lambda) $ covers 
$\tau(\lambda) $ and its $\lambda$-derivative $\dot c$ satisfies 
\begin{equation}  \label{eqn6D:5}
\dot c  \hook (\omega + \mathbf{d}H_\zeta \wedge \mathbf{d}\lambda) 
=  0.     
\end{equation}
 The terminology is justified by the following result, which shows
that (\ref{eqn6D:5}) is equivalent to  Hamilton's equations.  First note that the
tangent $\dot c$ to any curve $c$ in $\cp^\tau$ covering $\tau$ can be
uniquely split as
\begin{equation}  \label{eqn6D:6} 
\dot c  =  X + \zeta_\cp     
\end{equation}
 where $X$ is vertical in $\cp^\tau$.  Set $X_\lambda =  X \! \bigm|\!
\cp_\lambda$.

\begin{prop}\label{prop6.6}
A curve $c$ in $\cp^\tau$ is a dynamical
trajectory iff {\bfi Hamilton's equations\/}
\begin{equation}  \label{eqn6D:7} 
 X_\lambda \hook\, \omega_\lambda =  \mathbf{d}H_{\lambda, \zeta}     
\end{equation}
 hold at $c(\lambda) $ for every $\lambda \in \mathbb R$.
\end{prop}

\begin{proof} 
  With $\dot c$ as in (\ref{eqn6D:6}),  we compute
\begin{equation}  \label{eqn6D:8} 
\dot c \hook\, (\omega + \mathbf{d}H_\zeta \wedge \mathbf{d}\lambda)  = (X
\hook\,
\omega - \mathbf{d}H_\zeta) + (X[H_\zeta] +
\zeta_\cp[H_\zeta])\, d\lambda.     
\end{equation}
A one-form $\alpha$ on $\cp^\tau$ is zero iff the pull-back of $\alpha$ to
each $\cp_\lambda$ vanishes  and
$\alpha(\zeta_\cp) = 0$.  Applying this to (\ref{eqn6D:8})  gives 
\begin{equation*} 
 X_\lambda \hook \, \omega_\lambda = \mathbf{d}H_{\lambda, \zeta}      
\end{equation*}
 which is (\ref{eqn6D:7}), and 
\begin{equation}  \label{eqn6D:10}
 - \zeta_\cp[H_\zeta] + X[H_\zeta] +
\zeta_\cp[H_\zeta]  =  X[H_\zeta] = 0.     
\end{equation}
 But (\ref{eqn6D:7}) implies (\ref{eqn6D:10}), 
because $\omega_\lambda$ is skew-symmetric. 
\end{proof}

\begin{remark}[Remark \ ]
The difference between the two formulations (\ref{eqn6D:5}) and (\ref{eqn6D:7}) of
the dynamical equations is mainly one of outlook.  Equation (\ref{eqn6D:5})
corresponds to the approach usually taken in time-dependent mechanics
(\`a la Cartan), while (\ref{eqn6D:7}) is usually seen in the context of
conservative  mechanics (\`a la Hamilton), cf. Chapters  3 and 5 of
Abraham and Marsden [1978].  We use  both formulations here, since (\ref{eqn6D:5})
is most easily correlated with the covariant Euler--Lagrange equations
(see below), but (\ref{eqn6D:7}) is more appropriate for a study of the initial
value  problem (see \S 6E). 
\end{remark}

We now relate the Euler--Lagrange equations with Hamilton's equations in
the form (\ref{eqn6D:5}).   This will be done by relating 
the 2-form $\omega + \mathbf{d}H_\zeta \wedge \mathbf{d}\lambda$ on
$\cp^\tau$ with the 2-form $\Omega_\cl$ on
$J^{1}Y$.

Given $\phi \in \cy$, set $\sigma = \mathbb F\cl(j^1\ns\phi)$.  Using the
slicing, we map $\sig$ to a curve
$c_\phi$ in $\cp^\tau$ by applying the reduction map $R_\lambda$ to $\sig$ at
each instant $\lambda$; that is,
\begin{equation}  \label{eqn6D:11} 
 c_\phi(\lambda)   =  R_\lambda(\sig_\lambda)       
\end{equation}
 where $\sig_\lambda = \sigma \circ i_\lambda$ and $i_\lambda : \Sigma_\lambda \to X$ is the
inclusion. (That $c_\phi(\lambda)  \in
\cp_\lambda$ for each $\lambda$ follows from the commutativity of diagram
(\ref{eqn6C:6}).)  The curve $c_\phi$ is called the {\bfi canonical
decomposition\/} of the spacetime field $\phi$ with respect to the given
slicing.

The main result of this section  is the following, which asserts the
equivalence of the Euler--Lagrange  equations with Hamilton's equations.

 \begin{thm}\label{thm6.7}
 Assume {\bf A3} and {\bf A2}.\mbox{}
\begin{enumerate}
\renewcommand{\labelenumi}{\mbox{\rm(\roman{enumi})}}
\item 
Let the spacetime field
$\phi$ be a solution of  the Euler--Lagrange equations.  Then its
canonical decomposition $c_\phi$ with respect to  any slicing satisfies
Hamilton's equations.

\item 
 Conversely, every solution of Hamilton's equations is
the canonical decomposition  (with respect to some slicing) of a
solution of the Euler--Lagrange equations.
\end{enumerate}
\end{thm}

We observe that if $\phi$ is defined only locally (i.e., in a
neighborhood of a Cauchy surface)  and $c_\phi$ is defined in a
corresponding interval $(a, b) \in \mathbb R$, then the Theorem remains
true.

Recall from Theorem~3.1 that $\phi$ is a solution of the
Euler--Lagrange equations iff
\begin{equation}  \label{eqn6D:12} 
 (j^1\ns\phi)^*(\mathbf{i}_V \Omega_\cl)  =  0      
\end{equation}
for all vector fields $V$ on $J^{1}Y$ . 
Recall also that this statement remains valid if we require 
$V$ to be $\pi_{X,J^{1}Y}$-vertical.  Let $V$ be any such vector
field defined along $j^{1}\ns\phi$ and set $W = T\mathbb F\cl \cdot V$. 
For each
$\lambda \in \mathbb R$, define the vector $\cw_\lambda \in T_{c(\lambda) }\cp_\lambda$ by
\begin{equation}  \label{eqn6D:13} 
\cw_\lambda =  TR_\lambda\cdot(W \circ \sig_\lambda) .      
\end{equation}
 As $\lambda$ varies, this defines a vertical vector field $\cw$ on
$\cp^\tau$ along $c_\phi$.

\begin{lem}\label{lem6.8}
Let $V$ be a $\pi_{X,J^{1}Y}$-vertical vector field
on $J^{1}Y$  and $\phi \in
\cy$. With notation as above, we have
\begin{equation}  \label{eqn6D:14} 
\int_{c_\phi} \mathbf{i}_\cw (\omega 
+ \mathbf{d}H_\zeta \wedge \mathbf{d}\lambda)   = 
\int_X (j^1\ns\phi)^*(\mathbf{i}_V
\Omega_\cl ).      
\end{equation}
\end{lem}

\begin{proof}
The left hand side of (\ref{eqn6D:14}) is
$$
\int_{\mathbb R}\left
\{\mathbf{i}_{\dot c_\phi}\mathbf{i}_\cw(\omega + \mathbf{d}H_\zeta \wedge
\mathbf{d}\lambda) \right\}d\lambda ,
$$
 while the right hand side is
$$
\int_{\Sig \times \mathbb R} \fs^*_X (j^1\ns\phi)^*(\mathbf{i}_V \Omega_\cl
)  = 
\int_{\mathbb R}\left\{\int_\Sig 
\mathbf{i}_{\partial/\partial\lambda}\fs^*_X
(j^1\ns\phi)^*(\mathbf{i}_V\Omega_\cl)\right\} d\lambda.
$$
 Thus, to prove (\ref{eqn6D:14}), it suffices to show that
\begin{equation}  \label{eqn6D:15} 
\left(\omega + \mathbf{d}H_\zeta \wedge \mathbf{d}\lambda\right)
\left(\cw, \dot
c_\phi\right) =
\int_\Sig \mathbf{i}_{\partial/\partial\lambda}
 \fs^*_X (j^1\ns\phi)^*(\mathbf{i}_V\Omega_\cl ).      
\end{equation}

Using (3B.2), the right hand side of (\ref{eqn6D:15}) becomes
$$
\begin{aligned}
\int_\Sig \mathbf{i}_{\partial/\partial\lambda} 
\fs^*_X (j^1\ns\phi)^*(\mathbf{i}_V\mathbb F\cl^*\Omega)  
& = \int_\Sig \mathbf{i}_{\partial/\partial\lambda} \fs^*_X
\sigma^*(\mathbf{i}_W \Omega) 
=
\int_\Sig \tau^*_\lambda 
\left[\mathbf{i}_{\zeta_X} \sigma^*(\mathbf{i}_W\Omega)\right] \\[2ex] 
& = 
 \int_{\Sigma_\lambda} i^*_\lambda\left[\mathbf{i}_{\zeta_X}
\sigma^*(\mathbf{i}_W\Omega)\right]  = \int_{\Sigma_\lambda}
i^*_\lambda \sigma^*\left(\mathbf{i}_{T\sigma\cdot
\zeta_X} \mathbf{i}_W\Omega\right) \\[2ex]
& =
\int_{\Sigma_\lambda} \sigma^*_\lambda \left(\mathbf{i}_{T\sigma\cdot
\zeta_X} \mathbf{i}_W
\Omega\right) .
\end{aligned}
$$  
By adding and subtracting the same term, rewrite this as
\begin{equation} \label{eqn6D:16} 
\int_{\Sigma_\lambda} \sigma^*_\lambda \left(\mathbf{i}_{T\sigma\cdot
\zeta_X- \zeta_Z}
\mathbf{i}_W \Omega\right) +
\int_{\Sigma_\lambda} \sigma^*_\lambda \left(i_{\zeta_Z} \mathbf{i}_W
\Omega\right),      
\end{equation}
 where $\zeta_Z$ is the generating vector field of the induced
slicing of $Z$.

We claim that the first term in (\ref{eqn6D:16}) is equal to $\omega(\cw, \dot
c_{\phi})$.  Indeed, since $T\sigma\cdot
\zeta_X - \zeta_Z$ is $\pi_{X\! Z}$-vertical, (\ref{eqn5C:3}) and the fact
that
$R_\lambda$ is canonical give
$$
\begin{aligned}
\int_{\Sigma_\lambda} \sigma^*_\lambda & \left(\mathbf{i}_{T\sigma\cdot
\zeta_X- \zeta_Z} 
\mathbf{i}_W \Omega\right) \\[1.5ex] 
& =\Omega_\lambda(\sig_\lambda) (W \circ \sig_\lambda,  (T\sigma\cdot
\zeta_X - \zeta_Z) \circ \sig_\lambda) \\[2.5ex] 
& =  \omega_\lambda(c_\phi(\lambda) )(TR_\lambda\cdot
[W \circ \sig_\lambda],TR_\lambda\cdot [(T\sigma\cdot \zeta_X - \zeta_Z)
\circ
\sig_\lambda] ).
\end{aligned}
$$ 
Think of $\sig$ as a curve $\mathbb R \to \cn^\tau \subset \cz^\tau$
according to $\lambda \mapsto \sig_\lambda$.  The tangent to this curve
at time
$\lambda$ is $(T\sigma\cdot\zeta_X) \circ \sig_\lambda$ and, from
(\ref{eqn6A:6}), which states that $\zeta_{\cz}(\sig)= \zeta_Z \circ
\sig$, its vertical component is thus
$(T\sigma\cdot\zeta_X -
\zeta_Z)
\circ
\sig_\lambda$.  Since the  curve $\sig$  is mapped onto the curve
$c_\phi$ by
$R_\lambda$, it follows that $TR_\lambda\cdot [(T\sigma\cdot
\zeta_X - \zeta_Z) \circ \sig_\lambda$]  is the vertical component
$X_\lambda$ of
$\dot c_\phi(\lambda) $.   Thus in view of (\ref{eqn6D:13}),
 (\ref{eqn6D:6}), (\ref{eqn6D:1}), and (\ref{eqn6D:2}), the
above becomes
$$
\omega_\lambda(c_\phi(\lambda) )(\cw_\lambda,  X_\lambda)   
=  \omega(c_\phi(\lambda) )(\cw, \dot c_\phi),
$$
 as claimed.

Finally, we show that the second term in (\ref{eqn6D:16}) is just $\mathbf{d}H_\zeta
\wedge \mathbf{d}\lambda(\cw,
\dot c_\phi)$.  We compute at $c_\phi(\lambda)  
= R_\lambda(\sig_\lambda) $:
\begin{align*}
\mathbf{d}H_\zeta \wedge \mathbf{d}\lambda (\cw, \dot c_\phi) 
& = \cw [H_\zeta] 
=  \cw_\lambda[H_{\lambda, \zeta}] 
\\[2ex] 
& = - \cw_\lambda \left[\int_{\Sigma_\lambda}
\sigma^*_\lambda(\mathbf{i}_{\zeta_Z} \Theta)\right] 
= - \int_{\Sigma_\lambda}
         \sigma^*_\lambda(\text{\pounds}_W \mathbf{i}_{\zeta_Z}
\Theta)
\end{align*}
where we have used (\ref{eqn6D:3}), (\ref{eqn6C:10}) and (\ref{eqn6D:13}). 
By Stokes' theorem, this equals
\begin{equation*}
- \int_{\Sigma_\lambda} \sigma^*_\lambda(\mathbf{i}_W
\mathbf{d}\ps\mathbf{i}_{\zeta_Z} \Theta)
 = - \int_{\Sigma_\lambda}\sigma^*_\lambda(\mathbf{i}_W 
\text{\pounds}_{\zeta_Z}\Theta) - 
\int_{\Sigma_\lambda} \sigma^*_\lambda(\mathbf{i}_W
\mathbf{i}_{\zeta_Z} \Omega)
\end{equation*}
and the first term here vanishes since $\zeta_Z$ is a canonical lift
(cf. Remark 3 of \S 6A).
\end{proof}

\begin{proof}[Proof of Theorem~\ref{thm6.7}]

(i) 
First, suppose that $\phi$ is a solution of the Euler-Lagrange
equations.  From Theorem~3.1, the right hand side of (\ref{eqn6D:15})
vanishes.  Thus
\begin{equation}  \label{eqn6D:19} 
(\omega + \mathbf{d}H_\zeta \wedge \mathbf{d}\lambda) (\cw,
\dot c_\phi) =  0      
\end{equation}
  for all $\cw$ given  by (\ref{eqn6D:13}).  By {\bf A3} and 
Proposition~\ref{prop6.3}, every
vector on $\cp^\tau$ has the form
$\cw + f\dot c_\phi$ for some $\cw$ and some function $f$ on $\cp^\tau$. 
Since the form $\omega + \mathbf{d}H_\zeta
\wedge \mathbf{d}\lambda $ vanishes on $(\dot c_\phi, \dot c_\phi)$, it
follows from (\ref{eqn6D:19}) that $\dot c_\phi$ is in the ker\-nel of
$\omega +
\mathbf{d}H_\zeta
\wedge \mathbf{d}\lambda$.  The result now follows from
Proposition~\ref{prop6.6}.

(ii) 
Let $c$ be a curve in $\cp^\tau$.  By Corollary~\ref{cor6.4}
there exists a lift $\sig$ of $c$ to $\cn^\tau$; we think of $\sig$ as
a section of $\pi_{X\ns N}$.  It follows from (\ref{eqn6C:6}) that $\sigma =
\mathbb F\cl(j^1\ns\phi)$ for some $\phi \in \cy$.  Thus every such curve
$c$ is the canonical decomposition of some spacetime section $\phi$.

If $c$ is a dynamical trajectory,  then the right hand side of (\ref{eqn6D:14})
vanishes for every $\pi_{X,J^1Y}$-vertical vector field $V$ on
$J^{1}Y$ .  Arguing as in the proof of Theorem~3.1, it follows
that 
$\phi$ is a solution of the Euler--Lagrange equations. 
\end{proof}


\subsection{
Constraint Theory}

We have just established an important equivalence between solutions of
Hamilton's  equations as trajectories in $\cp^\tau$ on the one hand, and
solutions of the Euler--Lagrange equations as spacetime sections of $Y$
on the other.  This does {\it not\/} imply, however, that there is a
dynamical trajectory through every point in $\cp^\tau$.  Nor does it
imply that if such a trajectory exists it will be unique.  Indeed, two
of the novel features of classical field dynamics, usually absent in
particle dynamics, are the presence of both constraints on the choice
of Cauchy data and unphysical (``gauge") ambiguities in the resulting
evolution.  In fact, essentially every classical field theory of
serious interest---with the exception of pure Klein--Gordon type
systems---is both over- and underdetermined in these senses.  Later in
Part III, we shall use the energy-momentum map (as defined in \S 7F)
as a tool for understanding the constraints and gauge freedom of
classical field  theories.  In this section we give a rapid
introduction to the more traditional theory of initial value
constraints and gauge transformations following Dirac [1964] as
symplectically reinterpreted by  Gotay, Nester, and Hinds [1978].  An
excellent general reference is the book by Sundermeyer  [1982]; see
also Gotay [1979], Gotay and Nester [1979], and Isenberg and Nester [1980].
\medskip

We begin by abstracting the setup for dynamics in the instantaneous
formalism as presented  in \S\S 6A--D. Let $\cp$ be a manifold (possibly
infinite-dimensional) and let $\omega$ be a presymplectic  form on $\cp$. 
We consider differential equations of the form 
\begin{equation}  \label{eqn6E:1} 
\dot p =X(p)      
\end{equation}
 where the vector field $X$ satisfies
\begin{equation}  \label{eqn6E:2} 
\mathbf{i}_X\omega =  \mathbf{d} H      
\end{equation}
for some given function  $H$ on $\cp$.  Finding vector field
solutions $X$ of (\ref{eqn6E:2}) is an algebraic  problem at each point. 
When $\omega$ is symplectic, (\ref{eqn6E:2}) has a unique solution $X$. 
But when
$\omega$ is presymplectic, neither existence nor uniqueness of solutions
$X$ to (\ref{eqn6E:2}) is guaranteed.  In fact, $X$  exists at a point $p
\in \cp$ iff $\mathbf{d}H(p)$ is contained in the image of the map
$T_p\cp \to T^*_p\cp$ determined by $X \mapsto \mathbf{i}_X\omega$.

Thus one cannot expect to find globally defined solutions $X$ of
(\ref{eqn6E:2}); in general, if $X$ exists  at all, it does so only along
a submanifold
$\cq$ of $\cp$.  But there is another  consideration which is central
to the physical interpretation of these constructions:  we want
solutions $X$ of (\ref{eqn6E:2}) to generate (finite) temporal evolution
of the ``fields'' $p$ from the given ``Hamiltonian" $H$ via
(\ref{eqn6E:1}).  But this can occur on $\cq$ only if $X$ is {\it
tangent\/} to $\cq$.  Modulo considerations of well-posedness (see Remark
1 below), this ensures that $X$ will generate a flow on $\cq$ or, in
other words, that (\ref{eqn6E:1}) can be integrated. This additional
requirement further reduces the set on which (\ref{eqn6E:2}) can be
solved.

In Gotay, Nester and Hinds [1978]---hereafter abbreviated by GNH---a
geometric  characterization of the sets on which (\ref{eqn6E:2}) has
tangential solutions is presented.  The characterization  relies on the
notion of ``symplectic polar."  Let $\cq$ be a submanifold of $\cp$.  At
each $p
\in
\cq$, we define the {\bfi symplectic polar}
$T_p\cq^\bot$ of $T_p\cq$ in $T_p\cp$ to be   
$$
 T_p\cq^\bot = \left\{V \in T_p\cp \mid \omega(V, W) = 0 \quad \text{for
all} \quad  W \in T_p\cq\right\}.      
$$
 Set 
$$
 T\cq^\bot =\bigcup_{p \in \cq} T_p\cq^\bot.
$$
   Then GNH proves the following result, which provides the
necessary and sufficient conditions for the existence of tangential
solutions to (\ref{eqn6E:2}).

\begin{prop}\label{prop6.9}
The equation
\begin{equation}\label{eqn6E:3a}
(\mathbf{i}_X\omega - \mathbf{d}H)\!  \bigm| \! \cq  =  0 
\end{equation}
 possesses solutions $X$ tangent to $\cq$ iff the directional
derivative of $H$ along any vector in
$T\cq^\bot$ vanishes:
\begin{equation}\label{eqn6E:3b}
 T\cq^\bot [\ps H \ps]  =  0 . 
\end{equation}
\end{prop}

Moreover, GNH develop a symplectic version of Dirac's ``constraint
algorithm" which computes the {\it unique maximal\/} submanifold $\cc$
of
$\cp$ along which (\ref{eqn6E:2}) possesses solutions tangent to $\cc$. 
This {\bfi  final constraint submanifold\/} is the limit $\cc = 
\underset l \cap \,\cp^l$ of a string of sequentially defined {\bfi
constraint submanifolds\/}
\begin{equation}  \label{eqn6E:4} 
\cp^{l+1} = \left\{p \in \cp^l \mid (T_p\cp^l)^\bot [\ps H \ps] =
0\right\}      
\end{equation}
 which follow from applying the consistency
conditions (\ref{eqn6E:3b}) to (\ref{eqn6E:2}) beginning with $\cp^1 = \cp
$.  The basic facts are as follows. 

 \begin{thm}\label{thm6.10}
\begin{enumerate}
\item [\rm{(i)}]
 Equation 
{\rm  (\ref{eqn6E:2})} is consistent---that is, it admits
tangential solutions---iff  $\cc
\neq \varnothing$, in which case there are vector fields $X \in \Fx(\cc
)$ such that
\begin{equation}  \label{eqn6E:5} 
 (\mathbf{i}_X\omega - \mathbf{d}H) \! \bigm| \! \cc = 0.      
\end{equation}

\item [\rm{(ii)}]
 If $\cq \subset \cp$ is a submanifold along which 
{\rm (\ref{eqn6E:3a})} holds  with $X$
tangent to $\cq$, then $\cq  \subset \cc$.
\end{enumerate}
\end{thm}

The following useful characterization of the maximality of $\cc$
follows from ({ii}) above and  Proposition~\ref{prop6.9}.

\begin{cor}\label{cor6.11}
$\cc$ is the largest submanifold of $\cp$
with the property that
\begin{equation}  \label{eqn6E:6} 
T\cc^\bot [\ps H \ps] = 0.      
\end{equation}
\end{cor}

\begin{remarks}[Remarks \ 1.]
These results can be thought of as providing {\it formal\/}
integrability criteria for equation (\ref{eqn6E:1}), since they
characterize the existence of the vector field $X$, but do not imply that
it can actually be integrated to a flow.  The latter problem is a
difficult analytic one, since in  classical field theory (\ref{eqn6E:1})
is usually a system of hyperbolic PDEs and great care is required  (in
the choice of function spaces, etc.) to guarantee that there exist
solutions which  propagate for finite times.  We shall not consider this
aspect of the theory and  will simply assume, when
necessary, that (\ref{eqn6E:1}) is well-posed in a suitable sense.  See 
Hawking and Ellis [1973] and Hughes, Kato, and Marsden [1977] for some
discussion of  this issue.  Of course, in finite dimensions
(\ref{eqn6E:1}) is a system of ODEs and so integrability is automatic.

\paragraph{\bf 2.}
We assume here that each of the $\cp^l$  as well as $\cc$ are
smooth submanifolds of $\cp$.  In practice, this need {\it not\/}  be
the case; the $\cp^l$ and $\cc$ typically have quadratic singularities 
(refer to item {\bf 7} in the Introduction).  In such cases our
constructions and results must be  understood to hold at smooth points.  We
observe, in this regard, that the singular sets of  the $\cp^l$ and $\cc$
usually have nonzero codimension therein, and that constraint sets are 
``varieties" in the sense that they are the closures of their smooth
points.  For an  introduction to some of the relevant ``singular
symplectic geometry", see Arms, Gotay, and  Jennings [1990] and Sjamaar
and Lerman [1991].

\paragraph{\bf 3.}
In infinite dimensions, Proposition~\ref{prop6.9} and the
characterization (\ref{eqn6E:4}) of the $\cp^l$ are not valid without
additional technical qualifications which we will not enumerate here. 
See Gotay  [1979] and Gotay and Nester [1980] for the details in the
general case.

\paragraph{\bf 4.}
The above results pertain to the {\it existence\/} of solutions to
(\ref{eqn6E:2}).  It is crucial to realize that solutions, when they
exist, generally are not {\it unique\/}:  if $X$ solves (\ref{eqn6E:5}),
then so does 
$X + V$ for any vector field $V \in \ker \omega \cap \Fx(\cc )$. Thus,
besides being {\bfi overdetermined\/} (signaled by a {\it strict\/}
inclusion $\cc \subset \cp$), equation (\ref{eqn6E:2}) is also in general
{\bfi underdetermined\/}, signaling the presence of gauge freedom in the
theory.  
We will have more to say about this later.
\end{remarks}

We discuss one more issue in this abstract setting: the notions of first
and second class  constraints.  We begin by recalling the
classification scheme for submanifolds of presymplectic  manifolds
$(\cp, \omega)$.  Let $\cc \subset \cp$;  then $\cc$ is
\begin{enumerate}
\item [({i})]
 {\bfi isotropic\/} if $T\cc \subset T\cc^\bot$
\item [({ii})]
 {\bfi coisotropic\/} or {\bfi first class\/} if $T\cc^\bot
\subset T\cc$
\item [({iii})]
 {\bfi symplectic\/} or {\bfi second class\/} if $T\cc \cap
T\cc^\bot = \{0\}$.
\end{enumerate} 
These conditions are understood to hold at every point of
$\cc$.  If $\cc$ does not happen to fall into any of these categories,
then $\cc$ is called {\bfi mixed\/}.  Note as well that the classes are
not disjoint: a submanifold can be simultaneously isotropic and
coisotropic, in which case $T\cc = T\cc^\bot$ and $\cc$ is called {\bfi
Lagrangian\/}.

From the point of view of the submanifold $\cc$,  this classification
reduces to a  characterization of the closed 2-form $\omega_\cc$ obtained
by pulling $\omega$ back to $\cc$.  Indeed,
\begin{equation}  \label{eqn6E:7} 
\ker \omega_\cc = T\cc \cap T\cc^\bot.      
\end{equation}
 In particular, $\cc$ is isotropic iff $\omega_\cc = 0$ and symplectic
iff $\ker \omega_\cc = \{0\}$.  Our main interest will be in the
coisotropic case.

A {\bfi constraint\/} is a function $f \in \cf(\cp)$ which
vanishes on (the final constraint set) $\cc$.  The classification of
constraints depends on how they relate to $T\cc^\bot$.  A constraint
$f$ which satisfies
\begin{equation} \label{eqn6E:8} 
 T\cc^\bot [\ps f\ps]  =  0      
\end{equation}
 everywhere on $\cc$ is said to be {\bfi first class\/};  otherwise it
is {\bfi second class\/}.  (These definitions are due to Dirac [1964].)

\begin{prop}\label{prop6.12}
\begin{enumerate}
\item [{\rm({i})}]
 Let $f$ be a constraint.  Then the Hamiltonian vector
field $X_{\ns f}$ of $f$,  defined by 
$\mathbf{i}_{X_{\ns f}}\omega = \mathbf{d}f$,  exists along $\cc$ iff\/
$T\cp^\bot[\ps f \ps]\!\bigm|\! \cc = 0$. If it exists, then 
$X_{\ns f} \in
\Fx(\cc)^\bot$.
\item [{\rm({ii})}]
 Conversely, at every point of $\cc$, $T\cc^\bot$ is
pointwise spanned by the Hamiltonian vector fields of constraints. 
\item [{\rm({iii})}]
 Let $f$ be a first class constraint.  Then the
Hamiltonian vector field $X_{\ns f}$ of $f$ exists along $\cc$ and $X_{\ns
f} \in
\Fx(\cc) \cap \Fx(\cc)^\bot$.
\item [{\rm({iv})}]
 Conversely, at every point of $\cc$, $T\cc \cap
T\cc^\bot$ is pointwise spanned by the Hamiltonian vector fields of
first class constraints.
\end{enumerate}
\end{prop}

\begin{proof}

({i}) We study the equation
\begin{equation}  \label{eqn6E:9}
\mathbf{i}_{X_{\ns f}}\omega = \mathbf{d}f      
\end{equation}
 at $p \in \cc$. The first assertion follows immediately from 
Proposition~\ref{prop6.9} upon taking $\cq = \cp$. Then, if $X_{\ns
f}$ exists, $\omega(X_{\ns f},T_p\cc) = T_p\cc[\ps f\ps] = 0$ as
$f$ is a constraint, whence $X_{\ns f}(p) \in T_p\cc^\bot$.
%

({ii}) Let $V \in T_p\cc^\bot$ and set $\alpha =
\mathbf{i}_V\omega$.   Fix a
neighborhood $U$ of $p$ in $\cp$ and a Darboux chart 
$\psi : (U, \omega \! \bigm| \! U) \to (T_p\cp,\omega_p)$ such that  
\begin{enumerate} \item[] \begin{enumerate}
\item [\quad ({a})]
 $\psi(p) = 0$,  
\item [({b})]
 $T_p\psi = id_{T_p\cp}$ and 
\item [({c})]
 $\psi$ flattens $U \cap \cc$ onto $T_p\cc$.  
\end{enumerate} \end{enumerate}
Set $f =\alpha \circ \psi$ so that, by ({b}), 
$\mathbf{d}f(p)
=\mathbf{i}_V\omega$.  Then ({c}) yields
$$
 f(U \cap \cc ) = \alpha(\psi(U \cap \cc )) \subset \alpha(T_p\cc )  = 
\omega_p(V, T_p\cc )
$$
 which vanishes as $V \in T_p\cc^\bot$.  Thus $f$ is a constraint in
$U$ and the desired globally defined constraint is then $gf$,  where $g$ is a
suitable bump function.  

({iii}) 
Applying Proposition~\ref{prop6.9} to (\ref{eqn6E:9}) along
$\cc$ and taking (\ref{eqn6E:8}) into account, we see that $X_{\ns f}$ 
exists and is tangent to
$\cc$.  The result now follows from ({i}).

({iv}) Let $V \in T_p\cc \cap T_p\cc^\bot$.  We proceed as in
({ii}); it remains to show that $f$ is first class.  For any $q \in U
\cap
\cc$ and $W \in T_q\cc^\bot$, 
$$
\mathbf{d}f(q)\cdot W = (\alpha \circ T_q\psi)\cdot W 
= \omega_p(V, T_q\psi\cdot
W).
$$
 But $\psi$ is a symplectic map, and consequently $T_q\psi \cdot  W \in
T_p\cc^\bot$ in $T_p\cp$.  Therefore,
 $\omega_p(V, T_q\psi \cdot W) = 0$ as $V \in T_p\cc$.  
 Then $gf$ is
the desired globally defined first class constraint, where $g$ is a
suitable bump function. 
\end{proof}

\begin{remark}[Remark \ 5.]
Strictly speaking, $X_{\ns f}$ is defined only up to elements of
$\ker \omega = \Fx(\cp)^\bot$,  but we abuse the language and continue to
speak of ``the" Hamiltonian vector field $X_{\ns f}$ of the constraint
$f$.
\end{remark}

From this Proposition it follows that a second class submanifold can be
locally described  by the vanishing of second class constraints. 
Similarly, if $\cc$ is coisotropic, then all constraints are first
class.  In general, a mixed or isotropic submanifold will require both
classes of constraints for its local description.
\bigskip

We now apply the abstract theory of constraints, as just described, to
the study of classical  field theories.  To place these results into
the context of dynamics in the instantaneous formalism,  we fix an
infinitesimal slicing $(Y_\tau, \zeta)$.  
Then $(\cp, \omega)$ is
identified with the primary constraint submanifold $(\cp_\tau, 
\omega_\tau) $ of
\S 6C, $H$ with the Hamiltonian $H_{\tau,\zeta}$ and (\ref{eqn6E:2}) 
with Hamilton's
equations
\begin{equation}  \label{eqn6E:10} 
\mathbf{i}_X\omega_\tau = \mathbf{d}H_{\tau,\zeta},      
\end{equation}
 cf. \S 6D.  We have the sequence of constraint submanifolds
\begin{equation}  \label{eqn6E:11} 
\cc_{\tau,\zeta} \subset \dots \subset \cp^l_{\tau,\zeta} \subset \dots
\subset \cp_\tau \subset T^*\cy_\tau.     
\end{equation}
 {\it A priori\/}, for $l \geq 2$ the $\cp^l_{\tau,\zeta}$ depend upon
the evolution direction $\zeta$ through the consistency conditions
(\ref{eqn6E:4}), as $H_{\tau,\zeta}$ does.  We will soon see, however,
that the final constraint set is independent of $\zeta$.\footnote{\ In
fact, none of the $\cp^{\ps l}_{\tau,\zeta}$ depend upon $\zeta$, but
we shall not prove this here. We have already shown in
Corollary~\ref{cor6.4} that the primary constraint set is
independent of $\zeta$.}

The functions whose vanishing defines $\cp_\tau$ in $T^*\cy_\tau$ are
called {\bfi primary constraints\/};   they arise because of the
degeneracy of the Legendre transform.  Similarly, the functions whose 
vanishing defines $\cp^l_{\tau,\zeta}$ in $\cp^{l-1}_{\tau,\zeta}$ are
called $l$-{\bfi ary constraints\/} (secondary, tertiary, $\dots$). 
These constraints are generated by the constraint algorithm.  Sometimes,
for brevity, we shall refer to all $l$-ary constraints for $l \geq 2$ as
``secondary."  When we refer to the ``class" of a constraint, we will
adhere to the following conventions, unless otherwise noted.  The class
of a {\it secondary\/} constraint will always be computed relative
to 
$(\cp_\tau, \omega_\tau) $,  whereas that of a {\it primary\/} 
constraint relative
to  
$T^*\cy_\tau$ with its canonical symplectic form.  
Similarly, if $\cq_\tau \subset \cp_\tau$, the polar
$T\cq_\tau{\!}^\bot$ will be taken with respect to $(\cp_\tau, \omega_\tau)
$;  in
particular, $\cq_\tau$ is coisotropic, etc., if it is so relative to the
primary constraint submanifold.

These constraints are all {\bfi initial value constraints\/}.  Indeed,
thinking of $\Sigma_\tau$ as the  ``initial time," elements $(\varphi, 
\pi)
\in \cc_{\tau,\zeta}$ represent admissible initial data for the
$(n+1)$-decomposed field equations (\ref{eqn6E:10}).  Pairs $(\varphi,
\pi)$ which do not lie in
$\cc_{\tau,\zeta}$ cannot be propagated, even formally, a finite time
into the future.  The next series of results will serve to make these
observations precise.

Let $\operatorname{Sol}$ denote the set of all spacetime solutions of
the Euler--Lagrange equations.   (Without loss of generality, we will
suppose in the rest of this section that such solutions are globally
defined.)  Fix a Lagrangian slicing with parameter $\lambda$.  Referring
back to 
\S6D,  we define a map can : $\operatorname{Sol}\to
\Gamma(\cp^\tau) $  by assigning to each $\phi \in \operatorname{Sol}$ 
its canonical decomposition
$c_\phi$ with respect to the slicing.  Observe that, for each fixed
$\lambda \in \mathbb R$, can${}_\lambda(\phi) = c_\phi(\lambda)  \in
\cp_\lambda$ depends only upon
$\phi$ and the Cauchy surface $\Sigma_\lambda$,  but {\it not\/} on the
slicing.

\begin{prop}\label{prop6.13}
Assume {\bf A3} and {\bf A2}. Then, for each
$\lambda \in \mathbb R$,
$$
\text
{\rm  can$_\lambda$(Sol)}\subset \cc_{\lambda, \zeta}.
$$
\end{prop}

\begin{proof}
 Let $\phi \in \operatorname{Sol}$ and set $\lambda =
0$ for simplicity.  We will show that can$_0(\phi) = c_\phi(0) \in
\cc_{0,\zeta}$.  Define a curve $\gamma : \mathbb R \to \cp_0$ by
\begin{equation}  \label{eqn6E:12} 
\gamma(s) = f_{-s}(c_\phi(s))      
\end{equation}
 where $f_s$ is the flow of $\zeta_\cp$. We may think of $c_\phi$ in
$\cp^\tau$ as ``collapsing" onto $\gamma$ in $\cp_0$ as in Figure 6-6.

\begin{figure}[ht] \label{Gimmsy6-6}
\begin{center}
\includegraphics[scale=.9,angle=0]{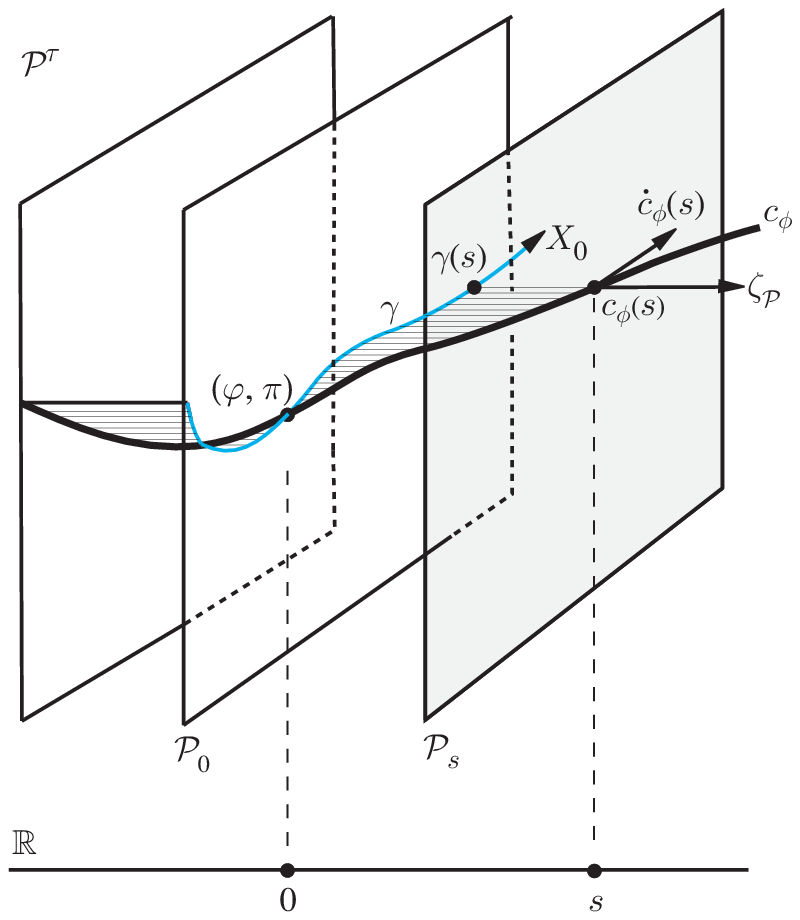}
\caption{Collapsing dynamical trajectories}
\end{center}
\end{figure}

Define a one-parameter family of curves  
$c^s: \mathbb R \to \cp^\tau$ by
\begin{equation*}  \label{eqn6E:13} 
c^s(t) = f_{-s}(c_\phi(s + t)) .      
\end{equation*}
By Theorem~\ref{thm6.7}({i}), $c_\phi$ is a dynamical trajectory. 
Using (6D.4) we see from  (6D.5) that each
curve $c^s$ is also a dynamical trajectory ``starting" at $c^s(0) =
\gamma(s)$.

The tangent to each curve $c^s(t)$ at $t = 0$ takes the form
$$
\left.\frac{d}{dt} c^s(t)\right|_{t=0} =  X_0(\gamma(s)) + \zeta_\cp
(\gamma(s)),
$$
 where $X_0$ is a vertical vector field on $\cp_0$ along $\gamma$.  From
(\ref{eqn6E:12}) it follows that $X_0(\gamma(s))$ is the tangent to
$\gamma$ at $s$.

Proposition~\ref{prop6.6} applied to each dynamical trajectory $c^s$ at
$t=0$ implies that
$X_0(\gamma(s))$ satisfies Hamilton's equations (\ref{eqn6E:10}) at each
point  
$\gamma(s)$.  Since $X_0$ is tangent to $\gamma$, 
Theorem~\ref{thm6.10}({ii}) shows that the image of $\gamma$ lies in
$\cc_{0,\zeta}$.  In particular, $\gamma(0) = c_\phi(0) \in \cc_{0,\zeta}$.
\end{proof}

\medskip

This Proposition shows that {\it only\/} initial data $(\varphi, \pi) \in
\cc_{\lambda, \zeta}$ can be extended to solutions of the Euler--Lagrange
equations.  The converse is true if we assume well-posedness.  We say
that the Euler--Lagrange equations are {\bfi well-posed\/} relative to a
slicing $\fs_Y$ if every
$(\varphi, \pi) \in \cc_{\lambda, \zeta}$ can be extended to a dynamical
trajectory $c :\,]\lambda -\varep,\lambda + \varep[
$ $\subset \mathbb R \to \cp^\tau$ with $c(\lambda) = (\varphi, \pi)$ and
that this solution trajectory depends continuously (in a chosen
function space topology) on $(\varphi, \pi)$. 
This will be a standing assumption in what follows.

\paragraph{A4 \ Well-Posedness}\enspace
{\it The Euler--Lagrange
equations are well-posed.
}  \\

In this notion of well-posedness, one has to keep in mind that we are
assuming that there is a \emph{given} slicing of the configuration bundle
$Y$. However, we will later prove (in Chapter 13) that well-posedness relative
to one slicing with a given Cauchy surface $\Sig$ as a slice will imply
well-posedness of any other (appropriate) slicing also containing $\Sig$ as a
slice. 

Well-posedness for theories without gauge freedom reduces, in
specific examples, to the well-posedness of a system of PDE's
describing that theory in a given slicing. These will be the
Hamilton equations that we have developed, written out in
coordinates. In the case of metric field theories, one typically would then
use theorems on strictly hyperbolic systems to establish well-posedness
(relative to a slicing by spacelike hypersurfaces).

The situation for theories with gauge freedom is a bit more subtle.
However, it has been established that well-posedness holds for
``standard'' theories such as Maxwell, Einstein, Yang-Mills and their
couplings. Here, very briefly, is how the argument goes for the case of the
Einstein equations (in the ADM formulation). To follow this argument, the
reader will need to be familiar with works on the initial value formulation of
Einstein's theory, such as Choquet--Bruhat [1962] and Fischer and Marsden
[1979b].

If one has a slicing $\fs_Y$ specified, and one gives initial data
$(\varphi,\pi)
\in \cc_{0,\zeta}$ over a Cauchy surface $\Sig_0$, then one
first takes this data and evolves it using a particular gauge or coordinate
choice in which the evolution equations form a strictly hyperbolic (or
symmetric hyperbolic) system. This then generates a
piece of spacetime on a tubular neighborhood $U$ of the initial
hypersurface and the solution $\phi$ so
constructed (in this case the metric) on this
piece of spacetime varies continuously with the choice of initial data. The
solution then satisfies the Euler--Lagrange equation. Since $\Sig_0$ is
compact, there exists an $\epsilon > 0$ such that
$\fs_X(]-\epsilon,\epsilon[\times \Sig) \subset U$. Thus $\phi$ induces the
required dynamical trajectory
$c_\phi:\; ]-\epsilon,\epsilon[ \; \to \cp^\tau$ with $c_\phi(0)
=(\varphi,\pi)$ relative to the given slicing. The argument for other
field theories follows a similar pattern.

As was indicated in the Introduction, the above notion of well-posedness
is not the same as the question of existence of solutions of the initial
value problem for a given choice of lapse and shift (or their
generalization, called atlas fields, to other field
theories) on a Cauchy surface. This is a more subtle question that we shall
address later in Chapter 13. The essential difference is that with a given
initial choice of lapse and shift, one still needs to construct the slicing,
whereas in the present context we are assuming that a slicing has been given.

There is evidence that well-posedness fails in both of the above
senses for many $R+R^2$ theories of gravity, as well as for most
couplings of higher-spin fields to Einstein's theory (with
supergravity being a notable exception; see Bao, Choquet--Bruhat,
Isenberg, and Yasskin [1985]).

This assumption together with  Proposition~\ref{prop6.13} yield:

\begin{cor}\label{cor6.14}
If {\bf A3} and {\bf A4} hold, then 
{\rm 
can}${}_\lambda${\rm(Sol)} $=
\cc_{\lambda, \zeta}$. 
\end{cor}

Since, as noted previously, ${\rm can}_\lambda$ depends only upon the
Cauchy surface $\Sigma_\lambda$,  we have:

\begin{cor}\label{cor6.15}
$\cc_{\lambda, \zeta}$ is independent of $\zeta$.
\end{cor}

Henceforth we denote the final constraint set simply by $\cc_\lambda$.  In
particular, this implies that the constraint algorithm computes the
same final constraint set regardless of which Hamiltonian
$H_{\lambda, \zeta}$ is employed, as the generator $\zeta$ ranges over all
compatible slicings (with
$\Sigma_\lambda$ as a slice).

Proposition~\ref{prop6.13} shows that every dynamical trajectory $c :
\mathbb R \to \cp^\tau$ ``collapses" to an  integral curve of Hamilton's
equations in $\cc_\lambda$ for each $\lambda$.  We now prove the converse; that
is,  every integral curve of Hamilton's equations on $\cc_\lambda$
``suspends" to a dynamical trajectory in 
$\cp^\tau$.

\begin{prop}\label{prop6.16}
Let $\gamma$ be an integral curve of a tangential
solution $X_\lambda$ of Hamilton's  equations on $\cc_\lambda$.  Then $c :
\mathbb R
\to \cp^\tau$ defined by
\begin{equation}  \label{eqn6E:14} 
 c(s) = f_s(\gamma(s))      
\end{equation}
 is a dynamical trajectory. 
\end{prop}

\begin{proof}
 Again setting $\lambda = 0$,  (\ref{eqn6E:14}) yields
\begin{equation}  \label{eqn6E:15}
\dot c(s) = X_s(c(s)) + \zeta_\cp (c(s))      
\end{equation}
 where $X_s = Tf_s\cdot X_0$.  Since $X_0(\gamma(s))$ satisfies
(\ref{eqn6D:7}) with $\lambda = 0$ for every $s$, (\ref{eqn6D:4}) implies
that $X_s(c(s))$ satisfies (\ref{eqn6D:7}) for every $s$.  The desired
result now follows from (\ref{eqn6E:15}) and Proposition~\ref{prop6.6}. 
\end{proof}
   
Combining the proof of Proposition~\ref{prop6.13} with
Proposition~\ref{prop6.16}, we have:

\begin{cor}\label{cor6.17}
The Euler--Lagrange equations are well-posed
iff every tangential solution
$X_\lambda$ of Hamilton's equations on $\cc_\lambda$ integrates to a
{\rm (}local in time{\rm )} flow for every $\lambda \in
\mathbb R$. 
\end{cor}
\medskip

It remains to discuss the role of gauge transformations in constraint
theory.  Just as initial  value constraints reflect the overdetermined
nature of the field equations, gauge transformations  arise when these
equations are underdetermined.

Classical field theories typically exhibit {\bfi gauge freedom\/} in the
sense that a given set of  initial data $(\varphi, \pi) \in \cc_\lambda$
does not suffice to uniquely determine a dynamical trajectory.  Indeed,
as noted in Remark 4, if $X_  \lambda$ is a tangential solution of Hamilton's
equations
\begin{equation}  \label{eqn6E:16} 
(X_  \lambda \hook \,\omega_\lambda 
 - \mathbf{d}H_{\lambda, \zeta}) \! \bigm| \! \cc_\lambda = 
0,     
\end{equation}
 then so is $X_  \lambda + V$ 
 for any vector field $V \in \ker \omega_\lambda \cap
\Fx(\cc_\lambda) $.  
For this reason we call vectors in $\ker \omega_\lambda \cap
T\cc_\lambda$ {\bfi kinematic\/} directions.

This is not the entire story, however; the indeterminacy in the
solutions to the field  equations is somewhat more subtle than
(\ref{eqn6E:16}) would suggest.  It turns out that solutions of
(\ref{eqn6E:16}) are  fixed only up to vector fields in
$\Fx(\cc_\lambda)  \cap
\Fx(\cc_\lambda) ^\bot$ which, in general, is larger than $\ker
\omega_\lambda \cap
\Fx(\cc_\lambda) $:
$$
\ker \omega_\lambda \cap \Fx(\cc_\lambda)  =
\Fx(\cp_\lambda) ^\bot \cap \Fx(\cc_\lambda)  \subset \Fx(\cc_\lambda) ^\bot \cap
\Fx(\cc_\lambda) .
$$

To see this, consider a Hamiltonian vector field $V \in \Fx(\cc_\lambda) 
\cap
\Fx(\cc_\lambda) ^\bot$; according to Proposition~\ref{prop6.12}(iv),
$\mathbf{i}_V
\omega_\lambda = \mathbf{d}f$ where $f$ is a first class constraint. 
Setting 
$X'_\lambda = X_  \lambda + V$, (\ref{eqn6E:16}) yields
\begin{equation}  \label{eqn6E:17} 
 (X'_\lambda \hook \,\omega_\lambda - \mathbf{d}(H_{\lambda, \zeta} +
f))\! \bigm| \! \cc_\lambda = 0.	      
\end{equation}
 Thus if $X_  \lambda$ is a tangential solution of Hamilton's equations
along $\cc_\lambda$ with Hamiltonian
$H_{\lambda, \zeta}$, then $X'_\lambda$ is a tangential solution of
Hamilton's equations along $\cc_\lambda$ with Hamiltonian $H'_{\lambda,
\zeta} = H_{\lambda, \zeta} + f$.

Physically, equations (\ref{eqn6E:16}) and (\ref{eqn6E:17}) are
indistinguishable.  Put another way, dynamics is  insensitive to a
modification of the Hamiltonian by the addition of a first class
constraint.  The  reason is that
$H'_{\lambda, \zeta} = H_{\lambda, \zeta}$ along $\cc_\lambda$ and it is
only what happens along $\cc_\lambda$ that matters for the physics;
distinctions that are only manifested ``off'' $\cc_\lambda$---that is,
in a dynamically inaccessible region---have no significance whatsoever. 
Thus the ambiguity in the solutions of Hamilton's equations is
param\-etr\-ized by
$\Fx(\cc_\lambda)  \cap \Fx(\cc_\lambda) ^\bot$.  For further discussion
of these points see GNH, Gotay and Nester [1979], and Gotay [1979,\! 1983].

\begin{remarks}[Remarks \ 6.]
We may rephrase the content of the last paragraph by saying
that what is really of central  importance for dynamics is not
Hamilton's equations per se, but rather their {\it pullback\/} to  
$\cc_\lambda$;  the pullbacks of (\ref{eqn6E:16}) and (\ref{eqn6E:17}) to
$\cc_\lambda$ coincide.  Furthermore, $\Fx(\cc_\lambda)  \cap
\Fx(\cc_\lambda) ^\bot$ is just the kernel of the pullback of $\omega_\lambda$ to
$\cc_\lambda$,  cf. (\ref{eqn6E:7}).

\paragraph{\bf 7.} 
Notice also that since $f$ is first class, (\ref{eqn6E:6}) and
(\ref{eqn6E:8}) guarantee that the constraint algorithm  computes the
same final constraint submanifold using either Hamiltonian $H_{\lambda,
\zeta}$ or
$H'_{\lambda, \zeta}$. 

\paragraph{\bf 8.}  
The addition of first class constraints to the Hamiltonian
(with Lagrange multipliers) is a familiar feature of the Dirac--Bergmann
constraint theory. 
\end{remarks}

The distribution $\Fx(\cc_\lambda)  
\cap \Fx(\cc_\lambda) ^\bot$ on  $\cc_\lambda$ is
involutive and so defines a foliation of $\cc_\lambda$.  Initial data
$(\varphi, \pi)$ and $(\varphi', \pi')$ lying on the same leaf of this
foliation are said to be {\bfi gauge-equivalent\/}; solutions obtained
by integrating gauge-equivalent initial data cannot be distinguished
physically.  We call $\Fx(\cc_\lambda)  \cap
\Fx(\cc_\lambda) ^\bot$ the {\bfi gauge algebra\/} 
and elements thereof {\bfi 
gauge vector  fields\/}.  The flows of such vector fields preserve this
foliation and hence map initial data to  gauge-equivalent initial data;
they are therefore referred to as {\bfi gauge transformations\/}.

Proposition~\ref{prop6.12} establishes the fundamental relation between
gauge transformations and  initial value constraints:  {\it first class
constraints generate gauge transformations\/}.  This encapsulates a
curious feature of classical field theory:  the field equations being
simultaneously overdetermined and underdetermined.  These
phenomena---{\it a priori\/} quite different and distinct---are
intimately correlated via the symplectic structure.  Only in special
cases (i.e., when
$\cc_\lambda$ is symplectic) can the field equations be overdetermined
without being underdetermined.  Conversely, it is not possible to have
gauge freedom without initial value constraints.


\begin{remark}[Remark \ 9.]
In Part III we will prove that the Hamiltonian
(relative to a $\cg$-slicing) of a parametrized field theory in which all
fields are variational vanishes on the final constraint set. Pulling
(\ref{eqn6E:16}) back to
$\cc_\lambda$ (cf. Remark 6), it follows that $X_  \lambda \in {\rm ker}\;
\omega_{\cc_\lambda}$---that is, \emph{the evolution is totally gauge!} We
will explicitly verify this in Examples {\bf a, c} and {\bf d} forthwith.
\end{remark}

A more detailed analysis using Proposition~\ref{prop6.12} (see also
Chapters 10 and 11) shows that the  first class {\it primary\/}
constraints correspond to gauge vector fields in $\ker \omega_\lambda
\cap
\Fx(\cc_\lambda) $, while first class {\it secondary\/} constraints
correspond to the remaining gauge vector fields in $\Fx(\cc_\lambda) 
\cap
\Fx(\cc_\lambda) ^\bot$, cf. GNH. In this context, it is worthwhile to
mention that second class constraints bear no relation to gauge
transformations at all.  For if $f$ is second class, then by
Proposition~\ref{prop6.12}, if it exists its Hamiltonian vector field
$X_{\ns f} \in T\cc_\lambda{}^{\ns\bot}$ everywhere along $\cc$, but 
$X_{\ns  f} \notin T\cc_\lambda$ at least at one point.  Thus $X_{\ns  f}$
tends to flow initial data {\it off\/}
$\cc_\lambda$,  and hence does not generate a transformation of
$\cc_\lambda$. An extensive discussion of second class constraints is
given by Lusanna [1991].

The field variables conjugate to the first class primary constraints
have a special property  which will be important later. We sketch the
basic facts here and refer the reader to Part IV for  further
discussion.

Consider a nonsingular first class primary constraint $f = 0$.  Let $g$ be
canonically conjugate to $f$ in the  sense that 
$$
\omega_{T^*\ns\cy_\lambda}(X_{\ns  f}, X_  g) = 1.
$$
 After a canonical change of coordinates, if necessary, we may write
\begin{equation}  \label{eqn6E:18} 
\omega_{T^*\ns\cy_\lambda} 
=  \int_{\Sigma_\lambda} [\mathbf{d}g \wedge \mathbf{d}f  + \dots ]
\otimes d^{\ps n}\ns x_0.      
\end{equation}
 Expressing the evolution vector field $X_  \lambda$ in the form 
$$
 X_  \lambda =  \frac{dg}{d\lambda} \frac{\delta}{\delta g}  + \dots 
$$
 and substituting into Hamilton's equations (\ref{eqn6E:16}), we see from
(\ref{eqn6E:18}) that Hamilton's equations place  no restriction on
$dg/d\lambda$.  Thus, the evolution of $g$ is completely arbitrary; i.e.,
$g$ is purely ``kinematic.''   Notice also from (\ref{eqn6E:18}) that
$$
\frac{\delta}{\delta g} =  X_{\ns f}  \in \ker \omega_\lambda \cap
\Fx(\cc_\lambda) ,
$$
 which shows that  ${\delta}/{\delta g}$ is a kinematic direction as
defined above.

This concludes our introduction to constraint theory.  In Part III we
will see how both the  initial value constraints and the gauge
transformations can be obtained ``all at once" from the 
energy-momentum map for the gauge group.

\startrule
\vskip-12pt
            \addcontentsline{toc}{subsection}{Examples}
\begin{examples}
\mbox{}\bigskip

{\bf a\ \  Particle Mechanics.}\quad
  We work out the details of the constraint algorithm for the 
relativistic free particle.  Now $\cp_\lambda 
\subset T^*\cy_\lambda$ is defined by the mass constraint
(\ref{6Cex:a4}):
$$
\Fh  =  g^{AB} \pi_A\pi_B + m^2  =  0 .
$$
 Then $\Fx(\cp_{\lambda})^{\bot} = \ker \omega_\lambda$ is spanned by the
$\omega_{T^*\ns\cy_\lambda}$-Hamiltonian vector field
\begin{equation}  \label{6Eex:a1} 
 X_  {\Fh}  =  2g^{AB} \pi_A \frac{\partial}{\partial q^B} - 
g^{AB}{}_{,C} \pi_A\pi_B
\frac{\partial}{\partial \pi_C}      
\end{equation}
 of the ``superhamiltonian'' $\Fh$.  For the Hamiltonian (\ref{6Cex:a5}), 
the consistency conditions (\ref{eqn6E:3b}) (cf. (\ref{eqn6E:4}) with
$l=1$) reduce to requiring that $X_  {\Fh}[H_{\lambda,\zeta}] = 0$.  A
computation gives
$$
 X_  {\Fh} [H_{\lambda, \zeta}]  =  (g^{AB}{}_{,C} \zeta^C - 2g^{AC}
\zeta^B{}_{,C}) \pi_A\pi_B = -2\zeta^{(A;B)} \pi_A\pi_B 
$$
 which vanishes by virtue of the fact that the slicing is
Lagrangian, so that $\zeta^A \partial/\partial q^A$ is a Killing
vector field, cf.  Example {\bf a} of \S6A.  Thus there are no
secondary constraints and so
$\cc_\lambda =
\cp_\lambda$.  The mass constraint is first class.

The most general evolution vector field satisfying Hamilton's equations
\linebreak
(\ref{eqn6E:16}) along  
$\cp_\lambda$ is $X_  \lambda = X + kX_  {\Fh}$, where $X$ is any particular solution
and $k \in \cf(\cc_\lambda) $ is arbitrary.  Explicitly, writing
$$
 X_  \lambda =  \left( \frac{dq^A}{d \lambda} \right) \frac{\partial}{\partial
q^A} + \left( \frac{d\pi_A} {d \lambda} \right) \frac{\partial}{\partial
\pi_A},
$$
 the space + time decomposed equations of motion take the form
\begin{equation}  \label{6Eex:a2} 
\begin{aligned}
\frac{dq^A}{d \lambda} & = - \zeta^A + 2kg^{AB} \pi_B \\[2ex]
\frac{d\pi_A} {d \lambda} & = \zeta^B{}_{,A} \pi_B - kg^{BC}{}_{,A}
\pi_B\pi_C.
\end{aligned}      
\end{equation}
 These equations appear complicated because we have written them
relative to an arbitrary (but
Lagrangian) slicing.  If we were to choose the standard slicing
$Y = Q \times \mathbb R$, then 
$\zeta^A = 0$ and (\ref{6Eex:a2}) are then clearly identifiable as the
geodesic equations on $(Q, g)$ with an arbitrary parametrization.

Since the equations of motion (\ref{6Eex:a2})
 for the relativistic free particle are \emph{ordinary} differential
equations, this example is  well-posed.

The gauge distribution $\Fx(\cp_\lambda)  \cap \Fx(\cp_\lambda) ^\bot$ is globally
generated by $X_  {\Fh}$.  The gauge  freedom of the relativistic free
particle is reflected in (\ref{6Eex:a2}) by the presence of the arbitrary
multiplier $k$, and obviously corresponds to time repara\-metrizations.
When $\zeta^A = 0$ the evolution is purely gauge, as predicted by Remark 9.
\bigskip

{\bf b\ \  Electromagnetism.}\quad
 Since $\mathfrak E^0=0$ is the only primary
constraint in Maxwell's theory, the polar $\Fx(\cp_{\lambda})^{\bot}$ is
spanned by $\delta/\delta A_0$.  From expression (\ref{6Cex:b8})
for the electromagnetic Hamiltonian, we compute that
$\delta H_{\lambda , \zeta}/\delta A_0=0$ iff
\begin{equation}  \label{6Eex:b1} 
\mathfrak E^i{}_{,i} =0,
\end{equation}
where we have performed an integration by parts. This is Gauss'
Law, and defines
$\cp_{\lambda,
\zeta}^2
\subset
\cp_{\lambda}$.  Continuing with the constraint algorithm, observe that
along with
${\delta}/{\delta A_0},\ \Fx(\cp_{\lambda ,\zeta}^2)^{\bot} $ is generated
by vector fields of the form $V=(D_{i}f){\delta}/{\delta A_i}$, where
$f :
\cp_{\lambda ,\zeta}^2 \rightarrow \cf(\Sig_\lambda) $ is arbitrary (cf.
(\ref{eqn5A:4})).  But then a computation gives
$$
 V[H_{\lambda ,\zeta}]= - \int_{\Sig_\lambda} \zeta^j f_{,j}
\mathfrak E^i{}_{,i}{}{}\,d^{\ps 3}\ns x_0
$$
which vanishes by virtue of (\ref{6Eex:b1}).
Thus the algorithm terminates
with $\cc_{\lambda} = \cp_{\lambda ,\zeta}^2$.  Note that $
\cc_{\lambda}$ is indeed independent of the choice of slicing generator
$\zeta$, as promised by Corollary~\ref{cor6.15}.  Moreover, it is obvious
from (\ref{6Eex:b1}) that $\Fx(\cc_{\lambda})^{\bot} \subset
\Fx(\cc_{\lambda})$ so
$\cc_{\lambda}$ is coisotropic and, in fact, all constraints are 
first class. Note, however, that $H_{\lambda,\zeta}\! \bigm| \!
\cc_\lambda \neq 0$ even though this theory is
parametrized; the reason is that the metric $g$ is not
variational.

Maxwell's equations in the canonical form (\ref{eqn6E:16}) are satisfied by
the vector field
$$
 X_  {\lambda}=\left(\frac{dA_0}{d\lambda}\right) \frac{\delta}{\delta A_0} +
\left(\frac{dA_i}{d\lambda}\right)\frac{\delta} {\delta A_i}+
\left(\frac{d\mathfrak E^i}{d\lambda}\right)
\frac{\delta}{\delta \mathfrak E^i} 
$$
 provided
\begin{align}
\frac{dA_i}{d\lambda} &= \zeta^0 N \gamma^{-1/2}\gamma_{ij}\mathfrak E^j
+ \frac{1}{N\sqrt{\gamma}}(\zeta^0 M^j + \zeta^j)\mathfrak F_{ji}
+(\zeta^\mu A_\mu-\chi)_{,i}
  \label{6Eex:b2} 
 \\
 \intertext{and}
\frac{d\mathfrak E^i}{d\lambda}& =
\Big( \zeta^0 \gamma^{ik}\gamma^{jm}
\mathfrak{F}_{km} +  \left[ (\zeta^0 M^i +
\zeta^i) \mathfrak{E}^j  - (\zeta^0M^j +
\zeta^j)\mathfrak{E}^i \right]
\Big)_{,j} .    
  \label{6Eex:b3} 
\end{align}

 Equation (\ref{6Eex:b2}) reproduces the definition (\ref{6Cex:b4}) of
the electric field density, while (\ref{6Eex:b3}) captures the dynamical
content of Maxwell's theory.  Note that $dA_0 / d{\lambda}$ is left
undetermined, in accord with the fact that ${\delta}/{\delta A_0}$  is a
kinematic direction.

Since the 4-dimensional form of the Maxwell equations in the Lorentz gauge
$A^\mu{}_{;\mu}=0$ reduce to wave equations for the $A^\nu$ (and hence are
hyperbolic), and the gauge itself satisfies the wave equation, this theory
is well-posed provided 
$\Sig_\lambda$ is spacelike.\footnote{\ In fact, to check well-posedness of
a theory with gauge freedom in a spacetime with closed Cauchy surfaces, it
is enough to verify this property in a particular gauge.} See Misner, Thorne,
and Wheeler [1973] and Wald [1984] for details here.

\medskip

On a Minkowskian background relative to the slicing (\ref{6Cex:b26}),
(\ref{6Eex:b2}) and  (\ref{6Eex:b3}) take their more familiar forms 
\begin{align}
\frac{dA_i}{d\lambda} &= \mathfrak E_i +A_{0,i} -\chi_{,i}      
  \label{6Eex:b24} 
 \\
 \intertext{and}
\frac{d\mathfrak E^i}{d\lambda}& = \Ff^{ij}{}_{,j}.      
  \label{6Eex:b25} 
\end{align}
\medskip

Of course, $X_  {\lambda}$ given by (\ref{6Eex:b2}) and  (\ref{6Eex:b3}) is
not uniquely fixed; one can add to it any vector field $V\in
\Fx(\cc_{\lambda})^{\bot}$. Such a $V$ has the form
\begin{equation*}  \label{6Eex:b4} 
 V=f_0\frac{\delta}{\delta A_0} + D_if\frac{\delta}{\delta A_i}    
\end{equation*}
 for arbitrary maps $f_0, f :
\cc_{\lambda} \rightarrow \cf(\Sig_\lambda) $.  The
first term in $V$ simply reiterates the fact that the evolution of $A_0$ is
arbitrary.  To understand the significance of the second term in $V$, it is
convenient to perform a transverse-longitudinal
 decomposition of the spatial $1$-form $\bold A =i_{\lambda}^*A$. (For
simplicity, we return to the case of a Minkowskian
background with the slicing (\ref{6Cex:b26}).) So split
$\bold A =\bold A_T +\bold A_L$, where $\bold A_T$ is divergence-free
and $\bold A_L$ is exact.  Then (\ref{6Eex:b24}) splits into two equations:
$$
\frac{d\bold A_T}{d{\lambda}} = \mathfrak E\quad\text{ and } \quad
\frac{d\bold A_L}{d{\lambda}}=\nabla A_0-\nabla\chi.
$$
 (Note that the electric field is transverse by virtue of
(\ref{6Eex:b1}).)  The effect of the second term in $V$ is to thus make
the evolution of the longitudinal piece $\bold A_L$ completely
arbitrary.  In summary, both the temporal and longitudinal components
$A_0$ and
$\bold A_L$ of the potential $A$ are gauge degrees of freedom whose
conjugate momenta are constrained to vanish, leaving the transverse
part $\bold A_T$ of $A$ and its conjugate momentum $\mathfrak E$ as the true
dynamical variables of the electromagnetic field.

\bigskip

{\bf c\ \  A Topological Field Theory.}\quad
From (\ref{6C39})  we have the instantaneous
primary constraint set 
$$
\mathcal P_\lambda =\left\{(A,\pi)\in T^*{{\mathcal Y}}_\lambda \mid
\pi^0=0 \:\mbox{ and }\: \pi^i  =  \epsilon^{0ij}A_j\right\}.
$$
It follows that $\Fx(\cp_{\lambda})^{\bot}$ is
spanned by the vector field ${\delta}/{\delta A^0}$.
With the Hamiltonian $H_{\lambda ,\zeta}$ given by (\ref{6C41}),
insisting that ${\delta}[H_{\lambda ,\zeta}]/{\delta A^0}=0$ produces
the spatial flatness condition (recall that $n=2$)
\begin{equation}  \label{6Eex:d2} 
F_{12}=0. 
\end{equation}
This equation defines  $\cp_{\lambda , \zeta}^2 \subset \cp_{\lambda}$.
Proceeding, we note that along with
${\delta}/{\delta A_0}$, $\Fx(\cp_{\lambda ,\zeta}^2)^{\bot} $ is generated
by vector fields of the form 
$$V=D_if\left( 
\epsilon^{0ij}\frac{\delta }{\delta \pi^j} - \frac{\delta }{\delta
A_i}\right),$$ 
where $f: \cp_{\lambda , \zeta}^2 \rightarrow \cf(\Sig_\lambda) $ is
arbitrary.  But then a computation gives 
$$
 V[H_{\lambda ,\zeta}]= \frac12 \int_{\Sig_\lambda} \epsilon^{0ij}\zeta^m
f_{,m}F_{ij}\,d^3x_0
$$
which vanishes in view of (\ref{6Eex:d2}).
Thus the constraint algorithm terminates
with $\cc_{\lambda} = \cp_{\lambda ,\zeta}^2$. 

Since $\Fx(\cc_\lambda)^\perp \subset \Fx(\cc_\lambda)$, $\cc_\lambda$ is
coisotropic in $\cp_\lambda$, whence the secondary constraint
(\ref{6Eex:d2}) is first class. The primary constraint $\pi^0=0$ is also
first class, while the remaining two primaries $\pi^i  - \epsilon^{0ij}A_j =
0$ are second class.

Next, suppose the vector field
$$
 X_  {\lambda}=\left(\frac{dA_0}{d\lambda}\right) \frac{\delta}{\delta A_0} +
\left(\frac{dA_i}{d\lambda}\right)\frac{\delta} {\delta A_i}+
\left(\frac{d\pi^i}{d\lambda}\right)
\frac{\delta}{\delta \pi^i} 
$$
satisfies the Chern--Simons equations in the Hamiltonian
form (\ref{eqn6E:16}). Then by (\ref{6C40}) we must have
\begin{equation}
\frac{dA_i}{d\lambda} = (\zeta^\mu A_\mu)_{,i},
  \label{6Eex:a} 
\end{equation}
and from (\ref{6C38}) we then derive
\begin{equation}
\frac{d\pi^i}{d\lambda} = \epsilon^{0ij}(\zeta^\mu A_\mu)_{,j}.      
  \label{6Eex:b} 
\end{equation}
 As in electromagnetism, ${\delta}/{\delta A_0}$  is a kinematic
direction with the consequence that  $dA_0 / d{\lambda}$ is left
undetermined. By subtracting $dA_i/d\lambda$ given by 
(\ref{6Eex:a}) from 
$$\dot A_i = \zeta^\mu A_{i,\mu}+A_\mu\zeta^\mu_{\;\; ,i}$$ obtained from
(\ref{eqn6B:2}) while taking (\ref{eq:ex6A:d2}) into account, we get 
$$\zeta^\mu(A_{\mu,i} - A_{i,\mu})=0$$
which, when combined with (\ref{6Eex:d2}), yields the remaining flatness
conditions
$F_{i0}=0$ in (3C.22). Equation (\ref{6Eex:b}) yields nothing
new. 

Finally, note that (i) when restricted to $\cc_\lambda$ the
Chern--Simons Hamiltonian (\ref{6C41}) vanishes by (\ref{6Eex:d2}), and
(ii) we may rearrange
$$
X_  \lambda = \left(\frac{dA_0}{d\lambda}\right)\frac{\delta} {\delta A_0} -
(\zeta^\mu A_\mu)_{,i}\left( 
\epsilon^{0ij}\frac{\delta }{\delta \pi^j} - \frac{\delta }{\delta
A_i}\right) \in \Fx(\cc
_\lambda)^\perp,
$$
so that the Chern--Simons evolution is completely gauge,
as must be the case for a parametrized field theory in which all fields
are variational.

One way to see that the Chern--Simons equations $F_{\mu \nu}=0$ make up a
well-posed system is to observe that 
if we make the gauge choices
$A_0=0$ and $\zeta_X = (1,{\bf 0})$, then the field equations  imply that
$\partial_0 A_\nu = 0$, which clearly determines a unique solution given
initial data consisting of $A_i$ satisfying $A_{[1,2]}=0$.

\bigskip

{\bf d\ \  Bosonic Strings.}\quad
 From (\ref{6Cex:e4}) and (\ref{6Cex:e8}) we see that
$\Fx(\cp_{\lambda})^{\bot}$ is spanned by the vector fields
${\delta}/{\delta h_{\sigma\ns\rho}}$ or, equivalently, 
${\delta}/{\delta h^{\sigma\ns\rho}}$.  Now demand that 
${\delta}[H_{\lambda ,\zeta}]/{\delta h^{\sigma\ns\rho}}=0$,  where
$H_{\lambda ,\zeta}$ is given by (\ref{6Cex:e7}).  For
$(\sigma,\rho) =(1,1)$, this yields
\begin{equation}  \label{6Eex:e1} 
\Fh=\pi^2 + \partial\varphi^2=0.      
\end{equation}
 Substituting this back into the Hamiltonian and setting $(\sigma ,
\rho)=(0,1)$, we get
\begin{equation}  \label{6Eex:e2} 
\Fj =\pi \cdot\partial\varphi =0.      
\end{equation}
 Setting $(\sigma , \rho)=(0,0)$ produces nothing new, so that
(\ref{6Eex:e1}) and (\ref{6Eex:e2}) are the only secondary constraints. 
Note that together they imply $H_{\lambda , \zeta}\! \bigm| \!\cp_{\lambda
,\zeta}^2 =0$, which of course reflects the fact that the bosonic string is
a param\-etr\-ized theory (and also that the slicing is a gauge slicing). 
As the notation suggests,
$\Fh$ and $\Fj$ are the analogues, for bosonic strings, of the
superhamiltonian and supermomentum, respectively, in ADM gravity.

For $N , M\in \cf (\Sigma_{\lambda})$, consider the Hamiltonian vector fields
\begin{equation}  \label{6Eex:e3} 
\begin{aligned} X_  {N\ns\Fh}&=2Ng^{AB}\pi_B \frac{\delta}{\delta\varphi^A}
+ 2g_{AB}
\partial(N\partial\varphi^B)
\frac{\delta}{\delta\pi_A}\\[2ex] X_  {M\Fj}&=M\partial\varphi^A
\frac{\delta}{\delta\varphi^A} +
\partial(M\pi_A)\frac{\delta}{\delta\pi_A}  
\end{aligned}      
\end{equation}
 of $N\Fh$ and $M\Fj$, respectively.  One verifies that $X_  {N\ns\Fh}$
and $X_  {M\Fj}$, together with the
${\delta}/{\delta h_{\sigma\ns\rho}}$, generate $\Fx(\cp_{\lambda , \zeta}^2)^{\bot} =
\Fx(\cp_{\lambda ,
\zeta}^2)^{\bot}\cap\Fx(\cp_{\lambda , \zeta}^2) \subset\Fx(\cp_{\lambda ,\zeta}^2)$.
Since in addition the Hamiltonian vanishes on $\cp_{\lambda , \zeta}^2$, it
follows that the constraint algorithm stops with $\cp_{\lambda , \zeta}^2
=\cc_{\lambda}$ and also that all constraints are first class.  

Writing the evolution vector field as
$$
 X_  {\lambda}=\left( \frac{d\varphi^A}{d\lambda} \right)
\frac{\delta}{\delta\varphi^A} +
\left( \frac{d\pi_A}{d\lambda} \right) \frac{\delta}{\delta\pi_A}
+\left(
\frac{dh_{\sigma\ns\rho}}{d\lambda} \right) \frac{\delta}{\delta
h_{\sigma\ns\rho}},
$$
Hamilton's
equations (\ref{eqn6E:16}) for the bosonic string are
\begin{align}
\frac{d\varphi^A}{d\lambda} 
&= -2Ng^{AB}\pi_B-M\partial\varphi^A      
  \label{6Eex:e4} 
\\[2ex]
\frac{d\pi_A}{d\lambda}
&= -2g_{AB}\partial
(N\partial\varphi^B)-\partial(M\pi_A).      
  \label{6Eex:e5} 
\end{align}
 Here the $dh_{\sigma\ns\rho} / d\lambda$ are undetermined, which is a
consequence of the fact that the
$h_{\sigma\ns\rho}$ are canonically conjugate to the first class primary
constraints $\rho^{\sigma\ns\rho}=0$, and hence are kinematic fields.

Since $X_  {\lambda} \in \Fx(\cc_{\lambda})\cap\Fx(\cc_{\lambda})^{\bot}$
the evolution is totally gauge.  The gauge transformations on the fields
$(\varphi^A,\pi_A)$ generated by the vector fields $X_  {N\ns\Fh}$ and
$X_  {M\Fj}$ for $N,M$ arbitrary express the invariance of the bosonic
string under diffeomorphisms of $X$ .  The complete indeterminacy of
the metric $h$ generated by the vector fields ${\delta}/{\delta
h_{\sigma\ns\rho}}$ is also a result of invariance under
diffeomorphisms---which in two dimensions implies that the conformal
factor is the only possible degree of freedom in $h$, cf. 
\S 3C.{\bf d}---coupled with conformal invariance---which implies that
even this degree of freedom is gauge. \hfill $\blacklozenge$
\renewcommand{\lozsymbol}{}
\end{examples}
\medskip

In our examples, we have encountered at most secondary constraints, and in
Example {\bf a} there were only primary constraints. This is 
typical: in mechanics it is rare to find (uncontrived) systems with
secondary constraints, and in field theories at most secondary
constraints are the rule. (Two exceptional cases are Palatini gravity,
which has tertiary constraints (see Part V), and the KdV equation,
which has only primary constraints (see Gotay [1988].) In principle,
however, the constraint chain (\ref{eqn6E:11}) can have arbitrary
length, but this has no physical significance.

\section{
The Energy-Momentum Map
}

In Chapter 4 we defined a covariant momentum mapping for a group $\cg$
of covariant canonical  transformations of the multisymplectic
manifold $Z$.  This chapter correlates those ideas with  momentum
mappings (in the usual  sense) on the presymplectic manifold
$\cz_\tau$ and the symplectic manifold
$T^*\cy_\tau$,  and  introduces the energy-momentum map on $\cz_\tau$.
We then show that this energy-momentum  map projects to a function
$\mathfrak E_\tau$ on the $\tau$-primary constraint set $\cp_\tau$, and that
under certain  circumstances, $\mathfrak E_\tau$ is identifiable with the
negative of the Hamiltonian. This is the key result  which enables us
in Part III to prove that the final constraint set for first class
theories coincides  with $\mathfrak E^{-1}_\tau(0)$,  when $\cg$ is the gauge
group of the theory. 
\bigskip

\subsection{
Induced Actions on Fields}

We first show how group actions on $Y$ and $Z$,  etc., can be extended
to actions on fields.  Given a left action of a group $\cg$ on  a
bundle $\pi_{X\ns K} : K \to X$ covering an  action  of   
$\cg$ on $X$,  we get an induced left action of $\cg$ on the space
$\ck$ of sections of $\pi_{X\ns K}$  defined by
\begin{equation}  \label{eqn7A:1}
\eta_\ck (\sigma)  =  \eta_K \circ \sigma \circ \eta^{-1}_X      
\end{equation}
 for $\eta \in \cg$ and $\sigma \in \ck$,  which generalizes the usual
push-forward operation on  tensor  fields.  The infinitesimal generator
$\xi_\ck (\sigma)$ of this action is simply the (negative of the) Lie
derivative:
\begin{equation}  \label{eqn7A:2} 
\xi_\ck (\sigma)  =  - \pounds_\xi \sigma =  \xi_K  \circ \sigma -
T\sigma \circ  \xi_X.      
\end{equation}

We consider the relationship between transformations of the spaces $Z$,
$\cz$, and $\cz_\tau$.   Let $\eta_Z : Z \to Z$ be a covariant canonical
transformation covering $\eta_X : X \to X$  with the induced
transformation $\eta_\cz : \cz \to \cz$ on fields given by
(\ref{eqn7A:1}).  For each $\tau \in {\rm Emb}(\Sig, X)$, $\eta_\cz$
restricts to the mapping
\begin{equation*}
\eta_{\cz_\tau} : \cz_\tau \to \cz_{\eta_X \circ \tau}      
  \label{eqn7A:4} 
\end{equation*}
 defined by
\begin{equation}  \label{eqn7A:5} 
\eta_{\cz_\tau}\! (\sigma)  =  \eta_Z \circ  \sigma \circ
\eta^{-1}_\tau,      
\end{equation}
 where $\eta_\tau := \eta_X \! \bigm| \!\Sigma_\tau$ is the induced
diffeomorphism from $\Sigma_\tau$ to
$\eta_X(\Sigma_\tau) $.

\begin{prop}\label{prop7.1}
$\eta_{\cz_\tau}$ is a canonical
transformation relative to the two-forms  
$\Omega_\tau$ and $\Omega_{\eta_X \circ \tau}$; that is,
$$
 (\eta_{\cz_\tau}\ns )^*\Omega_{\eta_X \circ \tau}  = \Omega_\tau.
$$
\end{prop}

\begin{proof} 
 From equation (\ref{eqn7A:5})
\begin{equation} \label{eqn7A:6}
T\eta_{\cz_\tau} \cdot V = T\eta_Z \circ (V \circ \eta^{-1}_\tau) 
\end{equation}
for $V \in T_\sigma \cz_\tau$.  Thus,
\begin{align}
 (\eta_{\cz_\tau}\ns )^*  
 & \Omega_{\eta_X \circ \tau} (V, W) 
       \nonumber\\[2ex] 
& = \Omega_{\eta_X \circ \tau} \left(T\eta_Z \cdot V \circ
\eta^{-1}_\tau, T\eta_Z \cdot W \circ \eta^{-1}_\tau\right) 
     \tag*{(by (\ref{eqn7A:6}))}  
\\[2ex] 
& = \int_{\eta_X(\Sigma_\tau) }(\eta_Z  \circ \sigma \circ
\eta^{-1}_\tau)^*(\mathbf{i}_{T\eta_Z \cdot W
	\circ \eta^{-1}_\tau} \mathbf{i}_{T\eta_Z 
 \cdot V\circ \eta^{-1}_\tau} \Omega) 
    \tag*{(by (\ref{eqn5C:3}))} 
\\[2ex] 
& = \int_{\eta_X(\Sigma_\tau) }
(\eta^{-1}_\tau) ^*[\sigma^*\eta^*_Z (\mathbf{i}_{T\eta_Z \cdot W}
	\mathbf{i}_{T\eta_Z \cdot V} \Omega)] 
       \nonumber\\[2ex] 
& = \int_{\Sigma_\tau} (\sigma^*\eta_Z{}^*)
(\mathbf{i}_{T\eta_Z \cdot W}
\mathbf{i}_{T\eta_Z \cdot V} \Omega) 
       \tag*{(change of variables formula)} 
\\[2ex] 
& = \int_{\Sigma_\tau} \sigma^*(\mathbf{i}_W
\mathbf{i}_V \eta_Z{}^* \Omega) 
      \tag*{(by naturality of pull-back)}
\\[2ex] 
& = \int_{\Sigma_\tau} 
\sigma^*(\mathbf{i}_W\mathbf{i}_V \Omega) 
      \tag*{(since $\eta$ is covariant canonical)} 
\\[2ex]  
& = \Omega_\tau(V, W).	
     \tag*{(by (\ref{eqn5C:3}))}
\end{align}
\end{proof}

Similarly, one shows the following:
 \begin{prop}\label{prop7.2}
If $\eta_\cz : \cz \to \cz$ is a special
covariant canonical transformation, then $\eta_{\cz_\tau}$ is a special
canonical transformation.
\end{prop}

\subsection{
The Energy-Momentum Map}

Let $\cg$ be a group acting by covariant canonical transformations on
$Z$ and let
$$
 J : Z \to \fg^* \otimes \Lambda^nZ
$$
 be a corresponding covariant momentum mapping.  This induces the map
$E_\tau : \cz_\tau \to  {\fg}^*$ defined by
\begin{equation}  \label{eqn7B:1} 
\langle E_\tau(\sigma), \xi\rangle  =  \int_{\Sigma_\tau}
\sigma^*\langle J,\xi\rangle      
\end{equation}
 where $\xi \in {\fg}$ and $\langle J,\xi\rangle : Z \to \Lambda^nZ$ is
defined by
$\langle J,\xi\rangle(z) := \langle J(z), \xi
\rangle$.  While $E_\tau$ is not  a momentum map in the usual sense on
$\cz_\tau$---since $\cg$ does not necessarily act on $\cz_\tau$---it will
be shown later to be closely related to the Hamiltonian in the
instantaneous formulation of classical field theory.  For this reason
we shall call $E_\tau$ the {\bfi energy-momentum map\/}.  Further
justification for this terminology is given in the interlude following
this chapter.

For actions on $Z$ lifted from actions on $Y$,  using adapted
coordinates and (4C.7),   (\ref{eqn7B:1})  becomes 
\begin{align*}\label{eqn7B:2} 
\langle E_\tau(\sigma), \xi \rangle	
& = \int_{\Sigma_\tau}
    \sigma^*\big((p_A{}^\mu  \xi^A 
    +  p\,\xi^\mu)\, d^{\ps n}\ns x_\mu - p_A{}^\mu
\xi^\nu dy^A \wedge d^{\ps n-1}\ns x_{\mu\nu} \big) \\[2ex]
	& = \int_{\Sigma_\tau}\big((p_A{}^0\xi^A + p\,\xi^0) \, d^{\ps n}\ns x_0 - 
p_A{}^\mu
\xi^\nu \sigma^A{}_{,i}\,\sigma^*(dx^i \wedge d^{\ps n-1}\ns x_{\mu\nu})
\big)
\end{align*}      
where the integrands are regarded as functions of $x^i$ and where we
write, in coordinates,
$\sig(x^i) = (x^i, \sigma^A(x^i), p(x^i), p_A{}^\mu (x^i))$.  Since
$$ 
dx^i \wedge d^{\ps n-1}\ns x_{\mu\nu} = \delta^i_\nu \, d^{\ps n}\ns x_\mu -
\delta^i_\mu \, d^{\ps n}\ns x_\nu,
$$ 
the  expression above can be written in the form
\begin{equation}
\langle E_\tau(\sigma), \xi \rangle 
 = \int_{\Sigma_\tau} \big(p_A{}^0(\xi^A
    - \xi^i\sigma^A{}_{,i}) + (p + p_A{}^i \sigma^A{}_{,i})\xi^0 
     \big)\, d^{\ps n}\ns x_0;
  \label{eqn7B:3} 
\end{equation} 
that is, 
\begin{equation}
\langle E_\tau(\sigma), \xi \rangle  = \int_{\Sigma_\tau}\big( p_A{}^0(\xi^A - \xi^\mu \sigma^A{}_{,\mu})
+ (p + p_A{}^\mu 
\sigma^A{}_{,\mu})\xi^0 \big) \, d^{\ps n}\ns x_0,      
  \label{eqn7B:4} 
\end{equation}
where (\ref{eqn7B:4}) is obtained from (\ref{eqn7B:3}) by adding
and subtracting the term
$\xi^0 p_A{}^0\sigma^A{}_{,0}$.   (For this to make sense, we suppose
that $\sig$ is the restriction to $\Sigma_\tau$ of a section of
$\pi_{X\ns Z}$. Of course, (\ref{eqn7B:4}) is independent of this choice of
extension.)

\subsection{
Induced Momentum Maps on $\cz_\tau$}

To obtain a bona fide momentum map on $\cz_\tau$,  we restrict attention
to the subgroup  
$\cg_\tau$ of $\cg$ consisting of transformations which stabilize the
image of $\tau$;  that is,
\begin{equation}  \label{eqn7C:1}
\cg_\tau :=  \{ \eta \in \cg \mid  \eta_X(\Sigma_\tau)  = \Sigma_\tau
\}.      
\end{equation}
 We emphasize that the condition $\eta_X(\Sigma_\tau)  = \Sigma_\tau$
within (\ref{eqn7C:1}) does not mean that each point of $\Sigma_\tau$ is
left fixed by
$\eta_X$, but rather that $\eta_X$ moves the whole Cauchy surface
$\Sigma_\tau$  onto  itself.  

For any $\eta \in \cg_\tau$, the map $\eta_\tau := \eta_X \! \bigm| \!
\Sigma_\tau$ is an element of $\operatorname{Diff}(\Sigma_\tau) $.   It
follows from Proposition~\ref{prop7.1} that
\begin{equation}  \label{eqn7C:2} 
\eta_{\cz_\tau}\! (\sigma)  =  \eta_Z \circ \sigma \circ
\eta^{-1}_\tau      
\end{equation}
 is a canonical action of $\cg_\tau$ on $\cz_\tau$.  From 
(\ref{eqn7A:2}), the infinitesimal generator of this action is 
\begin{equation}  \label{eqn7C:3}
\xi_{\cz_\tau}\! (\sigma)  = \xi_Z \circ \sigma - T\sigma
\circ
\xi_\tau,      
\end{equation}
 where $\xi_\tau$ generates $\eta_\tau$. 

Being a subgroup of $\cg$, $\cg_\tau$ has a covariant momentum map which
is given by $J $ followed by the projection from ${\fg}^* \otimes
\Lambda^nZ$ to ${\fg}^*_\tau \otimes 
\Lambda^nZ$, where ${\fg}_\tau$ is the Lie algebra of $\cg_\tau$.  Note
that in adapted coordinates, $\xi 
\in {\fg}_\tau$ when $\xi_X^0 = 0$ on $\Sigma_\tau$.  From (\ref{eqn7B:1}),
the map
$J$ induces the map $J_\tau := E_\tau \!\bigm| {\fg}_\tau : \cz_\tau \to
{\fg}^*_\tau$ given by  
\begin{equation}  \label{eqn7C:4} 
\langle J_\tau(\sigma), \xi \rangle = \int_{\Sigma_\tau} \sigma^*\langle
J,\xi\rangle      
\end{equation}
 for $\xi \in {\fg}_\tau$.

\begin{prop}\label{prop7.3}
$J_\tau$ is a momentum map for the
$\cg_\tau$-action on $\cz_\tau$ defined by 
{\rm  (\ref{eqn7C:2})}, and it is $\operatorname{Ad}^*$-equivariant if
$J$ is.
\end{prop}

\begin{proof} 
 Let $V \in T_\sigma \cz_\tau$ and let $v$ be a $\pi_{XZ}$-vertical
vector field on $Z$ such that 
$V = v  \circ  \sig$.  If $f_\lambda$ is the flow of $v$, let $\sig_\lambda =
f_\lambda \circ \sig$ so that the curve $\sig_\lambda \in \cz_\tau$ has tangent
vector $V$ at $\lambda = 0$.  Therefore, with $J_\tau$ defined by
(\ref{eqn7C:4}),
we have
$$
\langle \mathbf{i}_V\mathbf{d}J_\tau(\sigma), \xi\rangle = 
\frac{d}{d\lambda}
\left.\left[\int_{\Sigma_\tau}\sigma^*_\lambda \langle J,\xi\rangle
\right]\right|_{\lambda=0} =
\int_{\Sigma_\tau}\sigma^*\text{\pounds}_v \langle J,\xi\rangle .
$$
But
$$
\int_{\Sigma_\tau} \sigma^*\pounds_v\langle J,\xi\rangle =
\int_{\Sigma_\tau}
\sigma^*(\mathbf{d}\ps\mathbf{i}_v\langle J,\xi\rangle + \mathbf{i}_v
\mathbf{d}\langle J,\xi\rangle),
$$
 and since $\Sig$ is compact and boundaryless,
$$ 
\int_{\Sigma_\tau} \sigma^*(\mathbf{d}\ps \mathbf{i}_v\langle J,\xi\rangle) 
= 
\int_{\Sigma_\tau} \mathbf{d}\sigma^*(\mathbf{i}_v\langle J,\xi\rangle) = 0
$$
 by Stokes' theorem.  Therefore, by the definition (4C.3) of a covariant
momentum mapping,
\begin{equation}  \label{eqn7C:6} 
\langle \mathbf{i}_V \mathbf{d}J_\tau(\sigma), \xi\rangle =
\int_{\Sigma_\tau}
\sigma^*(\mathbf{i}_v\mathbf{d}\langle J,\xi\rangle) =
\int_{\Sigma_\tau} \sigma^*[\mathbf{i}_v\mathbf{i}_{\xi_Z} \Omega].      
\end{equation}
 Note that $\xi_Z$ need not be $\pi_{X\ns Z}$-vertical, so we cannot yet
use Lemma~\ref{lem5.1}.  

Now for any
$w \in T\Sigma_\tau$,  we have
$$
\sigma^*(\mathbf{i}_v\mathbf{i}_{T\sigma\cdot w}\Omega) = -
\sigma^*(\mathbf{i}_{T\sigma\cdot w}\mathbf{i}_v\Omega) = -
\mathbf{i}_W\sigma^*(\mathbf{i}_v\Omega) = 0  
$$
  by the naturality of pull-back and the fact that
$\sigma^*(\mathbf{i}_v\Omega)$ vanishes since it is an
$(n + 1)$-form on an $n$-manifold.  In particular, for $w =
\xi_\tau$,   we have
\begin{equation*}  \label{eqn7C:7} 
\sigma^*(\mathbf{i}_v\mathbf{i}_{T\sigma\cdot \xi_\tau} \Omega) = 0.      
\end{equation*}
 Combining this result with (\ref{eqn7C:6}) and using the fact that $\xi_Z - T\sigma
\cdot \xi_\tau$ is $\pi_{X\ns Z}$-vertical, we get
\begin{align*} 
\langle \mathbf{i}_V\mathbf{d}J_\tau(\sigma), \xi\rangle & =
\int_{\Sigma_\tau} \sigma^*(\mathbf{i}_v\mathbf{i}_{\xi_Z - T\sigma
\cdot \xi_\tau}
\Omega)
\\[2ex] & = \Omega_\tau(\xi_{\cz_\tau}, V)      
  \label{eqn7C:8} 
\end{align*}
by (\ref{eqn7C:3}) and (\ref{eqn5C:3}). Thus $J_\tau$ is a momentum map.

To show that $J_\tau$ is Ad$^*$-equivariant, we verify 
that for $\eta \in
\cg_\tau$ and $\xi \in  {\fg}_\tau$, $J_\tau$ satisfies the condition
\begin{equation*}  \label{eqn7C:9} 
\langle J_\tau(\sigma), {\rm Ad}_{\eta^{-1}}\xi\rangle = \langle
J_\tau(\eta_{\cz_\tau}\! (\sigma)), \xi\rangle.      
\end{equation*}
 However, from (\ref{eqn7C:4}) and (4C.4), we have
$$ 
\langle  J_\tau(\sigma), {\rm Ad}_{\eta^{-1}} \xi \rangle =
\int_{\Sigma_\tau} \sigma ^{\ast}  \langle J,{\rm Ad}_{\eta^{-1}} \xi \rangle
= 
\int_{\Sigma_\tau} \sigma ^{\ast} \eta_Z{} ^{\ast} \langle J, \xi \rangle; 
$$
 whereas from (\ref{eqn7C:2}), (\ref{eqn7C:4}), and the change of
variables formula, we get
\begin{align*}
\langle J_\tau(\eta_{\cz_\tau}\! (\sigma)), \xi\rangle & =
\int_{\Sigma_\tau} (\eta_Z \circ \sigma \circ \eta^{-1}_\tau)^*\langle
J,\xi\rangle \\[2ex] &= 
\int_{\Sigma_\tau} (\eta^{-1}_\tau)^*\sigma^*\eta_Z{}^*\langle J,\xi\rangle
\\[2ex] &=
\int_{\Sigma_\tau}\sigma^*\eta_Z{}^* \langle J,\xi\rangle,
\end{align*}
  thereby establishing the desired equality. 
\end{proof}

\subsection{
Induced Momentum Maps on
$T^*\cy_\tau$}

We now demonstrate how the group actions and momentum maps carry over
from the  multisymplectic context to the instantaneous  formalism. 
Recall that the  phase space  
$(T^*\cy_\tau, \omega_\tau) $ is the symplectic quotient of the presymplectic
manifold $(\cz_\tau, \Omega_\tau) $ by  the map $R_\tau$.  The key observation is
that both the action of $\cg_\tau$ and the momentum map $J_\tau$  pass to
the quotient.

First consider a canonical transformation $\eta_{\cz_\tau} : \cz_\tau \to 
\cz_\tau$. Define a map  
$\eta_{T^*\ns\cy_\tau} : T^*\cy_\tau \to  T^*\cy_\tau$ as follows:  For each
$\pi \in T^*_\varphi\cy_\tau$,  set
\begin{equation}  \label{eqn7D:1} 
\eta_{T^*\ns\cy_\tau}\!(\pi) = R_\tau(\eta_{\cz_\tau}\! (\sigma))      
\end{equation}
 where $\sig$ is any element of $R^{-1}_\tau(\{\pi\})$.
	
\begin{prop}\label{prop7.4}
The map $\eta_{T^*\ns\cy_\tau}$ is a canonical
transformation.
\end{prop}

\begin{proof} 
 To begin, we must show that $\eta_{T^*\ns\cy_\tau}$ is
well-defined; that is
$$
 R_\tau(\eta_{\cz_\tau}\! (\sigma)) = R_\tau(\eta_{\cz_\tau}\! (\sigma'))
\quad\text{ whenever }\quad \sigma, \sigma' \in R^{-1}_\tau(\{\pi\}).
$$
 Since $\eta_{\cz_\tau}$ is a canonical transformation, it preserves
the kernel of $\Omega_\tau$. But this kernel equals the kernel of $TR_\tau$
by Corollary~\ref{cor5.3}(ii). Therefore,
$\eta_{\cz_\tau}$ preserves the fibers of $R_\tau$, and so
$\eta_{T^*\ns\cy_\tau}$ is well defined.

Since $\eta_{\cz_\tau}$ is a diffeomorphism and $R_\tau$ is a submersion,
$\eta_{T^*\ns\cy_\tau}$ is a  diffeomorphism.  That the map
$\eta_{T^*\ns\cy_\tau}$ preserves the symplectic form $\omega_\tau$ is a
straightforward  computation using (\ref{eqn7D:1}), Corollary~\ref{cor5.3},
and the definitions. 
\end{proof}
	
This Proposition shows that the canonical action of $\cg_\tau$ on
$\cz_\tau$ gives rise to a  canonical action of $\cg_\tau$ on $T^*\cy_\tau$
such that $R_\tau$ is equivariant; that is, for $\eta \in 
\cg_\tau$, the following diagram commutes:
\begin{equation*}  \label{eqn7D:2} 
\begin{CD}
\cz_\tau @> R_{\tau} >> T^*\cy_\tau \\ @V \eta_{\cz_\tau} VV @VV
\eta_{T^*\ns\cy_\tau}V \\
\cz_\tau @>> R_{\tau} > T^*\cy_\tau %
\end{CD}       
\end{equation*}
 
Regarding momentum maps, we have:

\begin{prop}\label{prop7.5}
If $J_\tau$ is a momentum map for the action
of $\cg_\tau$ on $\cz_\tau$, then  
$\cj_\tau : T^*\cy_\tau \to {\fg}^*_\tau$ defined by the diagram 
\begin{equation}  \label{eqn7D:3} 
\begin{picture}	
(190,40)(-30,100) 	
\put(50,134){$\cz_\tau$}               
\put(45,70){$T^*\cy_\tau$}                
\put(86,102){$\fg_\tau^*$}               
\put(37,102){$R_\tau$}               
\put(76,125){$J_\tau$}              
\put(76,80){$\cj_\tau$}               
\put(55,125){\vector(0,-1){40}}   
\put(62,130){\vector(1,-1){20}} 
\put(62,80){\vector(1,1){20}}   
\end{picture} 
%
\end{equation}
\vskip 32pt
\noindent is a momentum map for the induced action of $\cg_\tau$ on
$T^*\cy_\tau$.  Further, if $J_\tau$ is  {\rm Ad}$^*$-equivariant, then so
is
$\cj_\tau$.
\end{prop}
	
\begin{proof} 
 This is a consequence of the facts that $R_\tau$ is
equivariant and $R_\tau{}^*\omega_\tau =
\Omega_\tau$. 
\end{proof}
	
We emphasize that {\it the momentum map\/} $\cj_\tau$, which we have
defined on $T^*\cy_\tau$,  {\it corresponds to the action of $\cg_\tau$
only\/}.  For the full group $\cg$, the corresponding energy-momentum map
does {\it not\/} pass from $\cz_\tau$ to $T^*\cy_\tau$.  However, as we
will see in
\S7F, the energy-momentum map $E_\tau$ does project to the
\emph{primary constraint submanifold} in  
$T^*\cy_\tau$.

\subsection{
Momentum Maps for Lifted Actions}

For lifted actions we are able to obtain explicit formulas for the
energy-momen\-tum and momentum maps on  
$\cz_\tau$ and $T^*\cy_\tau$ and the relationship between them. Suppose the
action of $\cg$ on $Z$ is  obtained by lifting an action of $\cg$ on
$Y$.  Then $\eta \in \cg$ maps $\cy_\tau$ to $\cy_{\eta_X \circ \tau}$
according to
\begin{equation}\label{eqn7D:0}
\eta_{\cy_\tau}\! (\varphi) = \eta_Y \circ \varphi \circ \eta^{-1}_\tau 
\end{equation}
where $\eta_\tau = \eta_X \! \bigm| \! \Sigma_\tau$. This in turn
restricts to an action of $\cg_\tau$ on
$\cy_\tau$ given by the same formula, with the infinitesimal generator 
\begin{equation}  \label{eqn7E:1} 
\xi_{\cy_\tau}\! (\varphi) = \xi_Y \circ \varphi - T\varphi \circ
\xi_\tau      
\end{equation}
 where $\xi_\tau = \xi_X \! \bigm|\! \Sigma_\tau$.
 
\begin{cor}\label{7.6}
For actions lifted from $Y$: 
\begin{enumerate}
\renewcommand{\labelenumi}{\mbox{\rm(\roman{enumi})}}
\item 
 The energy-momentum map on $\cz_\tau$ is
\begin{equation}  \label{eqn7E:2}
\langle E_\tau(\sigma), \xi \rangle = \int_{\Sigma_\tau} \varphi^*(\mathbf{i}_{\xi_Y}\sigma)      
\end{equation}
 where $\xi \in {\fg}$, and $\varphi = \pi_{Y\! Z} \circ \sig$.  

\item 
 The induced $\cg_\tau$-action on $T^*\cy_\tau$ given by 
{\rm  (\ref{eqn7D:1})} is the usual cotangent  action; that is,  
\begin{equation*}  \label{eqn7E:3} 
\eta_{T^*\ns\cy_\tau}\!(\pi) = (\eta^{-1}_{\cy_\tau})^*\pi .      
\end{equation*}

\item 
 The corresponding induced momentum map $\cj_\tau$ on
$T^*\cy_\tau$ defined by 
{\rm  (\ref{eqn7D:3})} is the standard one; that is, 
\begin{equation}  \label{eqn7E:4}
\langle \cj_\tau(\varphi, \pi), \xi \rangle =  \langle \pi,
\xi_{\cy_\tau}\! (\varphi)\rangle =
\int_{\Sigma_\tau} \pi \left(\xi_{\cy_\tau}\! (\varphi)\right)      
\end{equation}
 for $\xi \in {\fg}_\tau$.  Moreover, the momentum maps $J, J_\tau$, and
$\cj_\tau$ are all    {\rm Ad}$^*$-equivariant. 
\end{enumerate}
\end{cor}

\begin{proof} 
 To prove (i), substitute
formula (4C.6) into (\ref{eqn7B:1}) and note that
\begin{equation}  \label{eqn7E:5} 
\sigma^*\langle J,\xi\rangle = \sigma^*\pi_{Y\! Z}{}^*
\mathbf{i}_{\xi_Y} \sigma = \varphi^*\mathbf{i}_{\xi_Y} \sig.      
\end{equation}

	 To prove (ii) let $\eta \in \cg_\tau$, $\pi
= R_\tau(\sigma) \in T^*_\varphi
\cy_\tau$ and $V \in T_{\eta_{\cy_{\tau}}\ns(\varphi)}\cy_\tau$.  Then
\begin{align}
 &\langle \eta_{T^*\!\cy_\tau}\ns (\pi), V\rangle 
             \nonumber\\[2ex] 
 &\quad =
\langle R_\tau(\eta_{\cz_\tau}\! (\sigma)), V\rangle 
            \tag*{(by (\ref{eqn7D:1}))} \\[2ex] 
&\quad = \int_{\Sigma_\tau}
(\eta_{\cy_\tau}\! (\varphi))^*[\mathbf{i}_V(\eta_{\cz_\tau}\! (\sigma))]
            \tag*{(by (\ref{eqn5D:1}))} \\[2ex]
&\quad = \int_{\Sigma_\tau} 
(\eta^{-1}_\tau) ^* \varphi^*
\eta_Y{}^*[\mathbf{i}_V(\eta_{\cz_\tau}\! (\sigma))] 
            \tag*{(by (\ref{eqn7D:0}))}\\[2ex] 
&\quad = \int_{\Sigma_\tau} \varphi^*
\eta_Y{}^*[\mathbf{i}_V(\eta_{\cz_\tau}\! (\sigma))] 
            \tag*{(by the change of variables formula)} \\[2ex] 
&\quad =
\int_{\Sigma_\tau} \varphi^*[\mathbf{i}_{T\eta^{-1}_Y\cdot V}
\eta_Y{}^*(\eta_{\cz_\tau}\! (\sigma))] \nonumber\\[2ex] 
&\quad  = \int_{\Sigma_\tau}
\varphi^*[\mathbf{i}_{T\eta^{-1}_Y\cdot V} \sig] 
            \tag*{(by (4B.3))} \\[2ex] 
&\quad = \langle R_\tau(\sigma), {T\eta^{-1}_Y\cdot V}\rangle
            \tag*{(by (\ref{eqn5D:1}))} \\[2ex] 
&\quad = \langle \pi, {T\eta^{-1}_Y\cdot V}\rangle \nonumber \\[2ex]
&\quad = \langle (\eta^{-1}_Y)^* \pi, V\rangle.     \nonumber
\end{align}

To prove (iii)  we compute, taking into account (\ref{eqn7D:3}),
(\ref{eqn7E:2}), and (\ref{eqn7E:1}),
$$
\begin{aligned}
\langle \cj_\tau(R_\tau(\sigma)), \xi\rangle & = \langle J_\tau(\sigma),
\xi\rangle =
\int_{\Sigma_\tau} \varphi^*(\mathbf{i}_{\xi_Y}\sig)	\\[2ex] & =
\int_{\Sigma_\tau}
\varphi^*(\mathbf{i}_{\xi_Y - T\varphi \cdot \xi_\tau} \sig)
=\langle R_\tau(\sigma),\xi_{\cy_\tau}\! (\varphi)\rangle,  
\end{aligned}
$$
 where we have used
$$
\varphi^*\mathbf{i}_{T\varphi\cdot \xi_\tau}\sigma = \mathbf{i}_{\xi_\tau}\varphi^*\sigma = 0
$$
 since $\varphi^*\sig$ is an $(n + 1)$-form on the $n$-manifold
$\Sigma_\tau$.

Finally, equivariance follows from  Propositions 4.3, \ref{prop7.3} and
\ref{prop7.5}. 
\end{proof}

\subsection{
The Hamiltonian and the Energy-Momentum Map}

In \S7B we defined the energy-momentum map $E_\tau$ on $\cz_\tau$.  Here we
show that for  lifted actions, $E_\tau$ projects to a well-defined
function
$$
\mathcal E_\tau : \cp_\tau \to {\fg}^* 
$$
 on the $\tau$-primary constraint set, which we refer to as the
``instantaneous energy-momentum map."   This is the central object for
our later analysis.

Let the group $\cg$ act on $Y$ and consider the lifted action of $\cg$
on $Z$. Using (4C.5) rewrite  formula (\ref{eqn7B:1}) as
\begin{equation*}  \label{eqn7F:1} 
\langle E_\tau(\sigma), \xi \rangle = \int_{\Sigma_\tau}
\langle \Fe_\tau(\sigma), \xi \rangle      
\end{equation*}
 for $\sigma \in \cz_\tau$ and $\xi \in {\fg}$, where
\begin{equation}  \label{eqn7F:2}
\langle \Fe_\tau(\sigma), \xi \rangle =
\sigma^*(\mathbf{i}_{\xi_Z}\Theta)      
\end{equation}
 defines the {\bfi energy-momentum density\/} $\Fe_\tau$.  

While $\Fe_\tau$ does not directly factor through the reduction map to give
an instantaneous energy-momentum density on $T^*\cy_\tau$, we 
nonetheless have:
\begin{prop}\label{prop7.7}
The energy-momentum density $\Fe_\tau$ 
induces an instantaneous energy-momentum density on $\cp_\tau
\subset T^*\cy_\tau$.
\end{prop}
\begin{proof} 
Given any
$(\varphi,\pi) \in \cp_\tau$, let $\sig$ be a holonomic lift of $(\varphi,
\pi)$ to $\cn_\tau$ (cf. \S6C). We claim that for any $x \in
\Sig_\tau$ and $\xi \in \fg$, the quantity
$$
\langle \Fe_\tau(\sig)(x), \xi \rangle \in
\Lambda^n_x\Sigma_\tau 
$$
depends only upon $j^1\ns\varphi(x)$ and $\pi(x)$.
Thus, setting
\begin{equation}  \label{eqn7F:4} 
\langle \Fe_\tau(\varphi, \pi)(x), \xi \rangle  =  \langle
\Fe_\tau(\sig)(x), \xi \rangle
  \end{equation}
defines the instantaneous energy-momentum density (which we denote
by the same symbol $\Fe_\tau$) on
$\cp_\tau$.

%

If $\xi_X(x)$ is transverse to
$\Sigma_\tau$, then (\ref{eqn7F:2}) combined with
(\ref{eqn6C:11}) gives
\begin{equation}  \label{eqn7F:5} 
 \langle \Fe_\tau(\sig)(x), \xi \rangle =  - \Fh_{\tau,\xi}(\varphi,
\pi)(x).     
\end{equation}
 On the other hand, if $\xi_X(x) \in T_x\Sigma_\tau$, 
%
then from (\ref{eqn7E:5}) we compute
\begin{align}  \label{eqn7F:6}
 \langle \Fe_\tau(\sig)(x), \xi \rangle  =
\varphi^*(\mathbf{i}_{\xi_Y(\varphi(x))} \sig(x))  =
\varphi^*(\mathbf{i}_{\xi_Y(\varphi(x)) - T_x\varphi \cdot\xi_X(x)}
\sig(x))
\end{align}
 where we have used the same `trick' as in the
proof of Corollary~\ref{7.6}(iii). Since $\xi_Y - T\varphi
\cdot\xi_X$ is
$\pi_{XY}$-vertical, we can now apply (\ref{eqn5D:2}) to obtain  
\begin{align}  \label{eqn7F:6.5}
 \langle \Fe_\tau(\sig)(x), \xi \rangle & =
\big\langle  R_\tau(\sig)(x),{\xi_Y(\varphi(x)) - T_x\varphi
\cdot\xi_X(x)}\big\rangle \nonumber
\\[1.5ex] 
& = \langle  \pi(x),\xi_{\cy_\tau}\! (\varphi)(x)\rangle. 
\end{align}

In either case, $\langle \Fe_\tau(\sig)(x), \xi \rangle$
depends only upon the values of
$\varphi$, its first derivatives, and $\pi$ along $\Sigma_\tau$.  Thus the
definition (\ref{eqn7F:4}) is meaningful for any $\xi \in
{\fg}$. 
\end{proof}

Integrating (\ref{eqn7F:4}), we get the {\bfi instantaneous
energy-momentum map\/}
$\mathcal E_\tau : \cp_\tau \to {\fg}^*$ defined by
\begin{equation}  \label{eqn7F:7} 
\langle \mathcal E_\tau(\sig), \xi \rangle =
\int_{\Sigma_\tau}
\langle\Fe_{\tau}(\varphi, \pi),\xi\rangle.      
\end{equation}
 Two cases warrant special attention:

\begin{cor}\label{cor7.8}
Let $\xi \in {\fg}$.
\begin{enumerate}
\renewcommand{\labelenumi}{\mbox{\rm(\roman{enumi})}}
\item 
 If $\xi_X$ is everywhere transverse to $\Sigma_\tau$, then
\begin{equation}  \label{eqn7F:8} 
\langle \mathcal E_\tau(\varphi, \pi), \xi \rangle  =  - H_{\tau, \xi}(\varphi,
\pi)      
\end{equation}

\item 
 If $\xi_X$ is everywhere tangent to $\Sigma_\tau$, then
\begin{equation}  \label{eqn7F:9} 
\langle \mathcal E_\tau(\varphi, \pi), \xi \rangle  =  \langle \cj_\tau(\varphi,
\pi), \xi \rangle.      
\end{equation}
\end{enumerate}
\end{cor}

\begin{proof} 
Assertion (i) 
follows from (\ref{eqn7F:5}) and 
(ii) 
is a
consequence of (\ref{eqn7F:6.5}) and  (\ref{eqn7E:4}). 
\end{proof}

\begin{remarks}[Remarks \ 1.]
In general, $\mathcal E_\tau$ is defined only on the primary
constraint set $\cp_\tau$, as $H_{\tau,\xi}$ is.  However, if $\cg =
\cg_\tau$, then
$\mathcal E_\tau = \cj_\tau$ is defined on all of $T^*\cy_\tau$. (It was not
necessary that
$\sig$ be a \emph{holonomic} lift for the proof of the second part of
Proposition~\ref{prop7.7}, corresponding to the case when $\xi_X(x)
\in T_x\Sigma_\tau.$)

\paragraph{\bf 2.} 
Although the instantaneous energy-momentum map can be
identified with the Hamiltonian  (when $\xi_X \pitchfork \Sigma_\tau$) and
the momentum map $\cj_\tau$ for $\cg_\tau$ (when $\xi_X \,\|\:
\Sigma_\tau) $, it is important to realize that $\langle \mathcal
E_\tau(\varphi, \pi), \xi
\rangle$ is defined for any $\xi \in  {\fg}$, regardless of whether or not
it is everywhere transverse or tangent to $\Sigma_\tau$.

\paragraph{\bf 3.}  
The relation (\ref{eqn7F:8}) between the instantaneous energy-momentum
map and the Hamiltonian is only asserted to be  valid in 
the context of lifted actions; for more general actions,
we do not claim such a relationship.   Fortunately, in
most examples, lifted actions are the appropriate ones to 
consider. 
\end{remarks}

The instantaneous energy-momentum map $\mathcal E_\tau$ on $\cp_\tau$ is the
cornerstone of our  work since, via (\ref{eqn7F:8}) above,
it constitutes the fundamental link between dynamics and
the gauge  group.  From it we will be able to correlate the
notion of ``gauge transformation" as arising from  the
gauge group action with that in the Dirac--Bergmann theory of
constraints.  This in turn will make it  possible to
``recover" the first class initial value constraints from
$\mathcal E_\tau$ because, according to
\S6E, they are the generators of gauge transformations.

\begin{remark}[Remark \ 4.]
Indeed, in Chapter 11 we will show that for
parametrized theories in which all fields are
variational, the final constraint set
$\cc_\tau
\subset
\mathcal E_\tau^{-1}(0).$
Combining this with the relation (\ref{eqn7F:8}), we see
that for such theories the Hamiltonian (defined relative
to a $\mathcal G$-slicing) must vanish ``on shell;'' that
is, $ H_{\tau, \xi} \!\bigm| \! \cc_\tau = 0$ as predicted in Remark 9 in
\S6E.
\end{remark}

Thus, in some sense, the energy-momentum map encodes in a single
geometric object  virtually {\it all\/} of the physically relevant
information about a given classical field theory: its dynamics, its
initial value constraints and its gauge freedom.  Momentarily, in
Interlude~II, we will see that $\mathcal E_\tau$ also incorporates the
stress-energy-momentum tensor of a theory.  It is  these properties of
$\mathcal E_\tau$ that will eventually enable us to achieve our main goal;
viz., to write the evolution equations in adjoint form.

\startrule
\vskip-12pt

            \addcontentsline{toc}{subsection}{Examples}
\begin{examples}
\mbox{}\bigskip

{\bf a\ \; Particle Mechanics.}\enspace 
  If $\cg = \operatorname{Diff}(\mathbb R)$
acts on $Y = \mathbb R \times Q$ by time 
repara\-metrizations, then from (4C.9) the
energy-momentum map on $\cz_t=\mathbb R \times T^*Q$ is 
$$
\langle E_t(p,q^1,\cdots , q^N,p_1,\cdots , p_N),\chi\rangle=p\chi(t).
$$
 But $p=0$ on $\cn$ by virtue of the time reparametrization-invariance 
of $\cl$, cf. example {\bf a} in \S 4D.  Thus
the instantaneous energy momentum map on
$\cp_t=R_t(\cn_t)$  vanishes. 
The subgroup $\cg_t$ consists of those diffeomorphisms
which fix
$\tau(\Sig) = t
\in \mathbb R$.  However,  the actions of $\cg_t$ on $\cz_t$ and on
$T^*\cy_t = T^*Q$ are trivial.

If $\cg = {\rm Diff}(\mathbb R) \times G$, where
$G$ acts only on the factor 
$Q$,  then $\cg_t  = G$.  In this case, $\cj_t$ reduces
to the usual momentum map on $T^*Q$.  
\bigskip 

{\bf b\ \; Electromagnetism.}\enspace 
 For electromagnetism on a fixed
background with $\cg = \cf(X)$, we find from  (4C.12) and
(\ref{eqn7B:1}) that in adapted coordinates,
\begin{equation*}   \label{7Fex:b1} 
\langle E_\tau(A, p, \Ff),\chi\rangle =
\int_{\Sigma_\tau} \Ff^{\nu 0}\chi{}_{,\nu}\,
d^{\ps 3}\ns x_0        
\end{equation*}
for $\chi \in \cf(X)$.  Now $\cg =  \cg_\tau$, so in this case $\cj_\tau$
and $\mathcal E_\tau$ coincide.  Using the expression above for $E_\tau$,
(\ref{eqn7D:3}), and $\Fe^\nu =
\Ff^{\nu 0}$,  we get
\begin{equation} 
\langle \cj_\tau(A, \Fe),\chi \rangle = \int_{\Sigma_\tau} \Fe^\nu \chi{}_{,\nu}
\,d^{\ps 3}\ns x_0
  \label{7Fex:b2} 
\end{equation}
on $T^*\cy_\tau$. Note that this agrees with formula (\ref{eqn7E:4}).  When
restricted to the primary constraint set $\cp_\tau \subset T^*\cy_\tau$
given by $\Fe^0 = 0$,   (\ref{7Fex:b2}) becomes
\begin{equation}   \label{7Fex:b3} 
\langle \ce_\tau(A, \Fe), \chi \rangle = \int_{\Sigma_\tau} \Fe^i
\chi{}_{,i} \,d^{\ps 3}\ns x_0.        
\end{equation}
\medskip

In the parametrized case, when $\cg = {\rm Diff}(X) \;\circledS \; \cf(X)$,
$E_\tau$ is replaced by $\tilde E_\tau$ where, with the help of
(4C.17),
\begin{align*}
\langle {\tilde E_\tau}(A,& \, p, \Ff;g),  (\xi,\chi)\rangle 
\\[2ex] & = \int_{\Sigma_\tau}  \big({\mathfrak{F}}^{\nu
0}(-A_{\mu}\xi^{\mu}_{\:\:\: ,\nu}  - A_{\nu,\mu} \xi^{\mu} +
\chi_{,\nu})  + (p+\mathfrak
F^{\mu\nu}A_{\mu,\nu})\xi^{0}\big)\, d^{\ps 3}\ns x_0. 
\label{7Fex:b4}
\end{align*}
Since elements of $\cg_\tau$ preserve $\Sig_\tau$, each $(\xi,\chi) \in
\tilde \fg_\tau$ satisfies $\xi^0 \!\bigm|\! \Sig_\tau =
0.$ Then
$\tilde E_\tau$ projects to the momentum map 
\begin{equation}
\langle {\tilde \cj_\tau}(A, \Fe;g),  (\xi,\chi)\rangle 
 = \int_{\Sigma_\tau} {\mathfrak{E}}^{\nu}
(-A_{\mu}\xi^{\mu}_{\:\:\: ,\nu}  - A_{\nu,i} \xi^{i} +
\chi_{,\nu})\, d^{\ps 3}\ns x_0 
\label{7Fex:b4}
\end{equation}
for the action of $\tilde \cg_\tau$ on $T^*\cy_\tau$.

On $\cp_\tau$, $\tilde E_\tau$  induces the instantaneous
energy-momentum map
\begin{align*}  
\langle {\tilde \ce_\tau}(A, & \, \Fe;g),(\xi,\chi)\rangle \\[2ex] & = 
\int_{\Sigma_\tau}  \left({\mathfrak{E}}^{i}(-A_{\mu}\xi^{\mu}_{\:\:\: ,i} 
- A_{i,\mu} \xi^{\mu} + \chi_{,i})   - \frac{1}{4}\mathfrak
F^{\mu\nu} F_{\mu\nu})\xi^{0}\right)d^{\ps 3}\ns x_0, 
\end{align*}
where we have used (3C.14). Adding and subtracting $-\Fe^iA_{\mu,i}\xi^\mu$
to the integrand and rearranging yields
\begin{align*}  
\int_{\Sigma_\tau} 
\left({\mathfrak{E}}^{i}(\chi - A_{\mu}\xi^{\mu})_{,i} 
 + \Fe^iF_{ij}\xi^j + \left(\frac{1}{2}\Fe^iF_{i0} -
\frac{1}{4}\mathfrak F^{ij}
F_{ij}\right)\xi^{0}\right)d^{\ps 3}\ns x_0. 
\end{align*}
Using (\ref{6Cex:b2b}) and (\ref{6Cex:b4}) to express $F_{i0}$ in terms of
$\Fe^i$ and $\Ff_{ij}$, this  eventually gives
\begin{align} 
\langle {\tilde \ce_\tau}(A, \Fe;& \, g),(\xi,\chi)\rangle 
= \nonumber \\[2ex]
\int_{\Sigma_\tau}
 \bigg[ & (\xi^\mu A_\mu - \chi)_{,i} \mathfrak E^i -
\frac{1}{N\sqrt{\gamma}}(\xi^0 M^i +
\xi^i)\mathfrak E^j\mathfrak F_{ij} 
\nonumber \\[2ex] &
 - \xi^0N\gamma^{-1/2}\Big(\frac{1}{2} \gamma_{ij}\mathfrak E^i
\mathfrak E^j + \frac{1}{4N^2} \gamma^{ik} \gamma^{jm}\mathfrak F_{ij}
\mathfrak F_{km}\Big) 
\bigg] d^{\ps 3 }\ns x_0
\end{align}
where we have again made use of the splitting (\ref{eqn6B9})--(\ref{eqn6B11}) 
of the metric
$g$.

\paragraph{\bf c\ \; A Topological Field Theory.}\enspace 
Since the Chern--Simons Lagrangian density is not
equivariant with respect to the $\cg = {\rm Diff}(X) \,
\circledS \; \cf(X)$ action, we 
are not guaranteed that our theory as developed above will apply. So we must
proceed by hand.

On $\cz_\tau$ the multimomentum map (4C.17) induces the map
\begin{align*}
\langle {E_\tau}(\sigma), & (\xi,\chi)\rangle 
\\[2ex] & = \int_{\Sigma_\tau}  \left(p^{\nu
0}(-A_{\mu}\xi^{\mu}_{\:\:\: ,\nu}  - A_{\nu,\mu} \xi^{\mu} +
\chi_{,\nu})  + (p+p^{\mu\nu}A_{\mu,\nu})\xi^{0}\right)
d^{\ps 2}\ns x_0. 
\label{7Fex:d2}
\end{align*}
Now  $E_\tau$ projects to the genuine
momentum map 
\begin{equation}
\langle {\ci_\tau}(A, \pi),  (\xi,\chi)\rangle 
 = \int_{\Sigma_\tau}  {\pi}^{\nu}
(-A_{\mu}\xi^{\mu}_{\:\:\: ,\nu}  - A_{\nu,i} \xi^{i} +
\chi_{,\nu})\, d^{\ps 2}\ns x_0 
\label{7Fex:d3}
\end{equation}
on $T^*\cy_\tau$. Similarly, from (3C.19), one verifies that $E_\tau$
projects to the ``ersatz'' instantaneous energy-momentum
map
\begin{align}
\langle {\ce_\tau}(A), & \, (\xi,\chi)\rangle \nonumber
\\[2ex] &
 = \int_{\Sigma_\tau}  \left( \eps^{0ij}A_j
(-A_{\mu}\xi^{\mu}_{\:\:\: ,i}  - A_{i,\mu} \xi^{\mu} +
\chi_{,i})\right. 
 + \left.\eps^{\mu\nu\rho}A_\rho A_{\nu,\mu}\xi^0\right)
d^{\ps 2}\ns x_0 \nonumber \\[2ex] &
 = \int_{\Sigma_\tau}  \eps^{0ij}\left( A_j
(\chi -A_{\mu}\xi^{\mu})_{,i}  + A_{j}F_{ik} \xi^{k} +
\frac{1}{2}A_0 F_{ij}\xi^0\right) d^{\ps 2}\ns x_0
\label{7Fex:d4}
\end{align}	 
on $\cp_\tau.$

Not surprisingly, $\langle \ce_\tau,(\xi,\chi)\rangle$ fails to coincide
with the Chern--Simons Hamiltonian (\ref{6C41}) (when
$\xi_X$ is transverse to
$\Sig_\tau$) because of the term involving $\chi$. Nonetheless, an
integration by parts shows that they agree on the \emph{final} constraint
set, cf. (\ref{6Eex:d2}). Indeed, the extra term in $\ce_\tau$ amounts to
adding  the first class constraint $F_{12} = 0$ to the Hamiltonian with
Lagrange multiplier $\chi$, and this is certainly permissible according to the
discussion at the end of \S6E. From a slightly different point of view,
since the action of $\cf(X)$ on $J^1Y$ leaves the Lagrangian density
invariant up to a divergence, its action on $T\cy_\tau$ will leave the
instantaneous Lagrangian (\ref{eqn6C:c20}) invariant. In fact,
(\ref{7Fex:d3}) is just the momentum map for this action (compare
(\ref{7Fex:b2})).


Alternately, we could proceed by simply dropping the $\cf(X)$-action. The
above formul\ae\ remain valid, provided the terms involving $\chi$ are
removed. In this context (\ref{7Fex:d4}) will now of course be a genuine
energy-momentum map.

\paragraph{\bf d\ \; Bosonic Strings.}\enspace 
 For the bosonic string, (4C.26)
eventually leads to the expression
\begin{align}
\langle E_{\tau}(\sigma), (\xi,\lambda) \rangle 
     =\int_{\Sigma_{\tau}}
     &\left(-p_A{}^0\varphi^A{}_{,\mu}\xi^{\mu}\right.
\nonumber\\[2ex]  
&+q^{\sigma\ns\rho \ps 0}
(2\lambda h_{\sigma\ns\rho} -h_{\sigma\nu}\xi^{\nu}{}_{,\rho} 
     -h_{\rho \nu} \xi^{\nu}{}_{,\sigma}
     -h_{\sigma\ns\rho,\nu}\xi^{\nu})
\nonumber\\[2ex]  
&\left. +\,(p +p_A{}^\mu\varphi^A{}_{,\mu}
+q^{\sigma\ns\rho \mu} h_{\sigma\ns\rho,\mu} )\xi^0 \right)
d^{\ps 1}\ns x_0         
  \label{7Fex:e1} 
\end{align}
 for the energy-momentum map on $\cz_\tau$.

Restricting to the subgroup $\cg_{\tau}$, (\ref{7Fex:e1}) reduces to 
\begin{align}
\langle\cj_{\tau}  (\varphi,h,& \, \pi ,\rho), (\xi,\lambda) \rangle=
\nonumber\\[2ex] 
& \int_{\Sigma_{\tau}} \left(-(\pi\cdot\partial\varphi)\xi^1 +
2\lambda\rho^{\sigma}{}_\sigma\right.
\left. -2\rho^{\sigma}{}_\rho \xi^\rho {}_{,\sigma}
-\rho^{\sigma\ns\rho}h_{\sigma\ns\rho,1}\xi^1\right)
d^{\ps 1}\ns x_0        
  \label{7Fex:e2} 
\end{align}
 on $T^*\cy_\tau$, where we have used $h$ to lower the index on $\rho$.

Finally, making use of (3C.24)--(3C.26) and (\ref{eqn6B9})--(\ref{eqn6B11})
in (\ref{7Fex:e1}), we compute on
$\cp_\tau$ 
\begin{align}\label{7Fex:e3} 
\langle \ce_{\tau}&(\varphi,h,\pi),(\xi,\lambda) \rangle  \nonumber
\\[2ex] & =\int_{\Sigma_{\tau}}
\left(\frac12 |h|^{-1/2}\frac{1}{h^{00}} \xi^0 (\pi^2
+\partial\varphi^2)\right.
\left.+\biggl(\frac{h^{01}}{h^{00}}
\xi^0-\xi^1\biggr) (\pi\cdot\partial\varphi) \right) d^{\ps 1}\ns x_0
\nonumber
\\[2ex] & =
- \int_{\Sigma_\tau}\left(\frac{1}{2\sqrt{\gamma}} \zeta^0 N(\pi^2
+\partial\varphi^2) + (\zeta^0M +
\zeta^1)(\pi\cdot\partial\varphi)\right) d^{\ps 1}\ns x_0.
\end{align}
When $\xi = (1,{\bf 0})$, this 
reduces to
\begin{align*}\label{7Fex:e3}
\langle \ce_{\tau}(\varphi,h,\pi),((1,{\bf 0}),\lambda) \rangle  = 
- \int_{\Sigma_\tau}\left(\frac{1}{2\sqrt{\gamma}}  N(\pi^2
+\partial\varphi^2) + M (\pi\cdot\partial\varphi)\right) d^{\ps 1}\ns
x_0
\end{align*}
from which one can read off the string {\bfi superhamiltonian\/ } 
$$\Fh = \frac{1}{2\sqrt{\gamma}}  (\pi^2
+\partial\varphi^2)$$ and the string {\bfi supermomentum\/} 
$$\Fj = \pi\cdot\partial\varphi.$$
Thus as claimed in the introduction to Part I we have $\ce = -(\Fh,\Fj)$,
that is, the superhamiltonian and supermomentum are the components of the
instantaneous \emph{energy-momentum} map. The supermomentum by itself
is a component of the \emph{momentum} map $\cj_\tau$ 
for the group
$\cg_\tau$ which does act in the instantaneous formalism, unlike $\cg$. 
\hfill $\blacklozenge$
\end{examples}

\newpage


\pagestyle{myheadings}
\section*{References}
\markboth{References}{References}
\addcontentsline{toc}{section}{References}

\begin{description}

\item[]
 R. Abraham and J. Marsden [1978],
{\it Foundations of Mechanics\/},
Second Edition,
Addison-Wesley,
Menlo Park, California.

\item[]
 J. L. Anderson [1967],
{\it Principles of Relativity Physics\/},
Academic Press,
New York.

\item[]
 J. M. Arms, M. J. Gotay, and G. Jennings [1990],
Geometric and algebraic reduction for singular
momentum mappings,
{\em Adv. in Math.\/},
{\bf  79},
43--103.

\item[]
 R. Arnowitt, S. Deser, and C. W. Misner [1962],
The dynamics of general relativity,
{\it  Gravitation, an Introduction to Current Research\/}
(L. Witten, ed.),
227--265,
Wiley, New York.

\item[]
 A. Ashtekar, L. Bombelli, and O. Reula [1991],
The covariant phase space of asymptotically flat gravitational fields,
 {\it Mechanics, Analysis and Geometry: 200 Years After Lagrange\/}
(M. Francaviglia, ed.),
417--450,
North-Holland,  Amsterdam.

\item[]
 D. Bao, Y. Choquet--Bruhat, J. Isenberg, and P. Yasskin [1985],
The well-posedness of $(N=1)$ classical supergravity,
{\em J. Math. Phys.\/},
{\bf  26},
329--333.


\item[]
 F. J. Belinfante [1940],
On the current and the density of the electric charge, the energy, the linear
momentum and the angular momentum of arbitrary fields,
{\em Physica\/},
{\bf  vii},
449--474.

\item[]
 P. R. Chernoff and J. E. Marsden [1974],
Properties of infinite dimensional Hamiltonian systems,
{\em Lecture Notes in Math.\/},
{\bf  425},
Springer-Verlag,
New York.

\item[]
 Y. Choquet--Bruhat  [1962],
The Cauchy problem,
{\it  Gravitation, an Introduction to Current Research\/}
(L. Witten, ed.),
130--168,
Wiley, New York.
%

\item[]
 Y. Choquet--Bruhat, A. E. Fischer, and J. E. Marsden  [1979],
Maximal hypersurfaces and positivity
of mass,
{\it  Isolated Gravitating Systems and General Relativity\/}
(J.~Ehlers, ed.),
322--395,
Italian Physical Society.

\item[]
\v{C}. Crnkovi\'{c} and E. Witten [1987],
Covariant description of
canonical formalism in geometrical theories,
{\it Newton's Tercentenary Volume\/}
(S.W. Hawking and W. Israel eds.),
666--684,
Cambridge University Press, Cambridge.

\item[]
 P. A. M. Dirac [1964],
{\it Lectures on Quantum Mechanics\/},
Academic Press,
New York.

\item[]
 D. G. Ebin and J. E. Marsden [1970],
Groups of diffeomorphisms and the motion of an
incompressible fluid,
{\em Ann. Math.\/},
{\bf  92},
102--163.

\item[]
 A. E. Fischer and J. E. Marsden [1979a],
Topics in the dynamics of general relativity,
{\it  Isolated Gravitating Systems in General Relativity\/} 
(J.~Ehlers, ed.),
Italian Physical Society,
322--395.

\item[]
A. E.  Fischer and J. E. Marsden [1979b],
The initial value problem and the dynamical formulation
of general relativity,
{\it  General Relativity\/}
(S.W. Hawking and W. Israel, eds.),
138--211,
Cambridge Univ. Press, Cambridge.

\item[]
 M. J. Gotay [1979],
{\it Presymplectic Manifolds, Geometric Constraint Theory and the 
Dirac--Bergmann Theory of Constraints,\/} Thesis,
University of Maryland, Technical Report 80--063.

\item[]
 M. J. Gotay [1983],
On the validity of Dirac's conjecture regarding first class secondary
constraints,
{\em J.~Phys.~A: Math. Gen.\/},
{\bf  16},
L141--145.

\item[]
 M. J. Gotay [1988],
A multisymplectic approach to the KdV equation,
 {\it Differential Geometric Methods in Theoretical Physics\/}
(K. Bleuler and M. Werner eds.),
NATO Advanced Science Institutes Series C: Mathematical and Physical Sciences,
{\bf 250},
295--305,
Kluwer,
Dordrecht.

\item[]
 M. J. Gotay [1991], 
 A multisymplectic framework for classical field theory and the calculus of
variations II. Space + time decompostion,
{\em Diff. Geom. Appl.\/},
{\bf  1},
375--390.

\item[]
 M. J. Gotay, R. Lashof, J. \'Sniatycki, and A. Weinstein [1983],
Closed forms on symplectic fibre
bundles,
{\em Comment. Math. Helvetici\/},
{\bf  58},
617--621.

\item[]
 M. J. Gotay and J. E. Marsden [1992],
Stress-energy-momentum tensors and the Belinfante--Rosenfeld formula,
{\em Contemp. Math.\/},
{\bf 132},
367--391.

\item[]
 M. J. Gotay and J. M. Nester [1979],
Presymplectic geometry, gauge transformations and the
Dirac theory of constraints,
{\it Lecture Notes in Physics\/},
{\bf 94},
272--279,
Springer-Verlag,
New York.

\item[]
 M. J. Gotay and J. M. Nester [1980],
Generalized constraint algorithm and special presymplectic
manifolds,
{\em Lecture Notes in Math.\/},
{\bf 775},
78--104,
Springer-Verlag,
New York.

\item[]
 M. J. Gotay, J. M. Nester, and G. Hinds [1978],
Presymplectic manifolds and the Dirac--Bergmann
theory of constraints,
{\em J.~Math. Phys.\/},
{\bf  19},
2388--2399.

\item[]
 S. W. Hawking and G. F. R. Ellis  [1973],
{\it The Large Scale Structure of Space-time\/},
Cambridge University Press, Cambridge.

\item[]
 G. Horowitz [1989],
Exactly soluble diffeomorphism invariant theories,
{\em Commun. Math. Phys.\/},
{\bf  129},
417--437.

\item[]
 T. Hughes, T. Kato, and J. Marsden  [1977],
Well-posed quasi-linear second-order hyperbolic
systems with applications to nonlinear elastodynamics and general relativity,
{\em Arch. Rat. Mech. Anal.\/},
{\bf  63},
273--294.

\item[]
 J. Isenberg and J. Nester [1980],
Canonical analysis of relativistic field theories,
{\it  General Relativity and Gravitation, Vol. I\/}
(A. Held ed.),
Plenum,
New York.


\item[]
 L. Lusanna [1991],
The second Noether theorem as the basis of the theory of singular
Lagrangians and Hamiltonian constraints,
{\em Riv. Nuovo Cimento},
{\bf 14}(3),
1--75.

\item[]
 J. E. Marsden and T. J. R. Hughes [1983],
{\it Mathematical Foundations of Elasticity\/},
Prentice--Hall,
Redwood City, California.

\item[]
 C. W. Misner, K. Thorne, and J. A. Wheeler [1973],
{\it Gravitation\/},
W. H. Freeman,
San Francisco.

\item[]
 R. Palais [1968],
{\it Foundations of Global Nonlinear Analysis\/},
Addison--Wesley,
Reading, Massachusetts.

\item[]
 T. Regge and C. Teitelboim  [1974], 
Role of surface integrals in the Hamiltonian formulation of
general relativity,
{\em Ann. Phys.\/},
{\bf  88},
286--318.

\item[]
 L. Rosenfeld [1940],
Sur le tenseur d'impulsion-\'energie,
{\em M\'em. Acad. Roy. Belg. Sci.\/},
{\bf  18},
1--30.

\item[]
 J. C. Simo and J. E. Marsden [1984],
On the rotated stress tensor and a material version of the
Doyle--Ericksen formula,
{\em Arch. Rat. Mech. Anal.\/},
{\bf  86},
213--231.

\item[]
 R. Sjamaar and E. Lerman [1991],
Stratified symplectic spaces and reduction,
{\em Ann. Math.\/}
{\bf 134},
375--422.

\item[]
 J. \'Sniatycki [1988],
Conservation laws in asymptotically flat spacetimes revisted,
{\em Rep. Math. Phys.\/},
{\bf  25},
127--140.

\item[]
 K. Sundermeyer [1982],
Constrained dynamics,
{\em Lecture Notes in Physics\/},
{\bf  169},
Springer-Verlag,
New York.


\item[]
R. M. Wald [1984],
{\it General Relativity,}
University of Chicago Press, Chicago.

\item[]
G. J. Zuckerman [1987],
Action principles and global geometry,
{\it Mathematical Aspects of String Theory,}
(S. T. Yau, ed.),
{\em Adv. Ser. Math. Phys.}, 
{\bf 1},
259--284,
World Scientific, Singapore.

\end{description}



\newcommand{\indsp}{\null\phantom{2\enspace}}

\newpage
\section*{\Large\bfseries Table of Contents for Parts I and III--V}
\addcontentsline{toc}{section}{Table of Contents for Parts I, III and IV}
\markboth{Table of Contents}{ Parts I, III and IV}

\bigskip
\section*{\large\bfseries I---Covariant Field Theory
}

\begin{enumerate}
\medskip
\item[\bf 1\quad] {\bf Introduction}	
     
\medskip
\item[\bf 2\quad] {\bf Multisymplectic Manifolds}
\\	
      2A\quad  The Jet Bundle\\	
      2B\quad  The Dual Jet Bundle\\	
      6C\quad  The Instantaneous Legendre Transform\\	
      6D\quad  Hamiltonian Dynamics\\	
      6E\quad  Constraint Theory

\medskip
\item[\bf 3\quad] {\bf Lagrangian Dynamics}
\\	
      3A\quad  The Covariant Legendre Transformation\\	
      3B\quad  The Cartan Form\\	
      3C\quad  The Euler--Lagrange Equations	
     
\medskip
\item[\bf 4\quad] {\bf Covariant Momentum Maps and Noether's Theorem}
\\	
      4A\quad  Jet Prolongations\\	
      4B\quad  Covariant Canonical Transformations\\	
      4C\quad  Covariant Momentum Maps\\	
      4D\quad  Symmetries and Noether's Theorem

\end{enumerate}

\medskip
\noindent{\bf Interlude I---On Classical Field Theory}
%
%
%
%
%

\section*{\large\bfseries 
III---Gauge Symmetries and Initial Value Constraints}

\noindent{\bf Interlude II---The Stress-Energy-Momentum Tensor}
\begin{enumerate}
\medskip
\item[\bf 8\quad] {\bf The Gauge Group}
\\	
      8A\quad  Principal Bundle Construction of the Gauge Group\\	
      8B\quad  Covariance, Localizability, and Gauge Groups\\	
      8C\quad  Gauge Transformations

\medskip
\item[\bf 9\quad] {\bf The Vanishing Theorem and Its Converse}
\\	
      9A\quad  Flexibility\\	
      9B\quad  The Vanishing Theorem\\	
      9C\quad  The Converse of the Vanishing Theorem

\medskip
\item[\bf 10\quad] {\bf Primary Constraints and the Momentum Map}
\\	
      10A\quad  The Foliation $\dot\cg_\tau$\\	
      10B\quad  The Primary Constraint Set Lies in the Zero Level of the \newline 
                \phantom{\null 10B\quad }Momentum Map\\	
      10C\quad  First Class Primary Constraints

\medskip
\item[\bf 11\quad] {\bf Secondary Constraints and the Energy-Momentum Map}
\\	
      11A\quad  The Final Constraint Set Lies in the Zero Level of the\newline 
                \phantom{\null 11A\quad }Energy-Momentum Map\\	
      11B\quad  First Class Secondary Constraints
\end{enumerate}

\medskip
\noindent{\bf Interlude III---Singularities in Solution Spaces of\newline 
         \phantom{Interlude III---}Classical Relativistic Field Theories}


\section*{\large\bfseries IV---The Adjoint Formalism  }

\begin{enumerate}
\medskip
\item[\bf 12\quad] {\bf The Dynamic and Atlas Fields}
\\	
      12A\quad  The Dynamic Bundle\\	
      12B\quad  Bundle Considerations\\	
      12C\quad  The Atlas Bundle

\medskip
\item[\bf 13\quad] {\bf The Adjoint Formalism}
\\	
      13A\quad  Linearity of the Hamiltonian\\	
      13B\quad  Model Bundles \\	
      13C\quad  The Adjoint Form, Reconstruction and Decomposition
\end{enumerate}

\medskip
\noindent{\bf Conclusions}


\section*{\large\bfseries V--- Palatini Gravity}

\begin{enumerate}
\medskip
\item[\bf 14\quad] {\bf  Application to Palatini Gravity}
\\	
      14A\quad  Covariant Analysis\\	
      14B\quad  Canonical Analysis\\	
      14C\quad  Energy-Momentum Map Analysis\\
      14D\quad  The Adjoint Formalism
\end{enumerate}



\end{document}